\documentclass[useAMS,usenatbib]{mn2e}
\usepackage{graphicx,natbib,amsmath,subfig,tabularx,float,rotating,amssymb}
\usepackage[T1]{fontenc} 
\usepackage{aecompl}
\usepackage[usenames,dvipsnames]{xcolor}
\input journals.def
\title[The initial conditions of observed star clusters
  I.]  {The initial conditions of observed star clusters - I. Method
  description and validation} 
\author[J.T. Pijloo et al.]
          {J.T.~Pijloo$^{1,2}$, S.F.~Portegies Zwart$^2$,
            P.E.R. Alexander$^3$, M. Gieles$^4$, S.S.~Larsen$^1$, \and
            P.J.~Groot$^1$, and B. Devecchi$^5$ \\ $^1$Department of
            Astrophysics/IMAPP, Radboud University, PO Box 9010, 6500
            GL Nijmegen, the Netherlands\\ $^2$Leiden Observatory,
            Leiden University, P.O. Box 9513, 2300 RA Leiden, the
            Netherlands\\ $^3$Institute of Astronomy, University of
            Cambridge, Madingley Road, Cambridge CB3
            0HA\\ $^4$Department of Physics, University of Surrey,
            Guildford GU2 7XH, UK\\ $^5$TNO Defence, Security and
            Safety\\} \date{Released 2014 Xxxxx XX}

\pagerange{\pageref{firstpage}--\pageref{lastpage}} \pubyear{2014}

\def\LaTeX{L\kern-.36em\raise.3ex\hbox{a}\kern-.15em
    T\kern-.1667em\lower.7ex\hbox{E}\kern-.125emX}

\begin{document}

\label{firstpage}

\maketitle

\begin{abstract}
We have coupled a fast, parametrized star cluster evolution code to a
Markov Chain Monte Carlo code to determine the distribution of
probable initial conditions of observed star clusters, which may serve
as a starting point for future $N$-body calculations. In this paper we
validate our method by applying it to a set of star clusters which
have been studied in detail numerically with $N$-body simulations and
Monte Carlo methods: the Galactic globular clusters M4, 47
Tucanae, NGC 6397, M22, $\omega$ Centauri, Palomar
14 and Palomar 4, the Galactic open cluster M67, and the M31 globular
cluster G1.\\ For each cluster we derive a distribution of initial
conditions that, after evolution up to the cluster's current age,
evolves to the currently observed conditions. We find that there is a
connection between the morphology of the distribution of initial
conditions and the dynamical age of a cluster and that a degeneracy in
the initial half-mass radius towards small radii is present for
clusters which have undergone a core collapse during their
evolution. We find that the results of our method are in agreement
with $N$-body and Monte Carlo studies for the majority of clusters. We
conclude that our method is able to find reliable posteriors for the
determined initial mass and half-mass radius for observed star
clusters, and thus forms an suitable starting point for modeling an
observed cluster\rq{}s evolution.
\end{abstract}


\section{INTRODUCTION}

What were the initial conditions of the star clusters we observe
today? Answering this question not only requires accurate observations
of the current conditions, but also proper modeling of star cluster
evolution over a large amount of time (for a globular cluster
typically 12\,Gyr), taking into account a number of physical
processes, such as two-body relaxation, three- and four body
interactions, the stellar evolution of single and binary stars and the
effects of a galactic tidal field \citep{1938UCHZAP.22..19A,
  1942psd..book.....C, 1958AJ.....63..109K, 1992PASP..104..981H,
  1960AnAp...23..668H, 1987ApJ...322..123L}. The required complexity
of the simulation depends on the type of cluster one wishes to model
and the physics question to be answered. In the Milky Way Galaxy we
know 157 globular clusters \citep{2010arXiv1012.3224H}, and about 2200
Galactic star clusters in the disc. Of the latter,
$\sim$2170 are relatively low-mass ($< 10^4\,M_{\odot}$) systems which
we traditionally classify as open clusters
(\citealt{2002A&A...389..871D} version 3.3 - jan/10/2013), while
$\sim$30 have masses and luminosities comparable to the type of
clusters that are typically found in studies of external galaxies,
classified as young massive ($> 10^4 M_{\odot}$) clusters
\citep*{2010ARA&A..48..431P}. The obvious differences between these
three cluster types lie in the age, the number of stars, the
(half-mass) density and the amount of gas currently present in the
cluster. Each factor complicates the simulation of the cluster
evolution, if its value is high. However, at the basic level, each of
these star clusters is subject to the same physical processes
\citep{2002IAUS..207..421L} and will eventually meet the same fate of
complete dissolution \citep{1973VA.....15...13A,2006astro.ph..5125B}.

\subsection{Star cluster physics}\label{subsec:physics}

Star clusters form in the collapse of giant molecular clouds (see
e.g. \citealt{2001Natur.409..159A,2014arXiv1402.6196M}). After the
complex phases of cluster formation and early evolution
with residual gas expulsion, cluster expansion and
  re-virialisation (see e.g. \citet{2007MNRAS.380.1589B},
  \citet{2012MNRAS.420.1503P}, \citet{2013ApJ...764...29B} or
  \citet{2014arXiv1401.4175L}), the star cluster is often assumed to
evolve as a roughly spherically symmetric gas-less system under the
influence of the following physical processes:
\begin{enumerate}
 \item the dynamics of the stars on both the large scale (two-body
   relaxation) and on the small scale (three- and four-body
   interactions, binary formation and evolution)
   \citep{1938UCHZAP.22..19A, 1942psd..book.....C,
     1958AJ.....63..109K};
 \item the stellar evolution of single stars and binaries
   \citep{1992PASP..104..981H};
 \item the interaction with the tidal field of the galaxy it resides
   in \citep{1960AnAp...23..668H, 1987ApJ...322..123L}.
\end{enumerate}
Two-body relaxation tends to drive the cluster to the unreachable
state of equipartition, where the kinetic energy of the stars in the
cluster is equalized, see e.g. \citet{2007MNRAS.374..703K} and the
references therein. This has two major effects: 1) mass segregation:
the most massive stars will sink to the cluster center, whereas the
lower-mass stars populate the halo; 2)
core-collapse: due to the decrease of kinetic energy
in the core, gravitational collapse is no longer supported. Hence the
core will shrink until high enough core densities are reached to
produce a new source of kinetic energy: binaries
\citep{1973VA.....15...13A}. The binaries will halt the process of
core-collapse and start the core expansion by
interacting with less massive stars, in which the former become more
tightly bound (\lq{}hard\rq{}) and the latter escape into the halo of
the cluster \citep{1973VA.....15...13A}. The half-mass radius will
increase driven by stellar evolution and hard binaries
\citep{2010MNRAS.408L..16G}.  Since the low-mass components
transferred to the halo have higher kinetic energy, this process will
also cause the preferential loss of low-mass stars. The escape of
stars over the tidal radius leads to a contraction of the cluster, i.e. the decrease
of the half-mass radius, at a (roughly) constant density
\citep{1961AnAp...24..369H,2011arXiv1103.3499H,2011MNRAS.413.2509G}. If, on the other hand, the cluster has a significant amount of primordial binaries ($\gtrsim 10\%$), either the core-collapse would be less deep in the sense that the core radius would not decrease as much as it would without the presense of primordial binaries, or there would be no core-collapse at all, see e.g. \citet{2003ApJ...582L..21B}.

\subsection{Simulation techniques}\label{subsec:techniques}

In computational astronomy great progress has been made to develop
dedicated codes to study the evolution of star
clusters. These dedicated codes can roughly be
divided in three groups \citep{2012MNRAS.422.3415A}. The first group
of methods are the $N$-body simulations. Direct $N$-body simulations
(e.g. \citealt{1973VA.....15...13A}; \citealt{1996ApJ...471..796M};
\citealt{1999JCoAM.109..407S}) have the smallest number of simplifying
assumptions. The time complexity of these simulations scales with
$N^2$, where $N$ is the number of stars in the simulation, and
solutions up to arbitrary precision can be obtained, see
\citet{2014ApJ...785L...3P} and \citet{2014arXiv1411.6671B}. The $N^2$
time complexity makes these simulations computationally
expensive. Globular clusters contain tens of thousands to several
millions of stars at the present day, and theoretical motivations
point out that they must have had an even greater initial number of
stars \citep{1973VA.....15...13A}. Because of this large initial
number of stars a direct $N$-body simulation of the evolution of a
million body globular cluster over its entire lifetime is yet to be
performed. Direct $N$-body simulations with fewer stars that are
scaled up afterwards to match the observed clusters, have been made
successfully, and also the direct $N$-body integration of small
globular clusters, or of open clusters, has been done with success,
see e.g. \citet{2003ApJ...589L..25B}, \citet{2011MNRAS.411.1989Z},
\citet{2014MNRAS.440.3172Z} and \citet{2005MNRAS.363..293H} for
$N$-body simulations of the evolution of G1, Palomar 14, Palomar 4 and
M67, respectively. For the less accurate, but more efficient,
tree-codes the computation time scales with $N$$\log$($N$); see
\citet{2012EPJST.210..201B} for a review of the history of the
collisional direct and collision-less tree-code $N$-body
methods.\\ \\ Other methods, such as Monte Carlo methods
(\citet{Meteopolis:1949}, \citet{2001A&A...375..711F},
\citet{2006MNRAS.371..484G} and \citet{2008MNRAS.388..429G}) and
Fokker Planck models (\citet{Kolmogoroff1931} and
\citet{1979ApJ...234.1036C}), scale both with $N$ and make some
simplifying assumptions such that the computational cost decreases,
but at the price of the decrease of the accuracy. With this second
group of methods, the computation of the evolution of globular
clusters with initially more than 10$^6$ stars has become possible,
but can still take a substantial amount of time and computer
power. Therefore the choice of initial conditions is very important
and a lot of effort is usually put into choosing a plausible set of
initial conditions that will evolve to a cluster with characteristics
similar to the observed cluster of interest. See
e.g. \citet{2009MNRAS.395.1173G} where a number of small scale runs,
i.e. simulations for clusters with a lower initial number of stars,
are performed to this end.\\ \\ The third group of methods are the
gaseous models \citep{1970MNRAS.147..323L} and the semi-analytical
methods, where the computation time is approximately
$N$-independent. The semi-analytical methods make use of parametric
equations for the evolution of the cluster variables, such as the
number of stars, $N$, and the half-mass radius, $r_{\rm hm}$, which
are then solved with a numerical integrator. These methods are
therefore faster again. One such code is the recently developed Evolve
Me A Cluster of StarS ({\sc emacss}) code, which is based on the flow
of energy in a cluster and allows one to compute the evolution of a
globular cluster over 12 Gyr in a fraction of a second
\citep{2012MNRAS.422.3415A,2014MNRAS.437..916G,2014MNRAS.442.1265A}. This
code has been tested against $N$-body simulations and has proven to be
an extremely powerful tool in understanding cluster evolution and
exploring large regions in initial parameter space.

\subsection{Goal of this work}\label{subsec:goal}

We aim to determine the most probable sets of initial conditions in
total cluster mass, $M$, and half-mass radius, $r_{\rm hm}$, for any
observed star cluster and explore possible degeneracies.  Finding the
initial conditions of individual star clusters has been done before,
see e.g. \citet{2003MNRAS.339..486G}, \citet{2003ApJ...589L..25B},
\citet{2005MNRAS.363..293H}, \citet{2008MNRAS.389.1858H},
\citet{2009MNRAS.395.1173G}, \citet{2011MNRAS.410.2698G},
\citet{2011MNRAS.411.1989Z}, \citet{2014arXiv1401.3657H} and
\citet{2014MNRAS.440.3172Z}. These studies used elaborate simulation
techniques to model cluster evolution and found suitable initial
conditions that, after evolving up to the cluster\rq{}s age, resembled
a number of observables of their star cluster of interest. However,
due to the versatility of their methods, these studies were only able
to investigate up to several tens of sets of initial
conditions iteratively. They could not investigate the uniqueness of
their sets of initial conditions or explore whether there are
multiple, significantly different,
sets of initial conditions which evolve to the
current observables, i.e. whether there are degeneracies. In this
study we aim to address this latter point as
well. We accomplish this by coupling the fast,
parametrized star cluster evolution code {\sc emacss}
\citep{2014MNRAS.442.1265A} to the Markov Chain Monte Carlo (MCMC)
code {\sffamily emcee} \citep{2013PASP..125..306F}.\\ \\ In this paper
we describe and demonstrate our method. We validate it by applying it
to nine star clusters that have been studied to great extent with
either $N$-body simulations or Monte Carlo methods and by comparing
our results to the results of those methods. The paper is organized as
follows: in Section~\ref{sec:Method} we explain our method and we
summarize the functionality of the underlying star cluster evolution
code. In Section~\ref{sec:Validation} we describe our validation
strategy and set out the relevant parts of the extensive work that has
been done on the validation clusters by other authors. In
Section~\ref{sec:Results} we show our results and we discuss them in
detail. Section~\ref{sec:Summary} summarizes the paper and discusses
future work.

\section{METHOD}\label{sec:Method}

\subsection{The parametrized star cluster evolution code}\label{subsec:EMACSS}

We evolve the star clusters using the parametrized star cluster
evolution code Evolve Me A Cluster of StarS ({\sc
  emacss})
(\citet{2012MNRAS.422.3415A,2014MNRAS.437..916G,2014MNRAS.442.1265A}). {\sc
  emacss} includes a prescription for mass segregation, the evolution
of the mean stellar mass, $\tilde{m} = M/N$, as the result of stellar
evolution, the resulting expansion of the cluster and the escape of
stars over the tidal radius, $r_{\rm t}$. After a
phase of \lq\lq{}unbalanced\rq\rq{} evolution, in which the evolution
and escape of stars are the dominant drivers of the evolution, a phase
of \lq\lq{}balanced\rq\rq{} evolution starts
\citep{2014MNRAS.442.1265A}. Here it is assumed that the core produces
the correct amount of energy to sustain the two-body relaxation
process. Balanced evolution is assumed to start after a fixed number
of (half-mass) relaxation times.\\ \\ A cluster is evolved on a
circular orbit, with constant orbital velocity, $v$, at a constant
galactocentric radius, R$_{\rm GC}$, about the galactic center. {\sc
  emacss} assumes a logarithmic potential, $\phi$, for the galaxy
which imposes a static tidal field on the cluster
\citep{2014MNRAS.437..916G,2014MNRAS.442.1265A}:
\begin{eqnarray}
 \phi &=& v^2\ln(R_{\rm GC}) \label{eq:tides1} \\
r_{\rm J}^3 &=& \frac{GMR_{\rm GC}^2}{2v^2} \label{eq:tides2}
\end{eqnarray}
in which $r_{\rm J}$ is the Jacobi radius and G is the gravitational
constant. Note that $r_{\rm J} = r_{\rm t}$ for the type of potential
used here.\\ \\ The code uses a Kroupa initial mass function
\citep{2001MNRAS.322..231K} with a lower mass limit of
0.1\,M$_{\odot}$ and an optional upper mass limit $m_{\rm up}$. It was
tested against $N$-body simulations for $m_{\rm up} =$ 15\,M$_{\odot}$
or 100\,M$_{\odot}$ with an initial mean mass of 0.64\,M$_{\odot}$ in
both cases \citep{2014MNRAS.442.1265A}. In each of our simulations we
use the upper mass limit of 100\,M$_{\odot}$ and an initial mean mass
of 0.64\,M$_{\odot}$. {\sc emacss} furthermore offers an indication of
core-collapse based on an \lq{}average\rq{} cluster,
which is adequate to determine the state of a cluster substantially
before or after core collapse, although is unreliable at times within
a factor $\sim$2 of the predicated collapse itself. The code does not
explicitly include a prescription for stellar interactions and binary
formation, nor does it account for the effects of a primordial binary
population. This most recent version of {\sc emacss} also allows one
to take into account the effects of dynamical friction. We have not
included this feature in the simulations in this paper, because the
studies we compare our results to have not included this effect
either. We will explore the effects of dynamical friction in
forthcoming work. The details of {\sc emacss} are described in
references mentioned above.\\ \\ The most recent version of {\sc
  emacss} \citep{2014MNRAS.442.1265A} is available on
GitHub\footnote{https://github.com/emacss/emacss} and in the
Astrophysical Multipurpose Software Environment (AMUSE,
\citet{2009NewA...14..369P}).\footnote{http://amusecode.org/}

\subsection{{\sc emacss-mcmc} method}\label{subsec:emcee}

We define a set of initial conditions of a star cluster as the
cluster\rq{}s initial mass and initial half-mass radius, i.e. ($M_i$,
$r_{\rm hm,i}$). We constrain the initial conditions for an observed
star cluster from the observed current mass, $M_{\rm obs}$, half-mass
radius, $r_{\rm hm,obs}$, age, $\tau_{\rm obs}$, galactocentric
radius, $R_{\rm GC,obs}$ and orbital velocity, $v_{\rm obs}$. For each
observed cluster we simulate $n$ clusters with different initial total
masses and half-mass radii, $M_i$ and $r_{\rm hm,i}$ respectively,
from $t$ = 0\,Gyr to $t$ = $\tau_{\rm obs}$. These clusters all start
out with the same initial galactocentric radius and initial orbital
velocity, equal to the currently observed values\footnote{In this
  section we describe the general use of the two-dimensional version
  of our method. However, since our aim is to validate our method, we
  choose the same values for the age, the galactocentric radius and
  the orbital velocity as the studies we compare to in
  our 2D simulations, but in our 5D simulations we
    will marginalize over these three parameters in the
    five-dimensional version of our method, see
    Section~\ref{subsubsec:Independent}.}, since both these
parameters do not change throughout the evolution with {\sc emacss}
without dynamical friction. After evolving the clusters, we compare
their final conditions in mass and half-mass radius, i.e. $M_f$ and
$r_{\rm hm,f}$ respectively, to the observed present day values and
assign a (posterior) probability to each initial condition. Note that
the zero age of the cluster, i.e. $t$ = 0\,Gyr, is defined in {\sc
  emacss} as the time when both the residual gas of
the giant molecular gas cloud, from which the star cluster formed, has
escaped the cluster and the cluster has reached virial
equilibrium.\\ \\
\begin{figure*}
 \centering
 \includegraphics[width=\textwidth]{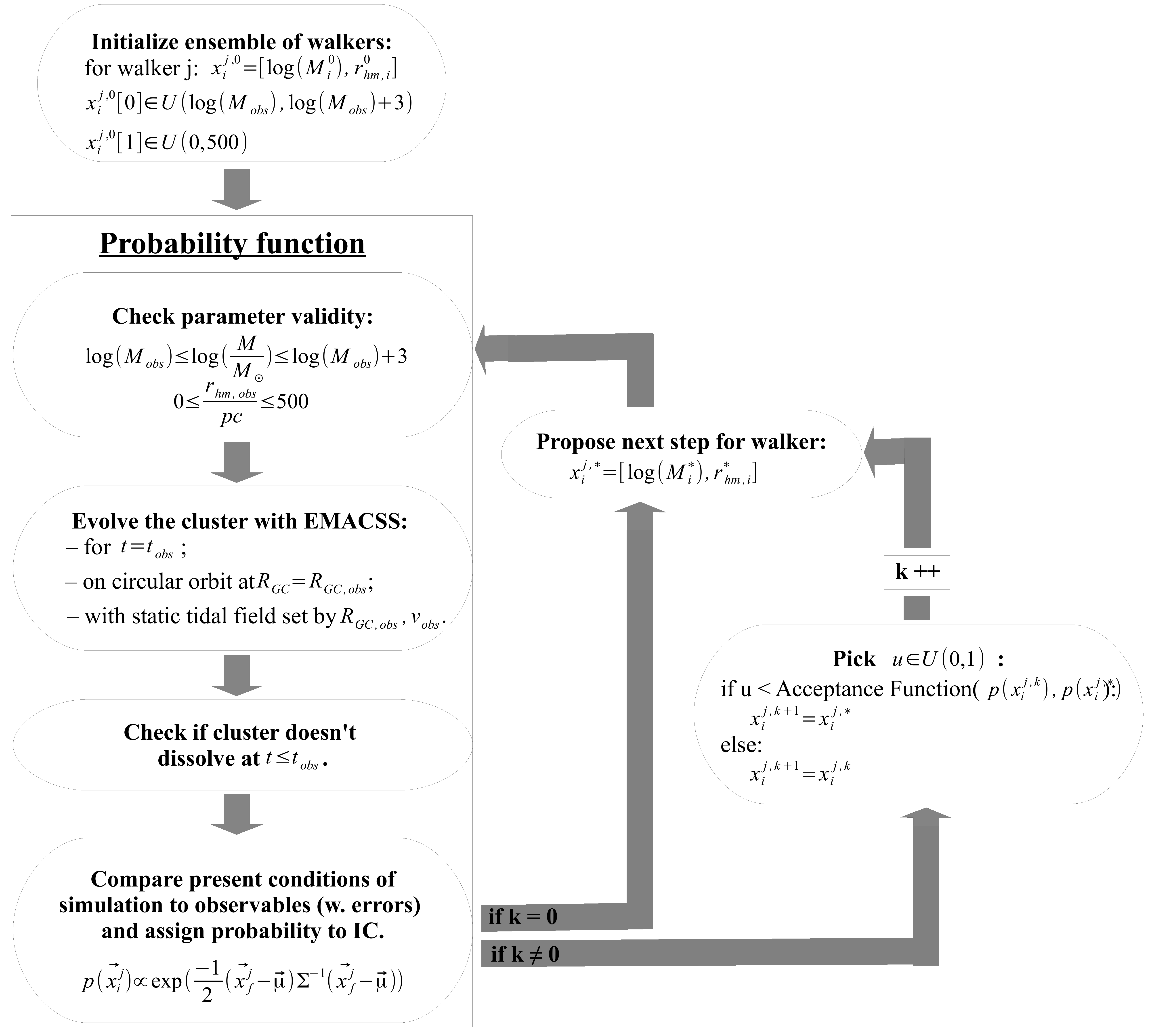}
 \caption{The schematic representation of our
     method. One round through this scheme represents
       one iteration, both the burn-in and subsequent chain iteration
       phase.\label{fig:model_description}}
\end{figure*}
For choosing the initial conditions in mass and half-mass radius, we
could have used a grid of $n$ mass and half-mass radius pairs,
e.g. evenly spaced in both half-mass radius and (the logarithm of)
mass. However, if one wants to determine the initial conditions in
more than two dimensions, which we do in our 5D
  simulations (see Section~\ref{subsubsec:Independent}), the grid approach is no longer feasible and one needs
to sample the initial conditions with a method that efficiently probes
and properly covers a multidimensional parameter space. We therefore
use the affine-invariant ensemble sampler for Markov Chain Monte Carlo
(MCMC) as coded up in the {\sffamily emcee} code
\citep{2013PASP..125..306F} based on
\citet{1189.65014}.\\ \\ MCMC is a procedure to
  generate a random walk in parameter space to obtain an approximation
  to the \textit{unknown} posterior density distribution function
  (PDF)
  \citep{2013PASP..125..306F,1953JChPh..21.1087M,HASTINGS01041970}.
  Sampling the PDF starts by initializing the \textit{walkers} accross
  parameter space according to some \textit{prior} distribution. The
  walkers then sample parameter space according to the specific MCMC
  algorithm one employs, and eventually converge towards those regions
  of parameter space with high posterior probability. Since in most
  cases one has no a priori knowledge what the PDF looks like, the
  simplest and most un-informative prior - a uniform distribution in
  each parameter/dimension - is used. The walkers \textit{burn-in} to
  probable regions of parameter space that can then be used as a
  starting point of the subsequent (chain iteration) phase
  \citep{2013PASP..125..306F}.\footnote{The \textit{burn-in} is defined as the first part of the Markov chain where the walkers reach a certain level of posterior probability, but is otherwise the same as the subsequent \textit{chain iteration phase}. Since the first part of the Markov chain usually contains lower probabilities, it is common practice to remove the data of the burn-in phase from the results such that further analysis is not biased by the low probabilities of the burn-in. See e.g. \citet{2009A&A...497..991P} for a nice explanation of the burn-in phase.}\\ \\ Figure~\ref{fig:model_description} presents the method we employ. One round through the
  scheme represents one iteration. For each of the observed clusters
we sample a two-dimensional parameter space, in
$\log(M)$ and $r_{\rm hm}$. We use $\log(M)$ instead of $M$, because
we experienced that the MCMC method is more efficient when it covers
the large range of several orders of magnitude in mass in logarithmic
space than in linear space. We use an ensemble of
$n_{\rm w}$ walkers, a burn-in phase of $n_{\rm b}$ iterations and
$n_{\rm c}$ subsequent (chain) iterations such that evolve a total of
$n = n_{\rm w}(1 + n_{\rm b} + n_c)$ clusters. A
walker $j$ at iteration $k$ is defined as $\vec{x}^{j,k}_i =
[\log(M^{j,k}_i), r^{j,k}_{\rm hm,i}]$; the subscript $i$ denotes that
it concerns an initial condition. We use the
  following boundary conditions for the parameters:
\begin{eqnarray}\label{eq:initialization}
\log(M_{\rm obs}) \leq &\log( \frac{M}{M_{\odot}})& \leq \log(M_{\rm
  obs}) + 3, \nonumber \\ 0 < &\frac{r_{\rm hm}}{\rm pc}& < 500,
\end{eqnarray}
in which the lower mass boundary comes from the fact that the initial
cluster must have been at least as massive as it is today. The other
boundaries were found to be reasonable values to not exclude any
possibly interesting regions in mass and half-mass radius and to get a
good balance between proper coverage and quick convergence. For the
initialization we have experimented with several different prior
distributions, see Section~\ref{subsec:Performance}. When the walkers
are initialized, they are evaluated in the probability function, see
Figure~\ref{fig:model_description}. The probability function
determines how suitable the sets of initial conditions are for the
observed cluster of interest by assigning a posterior
probability, $p$, to each initial
condition; $p$ takes value in the range 0 to 1 where
  0 denotes the lowest probability and 1 denotes the highest
  probability. This is done as follows:
\begin{enumerate}
\item \textbf{Initial condition check}: for a walker $j$ at iteration
  $k$ it checks whether its mass and half-mass radius, stored in
  $x^{j,k}_i[0]$ and $x^{j,k}_i[1]$ respectively, is within the ranges
  of Eq. (\ref{eq:initialization}). If a certain
  initial condition is in the requested range, it proceeds to step
  (ii). If this is not the case, steps (ii)-(iv) will
    be skipped and this initial condition is assigned a posterior
  probability $p = 0$.
\item \textbf{Evolution}: a cluster with this walker's initial mass
  and half-mass radius will be evolved with {\sc emacss}, at a
  constant galactocentric radius $R_{\rm GC} = R_{\rm GC,obs}$ and
  velocity $v=v_{\rm obs}$ from $t$ = 0\,Gyr until the observed
  cluster's age $t= \tau_{\rm obs}$ is reached. It proceeds to step
  (iii).
\item \textbf{Dissolution check:} it checks whether the cluster
  dissolved before reaching $t= \tau_{\rm obs}$. If for a certain set
  of initial conditions the cluster dissolved before reaching the age
  $t= \tau_{\rm obs}$, step (iv) will be skipped and
    this initial condition is assigned a posterior probability $p =
  0$. If the cluster stayed bound until $t= \tau_{\rm obs}$, it
  proceeds to step (iv).
\item \textbf{Posterior probability calculation:} the final conditions
  of the cluster - for a walker $j$ at iteration $k$ contained in
  $\vec{x}^{j,k}_f = [log(M^{j,k}_f), r^{j,k}_{\rm hm,f}]$, where the
  subscript $f$ denotes that it concerns a final condition - are
  compared to the observables and a posterior probability $p$ is
  calculated according to:
\begin{eqnarray}\label{eq:prob}
 ln(p) = -\frac{1}{2} \Bigl( (\vec{x}^{\rm j,k}_{\rm f} - \vec{\mu})^T
 \Sigma^{-1} (\vec{x}^{\rm j,k}_{\rm f} - \vec{\mu}) \Bigr),
\end{eqnarray}
in which $\vec{\mu} = [M_{\rm obs}, r_{\rm hm,obs}]$ are the present
day observed values and $\Sigma^{-1}$ is the inverse of the covariance
matrix which contains the errors in $M_{\rm obs}$ and $r_{\rm
  hm,obs}$. We assume that the errors in mass and half-mass radius are
not correlated such that the covariance matrix contains no nonzero
off-diagonal elements:
\begin{eqnarray}
 \Sigma = \Bigl( \begin{matrix} |\log(M_{\rm obs}) - \log(M_{\rm obs}
   - \Delta M_{\rm obs})| & 0\\ 0 & \Delta r_{\rm hm,obs}\end{matrix}
 \Bigr).
\end{eqnarray}
\end{enumerate}
We chose to use 10\% errors in both observables for all our clusters
in the majority of our simulation; this choice is arbitrary and used in this paper as a proof of concept. In further applications of our method one can use the actual observational errors. In
Section~\ref{subsubsec:Accuracies_of_observables} we investigate what
the effect of taking smaller (5$\%$) or larger (20$\%$) errors has on
the results.\\ \\ After an iteration $k$ each walker will be proposed
a new set of initial conditions for the next iteration $k+1$. Whether
the walker accepts or declines this proposed set is
determined as follows: for a walker $j$ a new set of initial
conditions $\vec{x}^{j,*}_i$ will be proposed according to the
\textit{stretch move} algorithm. This is an algorithm
  in which one simultaneously evolves an ensemble of walkers, whereby
  the proposal distribution, from which the proposed initial condition
  is drawn, is based on the initial conditions for the other $n_{\rm
    w} - 1$ walkers in the previous iteration ($k$), see
  \citealt{2013PASP..125..306F}. This set of initial conditions is
evaluated in the probability function, as explained above. After this
set of initial conditions is assigned a posterior probability, $p^*$,
$p^*$ will be compared to the posterior probability of the previous
iteration, $p$, and a probability $q$ will be calculated for the
acceptance of this proposed set of initial conditions, see
Eq. (9) of \citet{2013PASP..125..306F}. The values
of $q$ are in the range 0 to 1 and will approach the value 1 if
$p^*\gg p$ and the value 0 if $p^* \ll p$. Lastly, a random number $u$
will be drawn from a uniform distribution between 0 and 1; if $q > u$
the walker $j$ accepts the proposed set of initial conditions such
that $\vec{x}^{j,k+1}_i = \vec{x}^{j,*}_i$, i.e. the walker
\lq{}walks\rq{} to this proposed set. If $q < u$, then
$\vec{x}^{j,k+1}_i = \vec{x}^{j,k}_i$ and the walker \lq{}stays\rq{}
at its previous set of initial conditions. This
procedure is repeated for all the iterations, both
  the burn-in iterations and the subsequent chain iterations. See
\citet{2013PASP..125..306F} for further details of the code {\sffamily
  emcee}.\\ \\ In Section~\ref{subsec:Performance} we
first determine suitable values for the number of
walkers, $n_{\rm w}$, the number of burn-in iterations, $n_{\rm b}$,
and the number of subsequent chain iterations, $n_{\rm c}$, while
testing the performance of our method. Thereafter we
  investigate the effect of different prior distributions on the
  determined distribution of probable initial conditions. 

\subsection{Correcting observations for mass segregation}\label{subsec:Conversion}

When one wishes to compare observed structural parameters to the
simulated ones, it is of great importance that these parameters are
obtained in the same way. Since it is observationally more practical
to find the size of a cluster by determining the angular projected
half-light radius in arcminutes (arcmin), whereas
theoretically half-mass radii in parsec (\,pc) are
more functional, some conversions are needed before a proper
comparison between the simulations and the observations can be
made.\\ \\ The angular projected half-light radius,
$\theta_{\rm phl}$, in arcmin from the observations
can be converted to the projected half-light radius,
$r_{\rm phl}$, in\,pc once the distance from the
Earth to the cluster, $R_{\rm E}$, is known:
\begin{eqnarray}\label{eq:arcmin_to_pc}
r_{\rm phl} = \Bigl(\frac{R_{\rm E}}{\rm
  pc}\Bigr)\tan\Bigl(\frac{\theta_{\rm phl}}{\rm
  arcmin}\cdot\frac{\pi}{10800}\Bigr).
\end{eqnarray}
We calculate the \textit{observational} half-mass radius, $r_{\rm
  hm}$, for two extreme cases: with or without a
  correction for mass segregation (MS).
\begin{enumerate}
 \item Case 1: no correction for MS\\ We assume that
   the cluster experienced no MS at
     $t = \tau_{\rm obs}$ yet such that the 3D half-mass
   radius, $r_{\rm hm}$, is equal
   to the 3D half-light radius, $r_{\rm hl}$. In
   this case we convert the projected half-light radius to the 3D
   half-light radius by multiplying it with a geometrical factor 4/3,
   correcting for the projection:
\begin{eqnarray}\label{eq:rhl}
 r_{\rm hm} = r_{\rm hl} = (4/3)  r_{\rm phl}
\end{eqnarray}
 \item Case 2: with a correction for MS\\ We assume
   that the cluster did experience an amount of MS at
     $t = \tau_{\rm obs}$ in the form of a constant conversion factor
   between the projected half-light radius, $r_{\rm
     phl}$, and the 3D half-mass
   radius, $r_{\rm hm}$, which we
   read off from Figure 4 of \citet{2007MNRAS.379...93H}:
\begin{eqnarray}\label{eq:rhl_to_rhm}
 r_{\rm hm} = c_{\rm ms} r_{\rm \sc phl},
\end{eqnarray}
with $c_{\rm ms} = 1.9$ for clusters with ages $> 7$ Gyr and $c_{\rm
  ms} = 1.8$ for a cluster $\sim 4$ Gyr of age. Note that this
conversion includes the geometric conversion factor of $4/3$ and a
factor $\sim 1.425$ respectively $\sim 1.35$ to account for MS.
\end{enumerate}
By doing this we can compare the observational half-mass radii to the
half-mass radii from the simulations. See the third and fourth column
of Table~\ref{table:Validation_clusters2} for the calculated
observational half-mass radii without and with a
  correction for mass segregation, respectively.

\subsection{Confidence regions}\label{subsec:Confidence regions}

After a simulation we obtain sets of initial conditions, final
conditions and their corresponding posterior probabilities for
$n$ clusters.  This also includes proposed sets of
initial conditions, even if these were eventually rejected. From these
$n$ initial conditions we remove the initial conditions from the
initialization, from the burn-in and those outside
the ranges given in Eq.
(\ref{eq:initialization}). Hence we have each particular initial condition appear only once in our data. The remaining initial conditions include
those which survive until $t = \tau_{\rm obs}$ and
those which dissolve before reaching the observed
  cluster\rq{}s age. Besides analyzing the most probable regions in
initial total mass versus initial half-mass radius, we namely also
want to study the regions which are not suitable for producing the
currently observed clusters.\\ \\ Of the surviving initial conditions,
we determine the regions with a $68.3\%$ and $99.7\%$ confidence
level: 1) we calculate the normalized posterior
probability of each of the sets of initial conditions by dividing the
posterior probability of each set by the sum of all the posterior
probabilities; 2) from the set of initial conditions with the highest
normalized posterior probability downwards, we sum the normalized
posterior probabilities of each subsequent set of initial conditions
until this sum equals 0.997 (0.683). The sets of
initial conditions included in that sum are a subset of initial
conditions which make up $99.7\%$ ($68.3\%$) of the
total posterior probability, i.e. the $99.7\%$
($68.3\%$) confidence region. Note that our
definition of the $99.7\%$ confidence region is different from what is
usually meant with a $99.7\%$ confidence region, namely the region
containing $99.7\%$ of all the data points. The reason for our
alternative definition is that our aim is to show those regions of
initial conditions with high posterior probability and if we were to
use $99.7\%$ of all the data points, this region would contain a large
number of initial conditions with low posterior probability ($p <
0.01$).\\ \\
\begin{figure*}
 \centering
 \includegraphics[width=\textwidth]{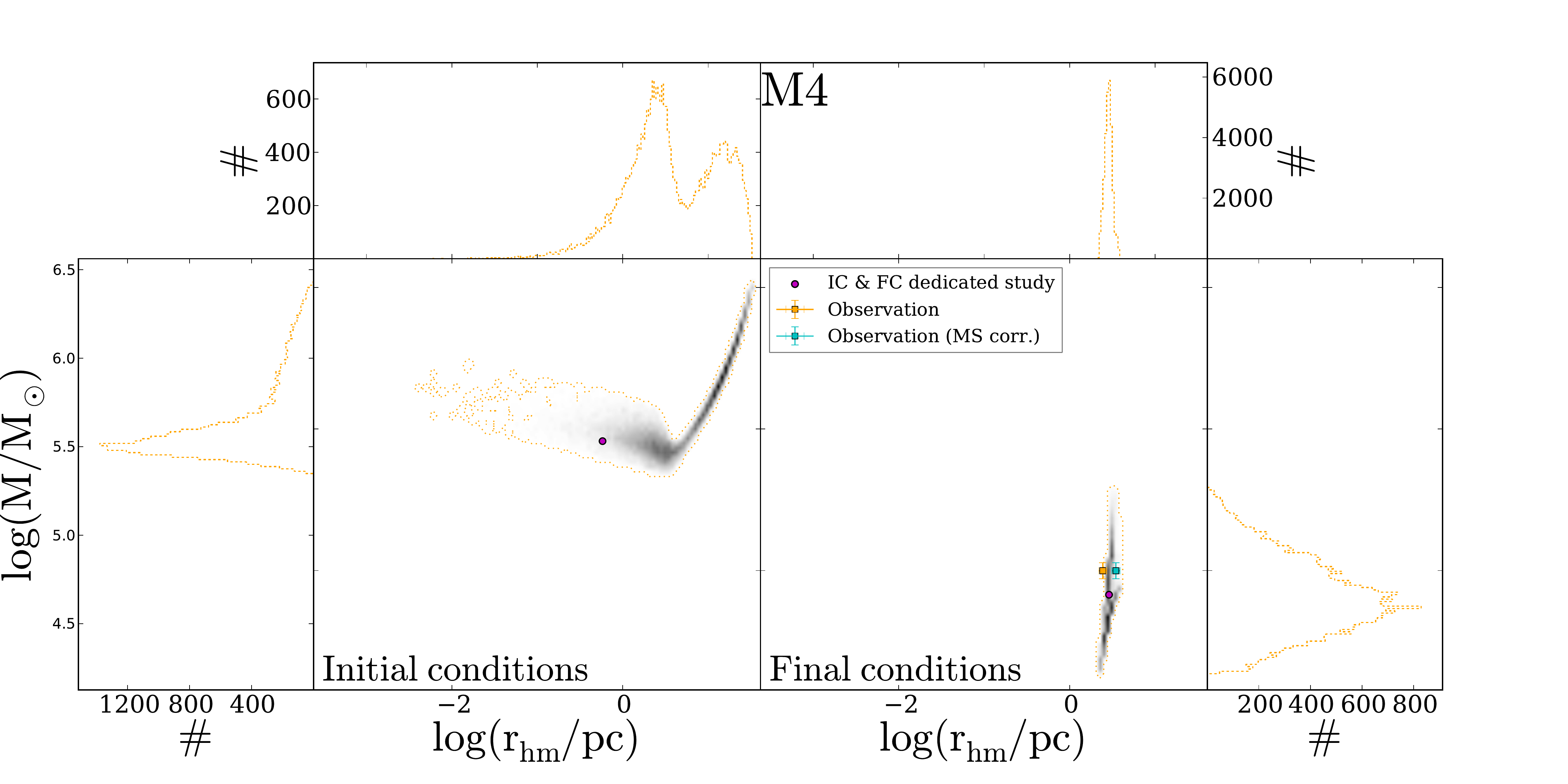}
 \caption{A simple example of the results for the cluster M4. Initial and final condition
          distributions in half-mass radius versus total mass, where the three panels on the left show the
          initial conditions and the three panels on the right show
          the final conditions. The square panels show the
          two-dimensional histograms in log(M) vs log($r_{\rm hm}$) of
          the $99.7\%$ confidence region of the simulations that fit
          the observables without a correction for mass segregation in
          a black-white density plot: the darker the area, the more
          initial conditions were sampled in this area. Over-plotted
          on these two-dimensional histograms are the outer contours
          of the two-dimensional histograms of the $99.7\%$ (yellow dotted line)
          confidence level initial conditions for
          both the simulations without a correction for mass
          segregation, with the projected histograms for log($r_{\rm
            hm}$) at the top and for log(M) on the left respectively
          right side. The yellow square with error bars
          show the observation of the cluster's current mass and
          half-mass radius when no correction for mass segregation is
          made, and the cyan square with error bars shows the
          observable with a correction. The pink filled circle in the
          left panel denotes the initial condition used by the
          dedicated study, which evolves to the final condition shown
          by the pink filled circle in the right panel.\label{fig:simple_results}}
\end{figure*}
In Figure~\ref{fig:simple_results} we show a simple example of our results for the cluster M4 without a correction for mass segregation, whereby the $99.7\%$ confidence region in both the initial and the final conditions are enclosed by the yellow, dotted contours.

\section{Validation}\label{sec:Validation}

\subsection{Validation strategy}\label{subsec:Validation_strategy}

We validate our method by applying it to nine star clusters that have
been studied to great extent with either $N$-body simulations or Monte
Carlo methods. These nine clusters include one Galactic open cluster,
seven Galactic globular clusters and one extragalactic globular
cluster; see Table~\ref{table:Validation_clusters} for the names of
the clusters, the cluster types, their host galaxy, the references to
papers in which these clusters were studied and with which simulation
technique.
\begin{table*}
 \centering
 \begin{tabular*}{0.74\textwidth}{ l l l l l}\hline
    Name          & Cluster type     & Galaxy    & Simulation technique & Reference\\ \hline \hline
    M67           & open cluster     & Milky Way & Direct $N$-body      & \citet{2005MNRAS.363..293H}\\
    NGC 6397      & globular cluster & Milky Way & Monte Carlo          & \citet{2009MNRAS.395.1173G}\\
    M4            & globular cluster & Milky Way & Monte Carlo          & \citet{2008MNRAS.389.1858H}\\ 
    M22           & globular cluster & Milky Way & Monte Carlo          & \citet{2014arXiv1401.3657H}\\ 
    Palomar 4     & globular cluster & Milky Way & Direct $N$-body      & \citet{2014MNRAS.440.3172Z}\\ 
    47 Tuc        & globular cluster & Milky Way & Monte Carlo          & \citet{2011MNRAS.410.2698G}\\
    Palomar 14    & globular cluster & Milky Way & Direct $N$-body      & \citet{2011MNRAS.411.1989Z}\\ 
    $\omega$ Cen  & globular cluster & Milky Way & Monte Carlo          & \citet{2003MNRAS.339..486G}\\ 
    G1            & globular cluster & M31       & Scaled $N$-body      & \citet{2003ApJ...589L..25B}\\ \hline
 \end{tabular*}
 \caption{The nine star clusters on which we apply our method to
   validate it. The first column lists the name of the cluster, the
   second column the cluster type and the third column the galaxy in
   which the cluster resides. The fifth column shows the references to
   the studies which already studied these clusters to great extent
   with the simulation technique mentioned in column four. The
   clusters are mentioned in the order of increasing current half-mass
   relaxation time that was taken from the Harris Catalogue
   \citep{2010arXiv1012.3224H}, except for the first and last
   mentioned cluster, where we took the relaxation times from the
   corresponding reference in this table. See
   Table~\ref{table:Validation_clusters2} for the values of these
   relaxation times.\label{table:Validation_clusters}}
\end{table*}
\begin{table*}
 \centering
 \begin{tabular*}{0.72\textwidth}{ l l l l l l l l}\hline
  Name         & $M_{\rm obs}$        & $r_{\rm hm,obs}$ & $r_{\rm hm,obs}$ (MS corr.) & $R_{\rm GC,obs}$ & $\tau_{\rm obs}$ & $v_{\rm obs}$ & log($t_{\rm rh})$\\
	           & [$10^4$ M$_{\odot}$] & [pc]      	   & [pc]      	                & [kpc]            & [Gyr]            & [km/s]        & [yr]\\ \hline \hline
  M67          & 0.14                 & 3.35 	           & 4.52                        & 8.0              & 4                & 220           & 8.48\\ 
  NGC 6397     & 6.6                  & 2.30 	           & 3.28                        & 1.99             & 12               & 220           & 8.6\\
  M4           & 6.3                  & 2.43 	           & 3.47                        & 1.68             & 12               & 220           & 8.93\\
  M22          & 33                   & 4.17 	           & 5.94                        & 5.28             & 12               & 220           & 9.23\\
  Pal 4        & 2.98                 & 24.7 	           & 35.2                        & 102.8            & 11               & 200           & 9.42\\ 
  47 Tuc       & 110                  & 4.87 	           & 6.94                        & 2.95             & 12               & 220           & 9.55\\
  Pal 14       & 1.1725               & 35.4 	           &  50.2                       & 71.6             & 11               & 220           & 10.02\\ 
  $\omega$ Cen & 390                  & 9.89	           & 14.09                       & 1.22             & 12               & 220           & 10.09\\
  G1           & 1500                 & 17.6 	           & 25.08                       & -                & 12               & -             & 10.7\\  \hline
 \end{tabular*}
 \caption{Observational data of the validation clusters as used in the
   references mentioned in Table~\ref{table:Validation_clusters}. The
   first column lists the cluster name and the second column the
   observed mass in $10^4$ M$_{\odot}$. The third and fourth columns
   lists the half-mass radius without a correction for mass
   segregation (i.e. equal to the 3D half-light radius) and the
   half-mass radius with a correction for mass segregation
   (MS) respectively. The galactocentric radius,
   age, orbital velocity of each cluster as used in
   the references mentioned in Table~\ref{table:Validation_clusters}
   are listed in column five, six and seven respectively. For the sake
   of comparison to these dedicated studies, we use these observables
   in our simulations, even though some cluster parameters are
   presently better constrained by more recent observations. See
   Section~\ref{subsec:Validation_clusters} for the reasoning behind
   choosing each of these parameter
   values.\label{table:Validation_clusters2}}
\end{table*}

\subsubsection{Direct comparison}\label{subsubsec:Direct_comparison}
The first step in the validation is to do a direct comparison of the
cluster evolution codes by starting from the same sets of initial
conditions and evolving the cluster under conditions which are as
similar as possible. Therefore, for each cluster we start at the
(best-fit) set of initial
condition put forward by the dedicated studies and we evolve this
cluster up to the same age under similar conditions, which we describe
in Section~\ref{subsec:Validation_clusters} and summarize in
Table~\ref{table:Validation_clusters2}.

\subsubsection{Independent {\sc emacss-mcmc} runs}\label{subsubsec:Independent}
The second step is to run our {\sc emacss-mcmc} method
(Section~\ref{subsec:emcee}) independently and determine which initial
masses and half-mass radii reproduce the cluster observables best,
explore possible degeneracies therein and observe whether the best-fit
initial conditions of the dedicated study are contained within our
confidence regions. For the sake of comparison to the mentioned
dedicated studies, we adopt the cluster observables that were used in
the references mentioned in Table~\ref{table:Validation_clusters},
even though some cluster parameters are presently better constrained
by more recent studies, see e.g. Table 1 of the recent study of
\citet{2010MNRAS.406.2000M}, where present day masses and half-mass
radii are listed for a number of clusters. Our goal in this paper is
to validate our method by comparing our results to the results which
were obtained with the dedicated studies. Therefore it is more
important to adopt the observables from these studies than to use the
currently best constrained observables. See
Section~\ref{subsec:Validation_clusters} and
Table~\ref{table:Validation_clusters2}.\\ \\ The next
  step is to exploit the power of MCMC. In the five-dimensional
version of our {\sc emacss-mcmc} method we add the galactocentric
distance, age and orbital velocity as nuisance parameters and
determine the most probable initial masses and half-mass radii. By
comparing this to the results of the two-dimensional
  runs explained above we can explore how dependent probable initial
masses and half-mass radii are on the choice of the galactocentric
radius, $R_{\rm GC}$, age, $\tau$,
and velocity, $v$. Adding these parameters as
nuisance parameters is observationally motivated,
  since all observables are always determined with some error. The
method here is similar to the description in
Section~\ref{subsec:emcee}, but the difference is that now at each
iteration we also sample a galactocentric radius, an age and an
orbital velocity from Gaussian distributions with the mean equal to
$R_{\rm GC, obs}$, $\tau_{\rm obs}$ and $v_{\rm obs}$ respectively, as
adopted from the references mentioned in
Table~\ref{table:Validation_clusters}, see
Table~\ref{table:Validation_clusters2}, and
a standard deviation equal to $10\%$ of the mean. For
    G1, though, we take $R_{\rm GC, obs} = 40$\,kpc as mentioned in
    \citet{2003ApJ...589L..25B} and $v_{\rm obs} = 230\,$km/s. We perform our 5D
runs only without a correction for mass
  segregation.\\ \\ The final step is to investigate
  the stability of our determined initial conditions against errors in
  the observed data. We do this in two ways. We first compare our 2D. The second thing we do, is checking
how
  increasing/decreasing the errorbars on the fitting parameters
  ($\log(M)$ and $r_{\rm hm}$) will effect the distribution of initial
  conditions. The second thing we do, is testing the stability of the
  initial condtions against varying the parameters $R_{\rm GC}$, $v$
  and $\tau$ within their errorbars, assuming that the observed mass
  and half-mass radius are unchanged. We take $10\%$ errors and look
  at the maximum difference, e.g. comparing a simulation with $R_{\rm
    GC} = R_{\rm GC,obs}$ to simulations with $R_{\rm GC} = R_{\rm
    GC,obs} + 0.1 R_{\rm GC,obs}$ and $R_{\rm GC} = R_{\rm GC,obs} -
  0.1 R_{\rm GC,obs}$ respectively.

\subsection{Validation clusters}\label{subsec:Validation_clusters}

\subsubsection{$\omega$ Centauri, M4, NGC 6397, 47 Tucanae and M22}\label{subsubsec:MonteCarlo}

In this section we describe the studies using the Monte Carlo code
MOCCA \citep{1998MNRAS.298.1239G, 2001MNRAS.324..218G,
  2006MNRAS.371..484G} modeling five Galactic globular clusters:
$\omega$ Cen (NGC 5139; \citealt{2003MNRAS.339..486G}), M4 (NGC 6121;
\citealt{2008MNRAS.389.1858H}), NGC 6397 \citep{2009MNRAS.395.1173G},
47 Tuc (NGC 104; \citealt{2011MNRAS.410.2698G}) and M22 (NGC 6656;
\citealt{2014arXiv1401.3657H}). Hereafter we refer to these five
studies as GH03-14.

\paragraph{Simulation Technique}\label{par:SimulationTechnique1}

For each of the globular clusters GH03-14 determined a set of initial
conditions - and a few sets of initial conditions in the case of M22 -
by means of small scale modeling, i.e. by performing simulations for
clusters with a lower initial number of stars. These initial
conditions were subsequently evolved up to the age of the cluster,
which was taken to be 12 Gyr for each of these
clusters. For M4, NGC 6397, 47 Tuc and M22 their
  simulations included prescriptions for single and binary stellar
  evolution, for the Galactic tidal field, for two-body relaxation and
  for binaries and their dynamical interactions. For $\omega$ Cen,
  which was studied with an early version of MOCCA, their simulation
  included simple prescriptions for single stellar evolution and for
  the dynamical evolution in the Galactic tidal field.
After evolution these clusters gave a satisfactory
  match to a number of observed characteristics, such as the surface
  brightness profile and the velocity dispersion profile. See the references mentioned in Section~\ref{subsubsec:MonteCarlo} for
the details of their MOCCA code.

\paragraph{Orbit and tides}\label{par:Orbitandtides1}

GH03-14 evolved each of these clusters on a circular orbit with a
circular velocity of 220 km/s\footnote{\citet{2003MNRAS.339..486G} do
  not mention which circular velocity they adopted, but we assume it
  was a value of 220 km/s as well.}. They set the tides by imposing an
initial tidal radius, $r_{\rm t,i}$. Given this tidal radius and their
initial total cluster mass, $M$, and assuming an isothermal model for
the Galaxy, we calculate per cluster with which galactocentric radius,
$R_{\rm GC}$, their model is consistent by combining
Eq.s (\ref{eq:tides1}) and (\ref{eq:tides2}). We use
these galactocentric radii as input for our simulations, see the
fifth column of
Table~\ref{table:Validation_clusters2}.\\ \\ They evolved their model
clusters with an initial tidal radius of 35\,pc, 86\,pc, 40\,pc,
89\,pc and 90\,pc, which are consistent with a galactocentric radius
of 1.68\,kpc, 2.95\,kpc, 1.99\,kpc, 5.28\,kpc and 1.22\,kpc for the
clusters M4, 47 Tuc, NGC 6397, M22 (we compare to their model B, see
Table 1 of \citet{2014arXiv1401.3657H}) and $\omega$
Cen, respectively. Given the fact that M4 is on an
eccentric orbit with (eccentricity $e$, perigalacticon $R_p$/kpc,
apogalacticon $R_a$/kpc) of about (0.8, 0.6, 5.9)
\citep{1999AJ....117.1792D}, this cluster experiences strong tides
near perigalacticon. Therefore their choice of the initial tidal
radius, and its corresponding galactocentric radius, are very
reasonable and better than evolving the cluster at its current
galactocentric radius of 5.9\,kpc \citep{2010arXiv1012.3224H}: the
averaged tidal field experienced by their cluster during its 12 Gyr
circular orbit with a galactocentric radius of 1.68\,kpc is comparable
to the tidal field experienced by M4 in its actual eccentric
orbit. Their choices for the initial tidal radii for NGC 6397 and M22
with ($e$, $R_p$, $R_a$) of about (0.34, 3.1, 6.3) and (0.53, 2.9,
9.3) respectively \citep{1999AJ....117.1792D} are also reasonable, but
their choices for $\omega$ Cen and 47 Tuc with ($e$,
  $R_p$, $R_a$) of about (0.67, 1.2, 6.2) and (0.17, 5.2, 7.3)
respectively, are slightly overestimating the effect of the tidal
field. However, in order to have a good comparison to the studies
mentioned above, for each of these clusters we evolve all our model
clusters on a circular orbit with a circular velocity of 220 km/s at
the galactocentric radius to which these models are consistent with,
see Table~\ref{table:Validation_clusters2}.

\paragraph{Observed mass and radius}\label{parObservedmassandradius1}

GH03-14 get the observational mass of M4 of $6.3 \cdot 10^4$
M$_{\odot}$ from \citet{2004AJ....127.2771R} and got the observed
half-light radius from \citet{1996AJ....112.1487H}, which in turn
lists angular half-mass radii, stated to be taken from the direct
average of \citet{1995AJ....109..218T} and
\citet{1991ApJ...375..594V}. Since both latter studies obtain their
half-light radii from projected data, we assume that
\citet{1996AJ....112.1487H} used a 1-to-1 relation between half-light
radii and half-mass radii, since \citet{2010arXiv1012.3224H} lists
half-light radii and these values are in many cases very similar to
the in \citet{1996AJ....112.1487H} mentioned \lq{}half-mass
radii\rq{}. The 2.3\,pc for the half-light radius mentioned in Table 2
of \citet{2008MNRAS.389.1858H} is a mistyped value (Heggie, private
communication), since the angular projected half-light radius
$\theta_{\rm phl}$ given in \citet{1996AJ....112.1487H} is 3.65 arcmin
(see also Table 1 of \citet{2009MNRAS.395.1173G}). In combination with
the distance between M4 and the Earth, $R_{\rm E} = 1.72$\,kpc
mentioned in \citet{2008MNRAS.389.1858H}, $\theta_{\rm phl}$ converts
to $r_{\rm phl} = 1.83$\,pc using Eq.
(\ref{eq:arcmin_to_pc}). We therefore use $r_{\rm phl} = 1.83$\,pc;
for the conversion to the 3D half-mass radius $r_{\rm hm}$
 with a correction for mass segregation (see
Section~\ref{subsec:Conversion}), we use the relation
(\ref{eq:rhl_to_rhm}) with $c_{\rm MS} = 1.9$.\\ \\ GH03-14 got the
observational mass for 47 Tuc of $1.1 \cdot 10^6$ M$_{\odot}$ from
\citet{1989A&A...214..106M}, for NGC 6397 of $6.6 \cdot 10^4$
M$_{\odot}$ from \citet{1995ApJS..100..347D} and for $\omega$ Cen of
$3.9 \cdot 10^6$ M$_{\odot}$ from \citet{1993ASPC...50..357P}. They do
not mention the observational cluster mass of M22, so we take the
cluster mass of $3.3 \cdot 10^5$ M$_{\odot}$ from
\citet{2008AJ....135.2141R}. GH03-14 got the observed radius for 47
Tuc of 2.79 arcmin and for NGC 6397 of 2.33 arcmin both from
\citet{1996AJ....112.1487H}; they do not mention the observational
radius for $\omega$ Cen and M22, so we get the observed radius for
$\omega$ Cen of 5.00 arcmin from \citet{1996AJ....112.1487H} and for
M22 of 3.36 arcmin from \citet{2010arXiv1012.3224H}
respectively. Using the same reasoning as mentioned for M4, we assume
the mentioned observed radii to be the projected (2D) angular
half-light radii $\theta_{\rm phl}$. We calculate $r_{\rm phl}$ in by
using Eq. (\ref{eq:arcmin_to_pc}) in combination
with distances of 47 Tuc, NGC 6397, $\omega$ Cen and M22 to the Earth,
$R_{\rm E}$, of 4.5\,kpc \citep{2011MNRAS.410.2698G}, 2.55\,kpc
\citep{2009MNRAS.395.1173G}, 5.1\,kpc \citep{1996AJ....112.1487H} and
3.2\,kpc \citep{2010arXiv1012.3224H} respectively. For the conversion
to the 3D half-mass radius $r_{\rm hm}$ with a
  correction for mass segregation, we use the relation
(\ref{eq:rhl_to_rhm}) with $c_{\rm MS} = 1.9$.

\paragraph{Model initial and final conditions}\label{par:InitialConditions}

The best-fit sets of initial conditions GH03-14 found for NGC 6397,
M4, 47 Tuc and M22 (model B) in (mass/\,M$_{\odot}$, half-mass
radius/pc) are listed in Table~\ref{table:results_2D}. For $\omega$
Cen \citet{2003MNRAS.339..486G} mention the initial and final mass of
their best-fit model (we list those in Table~\ref{table:results_2D}),
but they do not mention the initial and final half-mass radii of that
model, only their initial and final tidal radii. Assuming that the
tidal radius is the edge radius of the King model initially (Heggie,
private communication), we can calculate the initial half-mass radius
by using the ratio of tidal-to-half-mass radius taken from Figure
8.3. in \citet{2003gmbp.book.....H}. Using $r_{\rm t}/r_{\rm hm} \sim
9.65$ for a central potential $W_0 = 7.7$, we find $r_{\rm hm,i} =
9.33$\,pc. It is not clear whether this assumption is still valid
after 12 Gyr of evolution, so we cannot calculate their final
half-mass radius.

\subsubsection{Palomar 14 and Palomar 4}\label{subsubsec:DirectNbody_Pal}

In this section we describe the first published direct $N$-body
simulations of the two large and sparse Galactic globular clusters,
both residing in the outer halo: Pal 14 \citep{2011MNRAS.411.1989Z}
and Pal 4 \citep{2014MNRAS.440.3172Z}. They use the
  collisional $N$-body code NBODY6 \citep{2003gnbs.book.....A} on GPU
  computers. Hereafter we refer to these two studies by Z11-14.

\paragraph{Simulation Technique}\label{par:SimulationTechnique2}

Z11-14 simulate the evolution of Palomar 14 and
Palomar 4 and for Pal 14 they compute 65 models and
for Pal 4 a total of 20 models, divided in three categories: 1)
clusters with a \citet{2001MNRAS.322..231K} IMF in the range $0.08 <
M/M_{\odot} < 100$ (referred to as their \textit{canonical-NS} model),
2) clusters with a flattened IMF, 3) clusters with a
\citet{2001MNRAS.322..231K} IMF, but with primordial mass
segregation. For Pal 14 they computed one additional model with a
\citet{2001MNRAS.322..231K} IMF, but with primordial binaries. See the
references mentioned in Section~\ref{subsubsec:DirectNbody_Pal} for
the details of their simulations.

\paragraph{Orbit and tides}\label{par:Orbitandtides2}

For Pal 14 Z11-14 varied the initial half-mass
radius and mass and evolved each cluster for 11 Gyr on a circular
orbit in a logarithmic potential with a circular velocity of 220 km/s
at Palomar 14\rq{}s currently observed galactocentric radius, of which
the authors do not mention the value they adopted. After the
evolution, they fit the final values for the number of bright stars,
the projected half-light radius and the slope of the mass function in
the mass range 0.525-0.795 M$_{\odot}$ to the observed ones. To make a
fair comparison, we also evolve all of our model clusters for Pal 14
on a circular orbit at its current galactocentric radius, for which we
adopt a value of 71.6\,kpc from \citet{2010arXiv1012.3224H} with a
circular velocity of 220 km/s. To the best of our knowledge there are
no references reporting on the orbit of Pal
14. \citet{2009AJ....137.4586J} mention that the orbit could possibly
be eccentric, in which case evolving the cluster on a circular orbit
at the current galactocentric radius is an underestimation of the
tidal field.\\ \\ For Pal 4 Z11-14 also varied the
initial half-mass radius and mass and evolve each cluster for 11 Gyr
on a circular orbit with a circular velocity of 200 km/s at Palomar
4\rq{}s current galactocentric radius of 102.8\,kpc in an analytic
galactic background potential consisting of a bulge, a disc and a
logarithmic halo, which they adjusted to resemble the Milky Way.

\paragraph{Observed mass and radius}\label{parObservedmassandradius2}

Z11-14 got the observed total mass for Pal 14 of
about 12000 M$_{\odot}$ from \citet{2009AJ....137.4586J}
and for Pal 4 of about 29800 M$_{\odot}$ from
\citet{2012MNRAS.423.2917F}. For Pal 14 Z11-14 got the observed
projected angular half-light radius $\theta_{\rm phm}$ of 1.28 arcmin
from \citet{2006A&A...448..171H}, which they in turn convert to the
projected (2D) half-light radius $r_{\rm phl}$ of 26.4 $\pm$ 0.5\,pc
and a 3D half-light radius $r_{\rm hl}$ of 35.4 $\pm$ 0.6\,pc. For Pal
4 Z11-14 got the observed projected angular
half-light radius $\theta_{\rm phm}$ of 0.62 arcmin from
\citet{1966AJ.....71..276K} model fitting on WFPC2 data and broad-band
imaging with the Low-Resolution Imaging Spectrometer at the Keck II
telescope, which they convert to the projected (2D) half-light radius
$r_{\rm phl}$ of 18.4 $\pm$ 1.1\,pc and a 3D half-light radius $r_{\rm
  hl}$ of about 24\,pc. We converted these projected half-light radii
to the 3D half-mass radii $r_{\rm hm}$ with a
  correction for mass segregation (see
Section~\ref{subsec:Conversion}) by using the relation
(\ref{eq:rhl_to_rhm}) with $c_{\rm MS} = 1.9$.

\paragraph{Model initial and final conditions}\label{par:InitialConditions2}

For both Pal 14 and Pal 4 we compare to their canonical-NS models,
since these models are most comparable to {\sc emacss}. For Pal 14
these are the 28 models mentioned in Table 1 of
\citet{2011MNRAS.411.1989Z} with initial masses in the range
40000-60000\,M$_{\odot}$ (Zonoozi, private communication) and initial
half-mass radii in the range 15-25\,pc. For Pal 4 these are the seven
models mentioned in Table 1 of
\citet{2014MNRAS.440.3172Z} with initial masses in the range
50000-57000 \,M$_{\odot}$ and initial half-mass radii in the range
12-14.5\,pc (Zonoozi, private communication). The fact that these are
not their best-fit models is not a problem for validation
purposes. However, in our Figure~\ref{fig:IC_and_FC5}
  and \ref{fig:IC_and_FC7}, showing the results of the independent
  {\sc emacss-mcmc} runs in 2D, we also plot all the other models of
  \citet{2011MNRAS.411.1989Z} and \citet{2014MNRAS.440.3172Z} to see
  if their other (including best fitting) models are contained in our
  confidence regions.

\subsubsection{G1}\label{subsubsec:ScaledNbody}

In this section we describe the study using scaled $N$-body modeling
to investigate the evolution of M31\rq{}s largest globular clusters G1
(Mayall II; \citealt{2003ApJ...589L..25B}). They use
  the collisional $N$-body code NBODY4 \citep{1999PASP..111.1333A} on
  the GRAPE-6 computers \citep{2003PASJ...55.1163M}.

\paragraph{Simulation Technique}\label{par:SimulationTechnique3}

\citealt{2003ApJ...589L..25B} simulate the evolution of G1 by
 running simulations for star clusters with $N$ =
65536 stars, since direct $N$-body simulations with a number of stars
similar to the number of stars present in G1 ($N \sim 10^7$ according
to \citet{2003ApJ...589L..25B}) was, and still is, out of reach. They
used the same half-mass relaxation time $t_{\rm rh}$ for their model
as was inferred for G1 from observations (\citet{2001AJ....122..830M}
estimated $t_{\rm rh} \sim 50$ Gyr). They perform several dozen of
runs to determine their best fit to the surface density, velocity
dispersion, rotation and ellipticity
profiles. \citealt{2003ApJ...589L..25B} constructed two models: 1) a
single non-rotating cluster, and 2) a rotating merger product, where
the merger occurred during the formation process. They varied the
initial density profiles, half-mass radii, total masses, and global
mass-to-light ratios M/L and evolve each cluster for 13 Gyr. The
authors construct the final density and velocity profiles from 10
snapshots in the range 11.75 - 12.25 Gyr and give their final cluster
mass and half-mass radius at 12 Gyr. They use a
\citet{2001MNRAS.322..231K} mass function in the range $0.1 <
M/M_{\odot} < 30 $ and include the effects of stellar evolution and
two-body relaxation. Their simulations do not contain primordial
binaries, as G1 should still be far from
core-collapse. See the reference mentioned in
Section~\ref{subsubsec:ScaledNbody} for the details of their
simulation.

\paragraph{Orbit and tides}\label{par:Orbitandtides3}

\citealt{2003ApJ...589L..25B} do not include a tidal field, since they
argue that the tides would have a negligible effect on the
cluster\rq{}s evolution, since the cluster is currently at a distance
of 40\,kpc to the center of M31 \citep{2001AJ....122..830M}. We
therefore evolve all our model clusters for G1 for 12 Gyr without
including a tidal field (i.e. as an isolated
cluster); see Table~\ref{table:Validation_clusters2}.

\paragraph{Observed mass and radius}\label{parObservedmassandradius3}

\citet{2003ApJ...589L..25B} got a set of observed half-mass radii
$r_{\rm hm}$ in the range 12.3 - 15.0\,pc and observed total masses in
the range 7.3 - 17 $\cdot 10^6$ M$_{\odot}$ from
\citet{2001AJ....122..830M}, who in turn estimated the half-mass radii
and masses from the surface brightness profile from HST/WFPC2 images
and velocity dispersion profile from KECK/HIRES spectra in combination
with King model, King-Michie model and virial theorem estimates.  We
choose to compare our results to the observables which they obtain
with their King-Michie model nr 4, which gives somewhat average values
of the above mentioned ranges: $r_{\rm hm} = 13.2$\,pc, M = $15 \cdot
10^6$ M$_{\odot}$ and $t_{\rm rh} \sim 50$ Gyr. However, we have
checked that the \lq{}half-mass radius\rq{} mentioned in
\citet{2001AJ....122..830M} comes from their angular projected radii
in arcmin by using Eq. (\ref{eq:arcmin_to_pc}) in
combination with its distance to the Earth $R_{\rm E} = 770$\,kpc,
which they provided in their Table 1. They do not mention that they
corrected for projection (factor $4/3$) or that they did any
correction for mass segregation. We therefore assume that the radius
they refer to as the half-mass radius is actually the projected
half-light radius. For the conversion to the 3D half-mass radius
$r_{\rm hm}$ with a correction for mass segregation
  (see Section~\ref{subsec:Conversion}), we use the relation
(\ref{eq:rhl_to_rhm}) with $c_{\rm MS} = 1.9$.

\paragraph{Model initial and final conditions}\label{par:InitialConditions3}

For G1 we compare to their non-rotating model, since this model is
most comparable to {\sc emacss}. After scaling up, the cluster of this
model obtained a final mass of 7.6 $\cdot 10^6$ M$_{\odot}$ and a
final half-mass radius of 6.76\,pc \citep{2003ApJ...589L..25B}. They
found that during the evolution the cluster mainly expanded by a
factor of 1.75 due to stellar evolution and that it lost about $40\%$
of its mass over 12 Gyr (Baumgardt, private communication). From this
we calculated an initial mass of 1.27 $\cdot 10^7$
M$_{\odot}$ and an initial half-mass radius of 3.86\,pc.

\subsubsection{M67}\label{subsubsec:DirectNbody_M67}

In this section we describe the work using direct $N$-body modeling to
study the evolution of the rich and relatively old Galactic open
cluster M67 (NGC 2682;
\citealt{2005MNRAS.363..293H}). The authors simulate
  the evolution of M67 by using the collisional $N$-body code NBODY4
  \citep{1999PASP..111.1333A} on the GRAPE-6 computers
  \citep{2003PASJ...55.1163M}.

\paragraph{Simulation Technique}\label{par:SimulationTechnique4}

\citet{2005MNRAS.363..293H} modeled the evolution of M67 by performing
$N$-body simulations. They compared their modeled surface density
profile to the surface density profile of M67 of
\citet{2005A&A...437..483B} provided by Bonatto (private
communication), their modeled color-magnitude diagram (CMD) to the
observed CMD of \citet{1993AJ....106..181M} and their modeled
luminosity function and their structural parameters such as the
half-mass radius to the observational data from
\citet{1996AJ....112..628F}. They furthermore extensively study the
stellar populations in their simulation and especially focus on the
formation channels of blue stragglers (BSs) and compare their results
to observational data from \citet{1996AJ....112..628F},
\citet{1996ASPC...90..385L}, \citet{1992IAUS..151..475M} and
\citet{1996ApJ...470..521L}. For the single stars,
they use a \citet{1993MNRAS.262..545K} mass function in the range $0.1
< M/M_{\odot} < 50$ and their model fully accounts for the effects of
cluster dynamics as well as stellar and binary evolution, including a
significant fraction of primordial
binaries. \citet{2005MNRAS.363..293H} constructed two different
models, differing only in the initial mass, the Galactocentric radius
and the binary period distribution. Their second and favoredcolor
model are ran for a star cluster with $N = 36000$ stars. See
\citet{2005MNRAS.363..293H} for the details of their models and the
$N$-body code they used.

\paragraph{Orbit and tides}\label{par:Orbitandtides4}

We compare to the second model of \citet{2005MNRAS.363..293H}, because
it is their best-fit model. In this model they evolved the cluster for
4 Gyr on a circular orbit with a circular velocity of 220 km/s at a
galactocentric radius of 8.0\,kpc, which is a reasonable choice for a
cluster on an slightly eccentric orbit with a perigalaciticon of
6.8\,kpc and a apogalacticon of 9.1\,kpc \citep{1994A&A...288..751C}.

\paragraph{Observed mass and radius}\label{parObservedmassandradius4}

The half-mass radius of Main Sequence stars observed within 10\,pc
that \citet{2005MNRAS.363..293H} used was taken from
\citet{1996AJ....112..628F}, who determined it to be 2.5\,pc. However,
we checked that the \lq{}half-mass radius\rq{} mentioned in
\citet{1996AJ....112..628F} comes from converting their angular
projected radii in arcmin by using Eq.
(\ref{eq:arcmin_to_pc}) in combination with the cluster\rq{}s distance
to the Earth, $R_{\rm E} = 783$\,pc, calculated from their provided
distance modulus of 9.47 mag. They do not mention that they corrected
for projection (factor $4/3$) or that they did any correction for mass
segregation. We therefore assume that the radius they refer to as the
half-mass radius, is actually the projected half-light radius. For the
conversion to the 3D half-mass radius $r_{\rm hm}$ 
  with a correction for mass segregation (see
Section~\ref{subsec:Conversion}), we use the relation
(\ref{eq:rhl_to_rhm}) with $c_{\rm MS} = 1.8$. Both studies took the
total luminous cluster mass from \citet{1996AJ....112..628F}, which
determined that to be 1000\,M$_{\odot}$. \citet{2005MNRAS.363..293H}
estimated that this luminous mass represented a total cluster mass of
about 1400\,M$_{\odot}$.

\paragraph{Model initial and final conditions}\label{par:InitialConditions4}

The best-fit initial conditions \citet{2005MNRAS.363..293H} found for
M67 (model 2) in (mass/\,M$_{\odot}$, half-mass radius/pc) are listed
in Table~\ref{table:results_2D}.

\section{RESULTS}\label{sec:Results}

\subsection{Performance}\label{subsec:Performance}

\begin{figure*}
 \centering
 \includegraphics[width=0.9\textwidth]{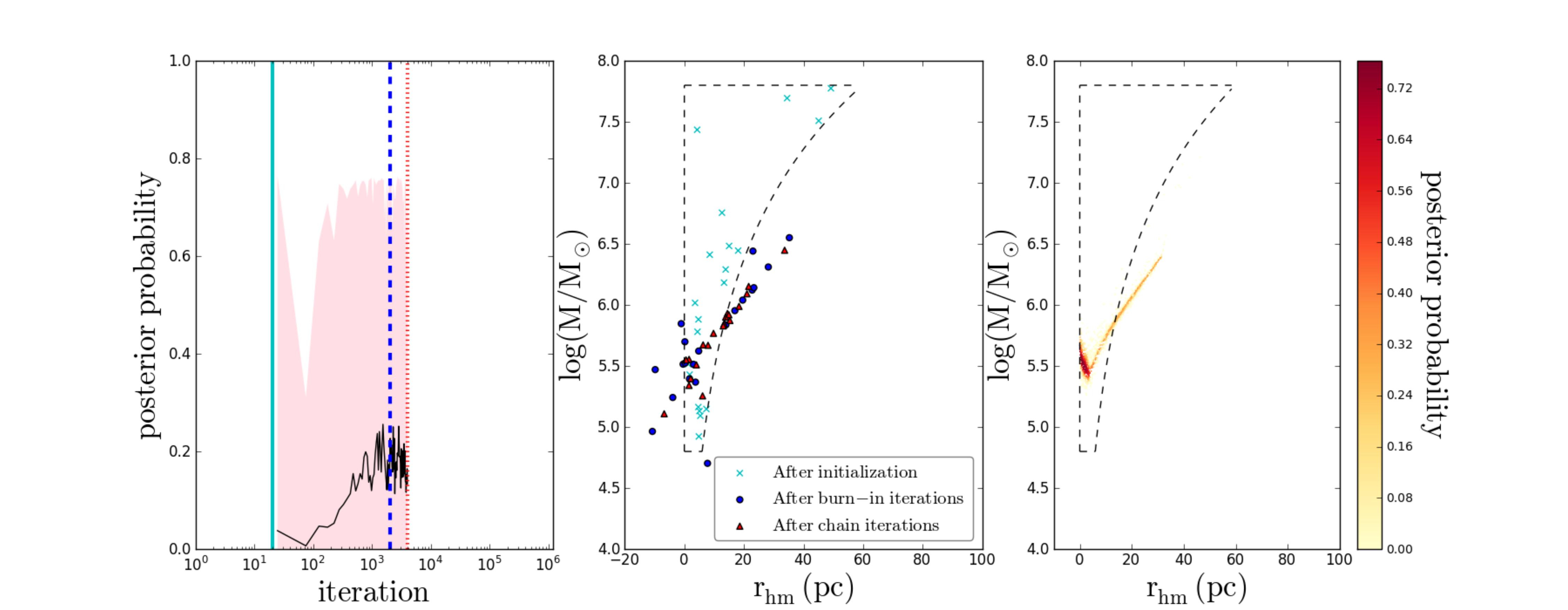}\\ 
 \includegraphics[width=0.9\textwidth]{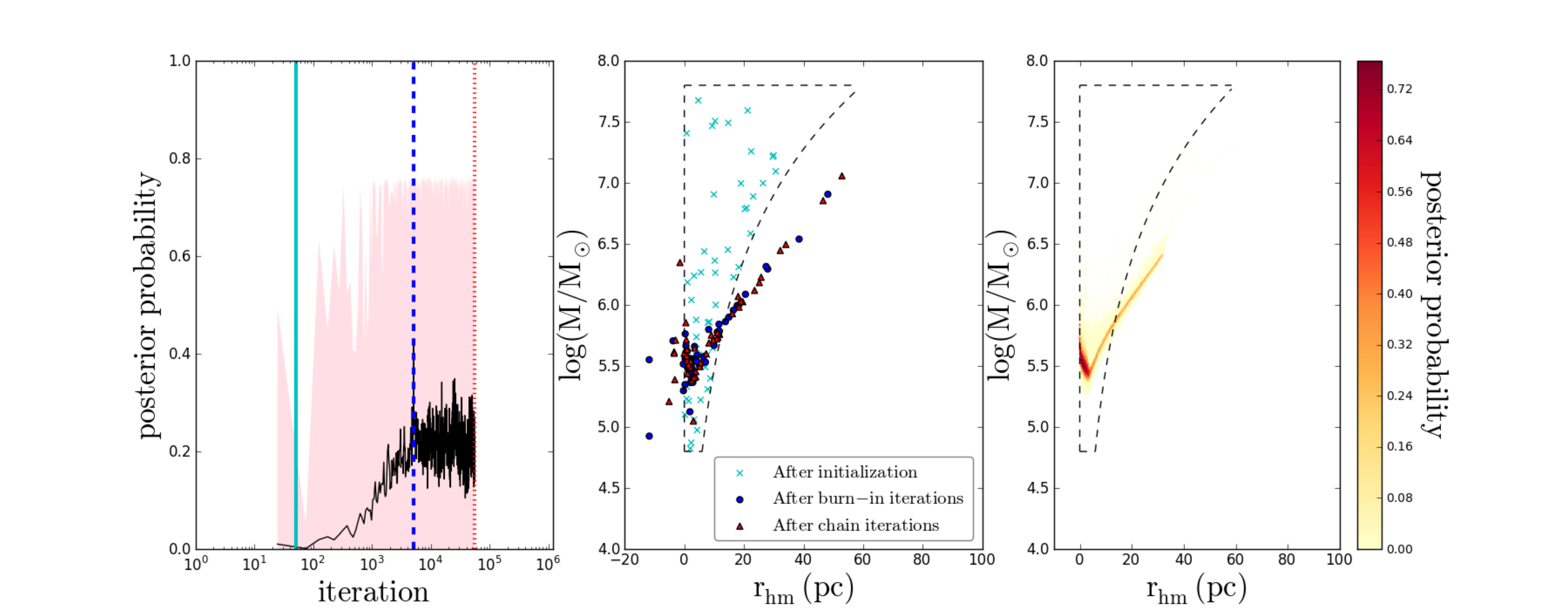}\\ 
 \includegraphics[width=0.9\textwidth]{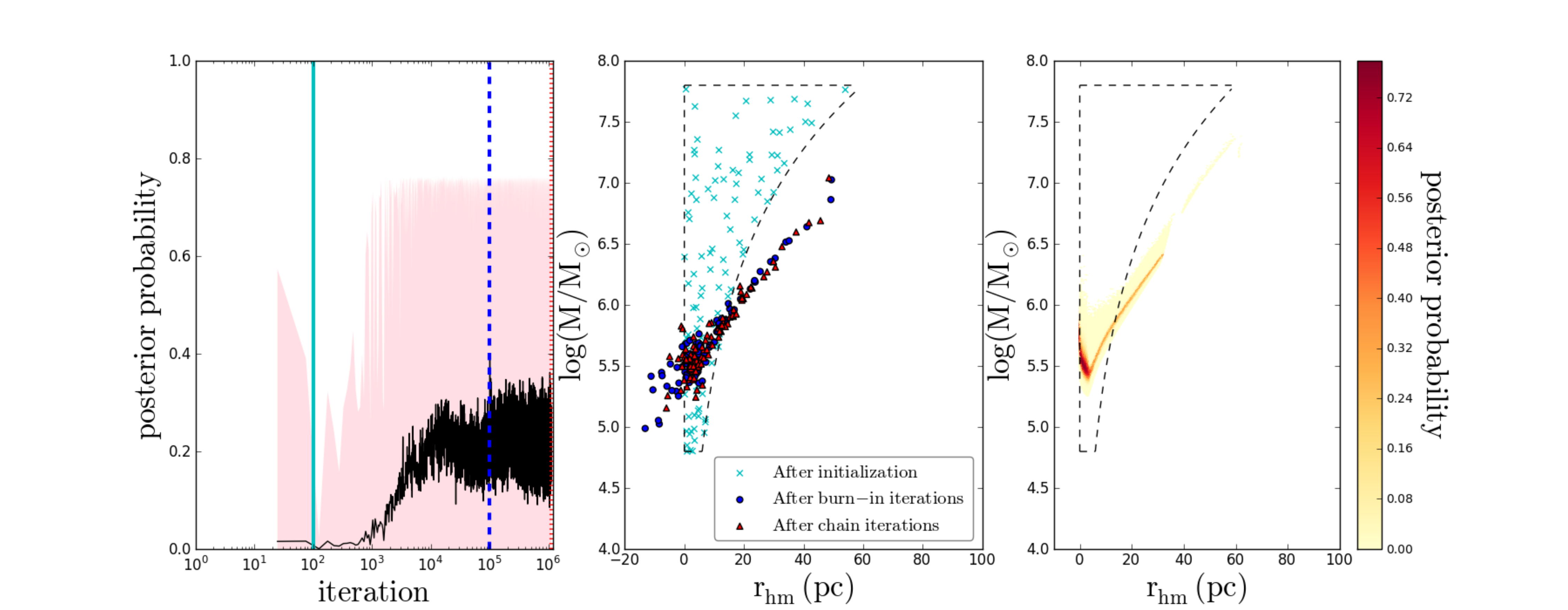}
 \caption{MCMC performance in terms of convergence
     for the simulation of M4 without a correction for mass
     segregation. The left column shows the posterior probability of
     an iteration as a function the iteration number, averaged over
     bins of 50 iterations. The solid black line shows the mean
     probability of each bin, the pink area marks the region between
     the minimum and the maximum probability in each bin and the
     vertical solid cyan, dashed blue and dotted red lines, mark the
     phases after initialization, the burn-in phase and the subsequent
     chain iterations, respectively. The middle column shows the
     distribution of the walkers after initialization (our prior; cyan
     crosses), burn-in (blue circles) and the chain iterations (red
     triangles), respectively. The right column shows the distribution
     of all the sampled initial conditions with a non-zero posterior
     probability, colour coded by the value of the posterior
     probability as indicated by the color bar. The first row is for
     the simulation with ($n_{\rm w}$, $n_{\rm b}$, $n_{\rm c}$) =
     (20, 100, 100), the second row for the simulation with ($n_{\rm
       w}$, $n_{\rm b}$, $n_{\rm c}$) = (50, 100, 1000) and the third
     row for the simulation with ($n_{\rm w}$, $n_{\rm b}$, $n_{\rm
       c}$) = (100, 1000, 10000). \label{fig:performance}}
\end{figure*}

\begin{figure*}
 \centering
 \includegraphics[width=0.9\textwidth]{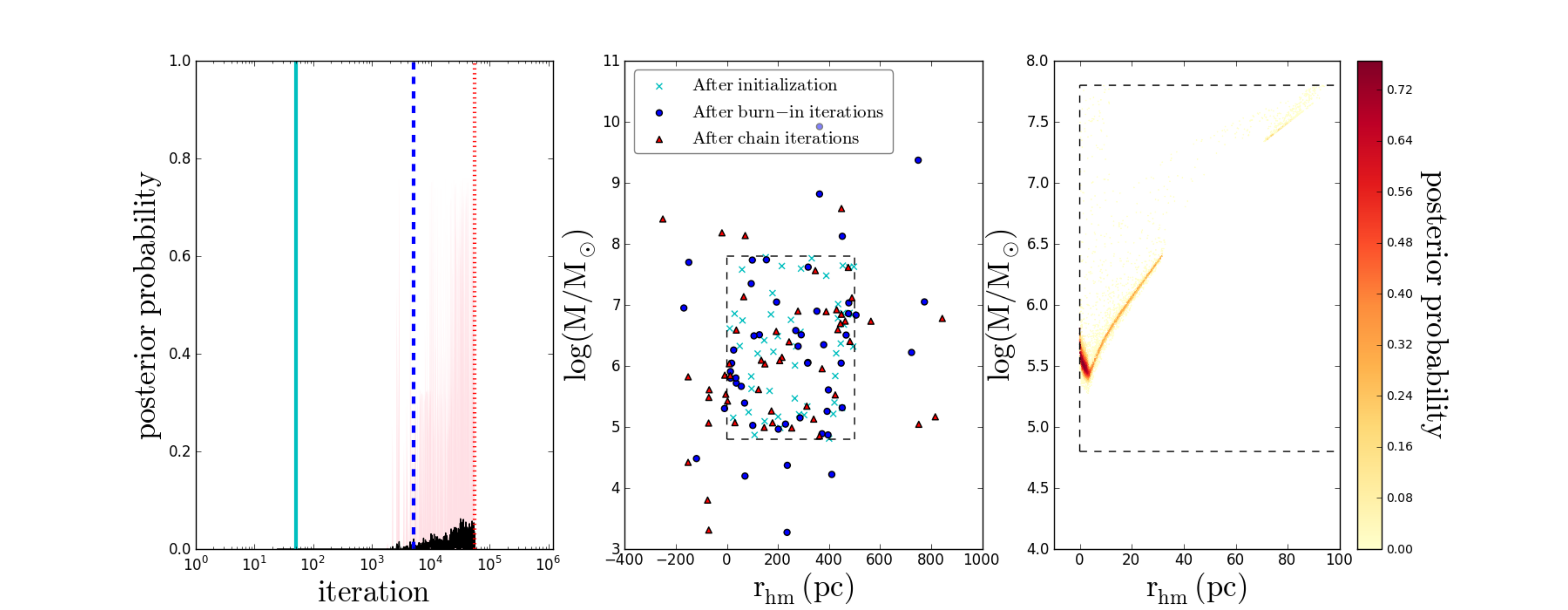}\\ 
 \includegraphics[width=0.9\textwidth]{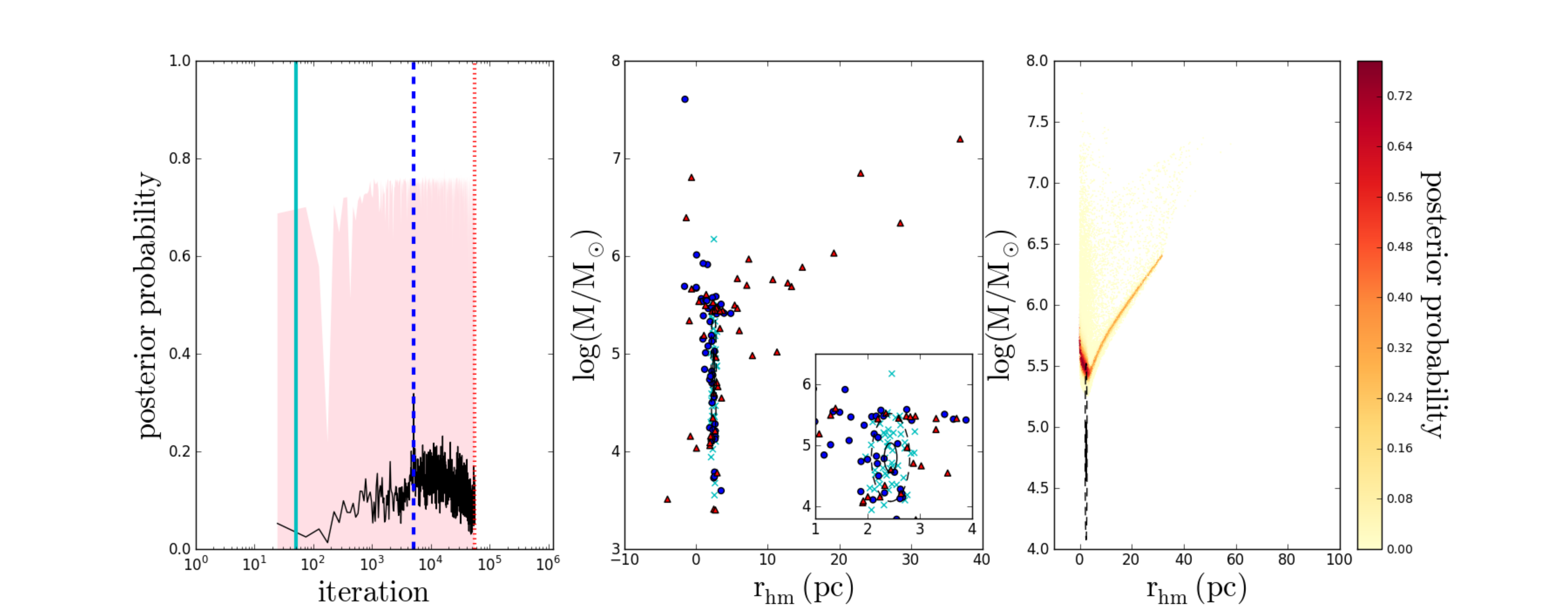}\\ 
 \includegraphics[width=0.9\textwidth]{performance_M4_w50_b100_c1000_2D.pdf}
 \caption{Similar to Figure~\ref{fig:performance},
     but this time showing the dependence of the initial conditions
     and performance in terms of convergence and coverage on the
     choice of prior distribution. The top panel shows the performance
     after choosing a uniform prior distribution indicated by the
     square box. The middle panel shows the performance after choosing
     a normal prior distribution, with the current mass and half-mass
     radius as the mean and a tenth of these values as the standard
     deviation; see the ellipses indicating the one sigma (solid black
     line) and the three sigma lines of the prior distribution in the
     zoomed in smaller panel. The third panel shows the performance
     after choosing a semi-uniform distribution, i.e. the mass is
     drawn uniformly between two boundaries, but the radius is
     dependent on the Jacobi radius, which is mass dependent; see the
     semi-uniform prior indicated by the black dashed
     line.\label{fig:prior}}
\end{figure*}

In this section we test the performance of the {\sc emacss-mcmc}
  method.  We first determined a suitable number of the walkers,
$n_{\rm w}$, burn-in iterations, $n_{\rm b}$, and subsequent chain
iterations, $n_{\rm c}$, such that we have a good balance between
proper coverage and quick convergence. To this end we ran a dozen
simulations for the cluster M4 varying these three numbers and
plotting the posterior probability of an iteration as a function of
the iteration number. We divided all the iteration in bins of 50
iterations and calculated the minimum, maximum and mean probability
per bin, see Figure~\ref{fig:performance}. In these simulations we
  used a prior distribution which is uniform in mass (similar to the
  first line of Eq. (\ref{eq:initialization})), but semi-uniform in
  half-mass radius: the upper limit of the half-mass radius is
  mass-dependent through the dependence on the Jacobi radius, $r_{\rm
    J}$:
\begin{eqnarray}\label{eq:prior}
\log(M_{\rm obs}) \leq &\log( \frac{M}{M_{\odot}})& \leq \log(M_{\rm
  obs}) + 3, \nonumber \\ 0 < &\frac{r_{\rm hm}}{\rm pc}& < 0.3 r_{\rm
  J}.
\end{eqnarray}
This mass-dependent upper limit of the initial half-mass radius is
motivated by the fact that most clusters with initial half-mass radii
larger than $30\%$ of the Jacobi radius will quickly dissolve
\citep{2014MNRAS.442.1265A}. As we show later on in this section,
initializing the walkers according to this prior does not exclude the
parameter space with $r_{\rm hm}/r_J > 0.3$, but it does accelerate
the convergence.\\ \\
In order to judge whether the walkers have converged, we look at the
instance that both the maximum and the mean posterior probability have
stabilized, which means that its value does not increase or decrease
by a significant amount, say $100 \%$, and thus only
show the variation caused by the scatter. The scatter in the mean
posterior probability is a natural feature of any MCMC sampler,
because it is important that even though a (local) maximum in
posterior probability has been found, the walkers continue to explore
other regions of parameter space, in order to locate possible other
maxima.\\ \\ From the first row of Figure~\ref{fig:performance},
showing the simulation with ($n_{\rm w}$, $n_{\rm b}$, $n_{\rm c}$) =
(20, 100, 100), we see that the walkers already start to converge
after about 2000 iterations, marking the end of the burn-in
phase. However, from the right column of the first row, we see that
the coverage of the $M-r_{\rm hm}$ plane is still poor, meaning that we can observe by eye that the two-dimensional parameter space is not well-sampled and/or that the sampled region does not cover a large region of that parameter space. In the second
row of Figure~\ref{fig:performance} we see that both convergence and
proper coverage, by eye, seem to have been established for the simulation with
($n_{\rm w}$, $n_{\rm b}$, $n_{\rm c}$) = (50, 100, 1000): the walkers
probed a wide range of initial masses and half-mass radii in proposing
initial conditions. Nevertheless, we have chosen to be conservative
and to use ($n_{\rm w}$, $n_{\rm b}$, $n_{\rm c}$) =
  (50, 100, 1000) only for \textit{test
    simulations}\footnote{Testing the effect of the
      prior distribution or observational errors on the determined
      probable initial condition distributions.} and to use ($n_{\rm
    w}$, $n_{\rm b}$, $n_{\rm c}$) = (100, 1000, 10000) for all main
  simulations in this work, see the third row of
Figure~\ref{fig:performance}.\\ \\ We secondly
  determined whether the choice of a prior distribution would effect
  the ranges of parameter space that are covered and thus the
  determined probable initial condition distribution. To test this we
  ran three simulations to determine the initial conditions for the
  cluster M4 with ($n_{\rm w}$, $n_{\rm b}$, $n_{\rm c}$) = (50, 100,
  1000) using three different priors, but otherwise being the
  same. These three priors were:
\begin{enumerate}  
\item[1] a uniform distribution in both (logaritmic) mass and
  half-mass radius, according to Eq. (\ref{eq:initialization});
\item[2] a normal distribution in both parameters with the mean equal to $\log(M_{\rm obs})$ and $r_{\rm hm,obs}$ respectively, and a standard deviation equal to $10\%$ of the mean;
\item[3] a semi-uniform distribution, according to
  Eq. (\ref{eq:prior}).
\end{enumerate}
The results of these three simulations are shown in
Figure~\ref{fig:prior}. From this figure we can see that the covered
area in the $M-r_{\rm hm}$ plane of the probable initial conditions is
similar. If we consider all sampled initial conditions (also those
outside of the boundaries given by Eq. (\ref{eq:initialization}), e.g. negative half-mass radii, with
posterior probabilities equal to zero, which is not shown in
Figure~\ref{fig:prior}), the sampled area is significantly different:
the simulation with a uniform prior covers a larger range in initial
half-mass radius ($-500 \la$ (r$_{\rm hm,i}/pc) \la 1000$) and
slightly different, but overlapping range in logarithmic mass ($2.0
\la \log(M_{\rm i}/M_{\odot}) \la 10.5$) compared to the semi-uniform
prior ($-68.9 \la$ (r$_{\rm hm,i}/pc) \la 150$ and $2.9 \la
\log(M_{\rm i}/M_{\odot}) \la 9.8$) and the normal prior ( $-26.9 \la$
(r$_{\rm hm,i}/pc) \la 60.4$ and $1.1 \la \log(M_{\rm i}/M_{\odot})
\la 9.1$). If we consider initial conditions with posterior
probability greater than zero, then the covered area is slightly more
similar: the walkers initialized according to a uniform prior reach
larger initial half-mass radii, but the initial mass range is
comparable (uniform: $10^{-3} <$ (r$_{\rm hm,i}/pc) < 95.7$ and $5.3 <
\log(M_{\rm i}/M_{\odot})< 7.8$; semi-uniform: $10^{-4} <$ (r$_{\rm
  hm,i}/pc) < 63.8$ and $5.3 < \log(M_{\rm i}/M_{\odot})< 7.8$;
normal: $10^{-4} <$ (r$_{\rm hm,i}/pc) < 58.0$ and $5.3 < \log(M_{\rm
  i}/M_{\odot})< 7.7$). We observe in the top panel of
Figure~\ref{fig:prior} that there is an area of initial conditions
with high initial mass and high initial radius, which is not probed
with the simulations with the other two priors. However, this is not a
favourable region in terms of posterior probability and thus we
conclude that simulations with different prior distributions properly
cover the relevant ranges of parameter space and that the derived
initial conditions are prior independent. Similar behaviour is seen
for the other clusters in our sample. We use the semi-uniform prior distribution for the rest of our simulations in this work.\\ \\
The final performance characteristic to test is the sampling of
initial conditions by the walkers. The results of all the simulations
for both the 2D and 5D runs are shown in
Figures~\ref{fig:IC_and_FC1}-\ref{fig:IC_and_FC9},
  Figure~\ref{fig:IC_and_FC_5D1} and
  Figures~\ref{fig:IC_and_FC_5D2}-\ref{fig:IC_and_FC_5D9}. By
comparing the two-dimensional histograms with the confidence contours
in these figures, we see that the most sampled areas overlap with the
high posterior probability regions. This is what one would expect for
an MCMC method with a sufficient number of iterations, and hence shows
the proper performance of our method. We note that
the number of sets of initial conditions in the $99.7\%$ and $68.3\%$
confidence regions is less than
$\sim$$99.7\%$, respectively
$\sim$$68.3\%$, of the total number of surviving
initial conditions, and that increasing the number of iterations does
not increase these percentages. For instance, for M4
the number of sets of initial conditions in the $99.7\%$ and $68.3\%$
confidence regions is $\sim$$80\%$ respectively
$\sim$$40\%$ of the total number of surviving initial conditions for
both the simulation with ($n_{\rm w}$, $n_{\rm b}$, $n_{\rm c}$) =
(50, 100, 1000) and the simulation with ($n_{\rm w}$, $n_{\rm b}$,
$n_{\rm c}$) = (100, 1000, 10000). This is again
  because even though the walkers have converged to high posterior
  probability regions of parameter space, they continue to sample
  unexplored regions as well.

\subsection{Direct comparison}\label{subsec:Direct_comparison_results}

\begin{figure*}
  \centerline{\includegraphics[width=200mm]{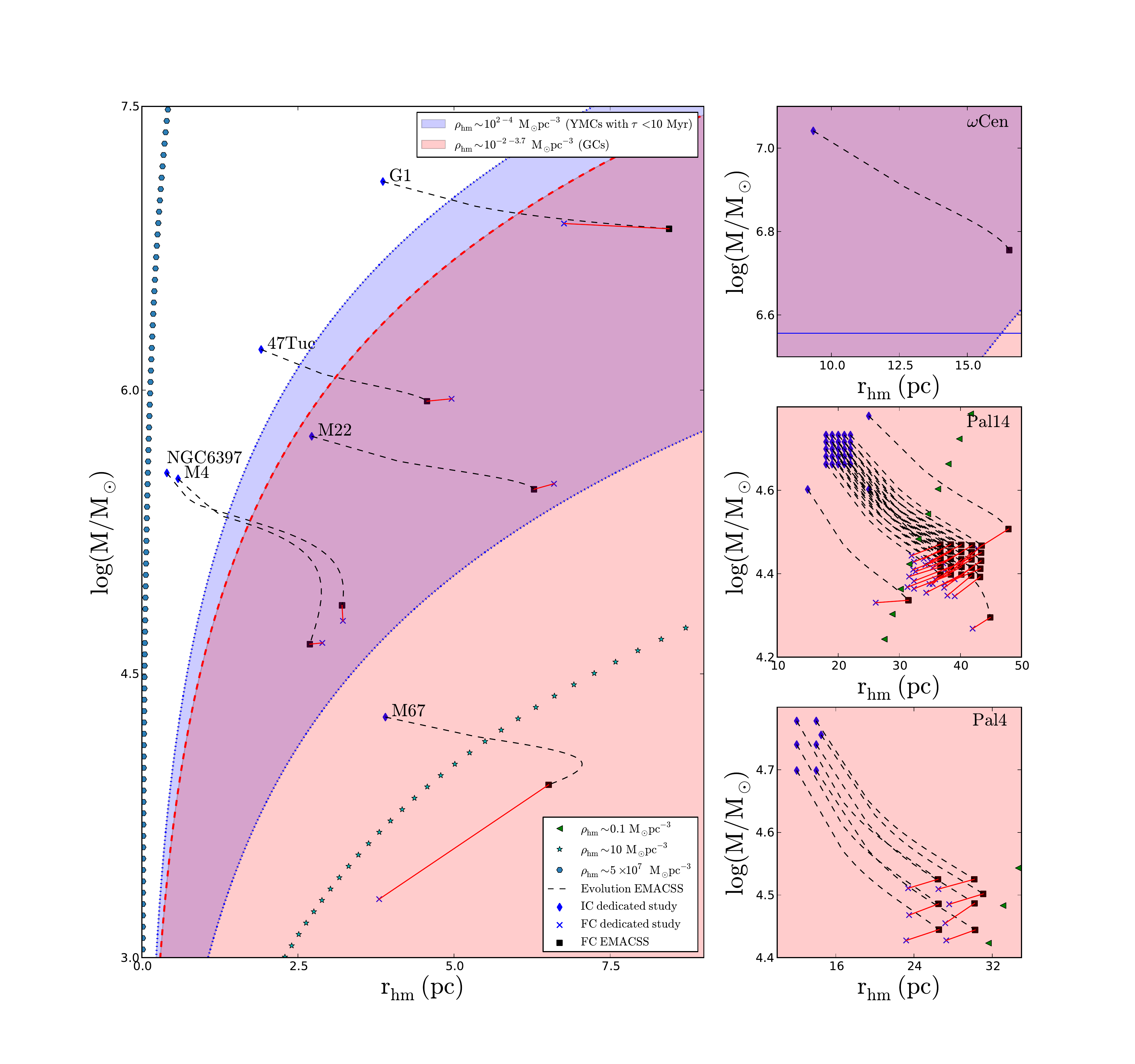}}
  \centering
      \caption{Direct comparison between {\sc emacss}
          and the dedicated studies (DSs) by running {\sc emacss} from
          the best-fit initial condition of each DS. Each panel shows
          the total mass versus half-mass radius for one of the nine
          validation clusters. The blue diamond(s) and cross(es) show
          the best-fit initial and final condition(s) respectively of
          the DS; the black dashed line shows the evolution with {\sc
            emacss} when started from the initial condition of the DS
          and the black square shows the final condition of this
          evolution with {\sc emacss}. For $\omega$ Cen we show a blue
          line, indicating the final mass of the DS, since their final
          half-mass radius was not given. For Pal 4 and Pal 14 we
          compare to more than one model, so we connect the final
          conditions of the DS and {\sc emacss} with a red solid line
          for clarity. As a comparison we also plot the minimum
          densities a cluster needs to have in order to be stable
          against the tidal disruption of a galaxy ($\rho_{\rm hm}
          \sim 0.1$ M$_{\odot}$pc$^{-3}$ \citep{1934HarCi.384....1B},
          green triangle line) and against passing giant molecular
          clouds ($\rho_{\rm hm} \sim 10$ M$_{\odot}$pc$^{-3}$
          \citep{1958ApJ...127...17S}, cyan star line). The turquoise
          hexagon line shows a mean density of $\sim 5 \cdot 10^7$
          M$_{\odot}$pc$^{-3}$ that might be required for a cluster
          with a half-mass radius of 0.2 pc to form an IMBH via a
          runaway merger \citep{2008CQGra..25k4007M}. We furthermore
          show the range of observed half-mass densities for globular
          clusters, $\rho_{\rm hm} \sim 10^{-2-3.7}$
          M$_{\odot}$pc$^{-3}$, (red shaded region) and the range of
          present-day observed half-mass densities for clusters
          younger than 10 Myr ($\rho_{\rm hm} \sim 10^{2-4}$
          M$_{\odot}$pc$^{-3}$ \citep{2010ARA&A..48..431P} (blue
          shaded region); the purple shaded region is the overlap of these blue and red regions. \label{fig:direct_comparison}}
\end{figure*}

Figure~\ref{fig:direct_comparison} shows the direct comparison between
{\sc emacss} and the dedicated studies by running {\sc emacss} from
their best-fit initial condition. All results presented in this
section are summarized in Table~\ref{table:results_2D}, where we
compare the best-fit results of the dedicated study (DS) and {\sc
  emacss}\rq{} result when starting from the initial condition of the
dedicated study in row two and three for each cluster.\\ \\
From Figure~\ref{fig:direct_comparison} we see that the {\sc emacss}
results compare quite well to the Monte Carlo and the $N$-body results
for NGC 6397, M4, M22, Pal 4, 47 Tuc,
Pal 14 and G1, with a difference in final conditions
$<\, 25\,\%$ with respect to the dedicated studies in both
(linear) mass and half-mass radius for each of these
clusters. For M4, M22 and 47 Tuc, we see that the evolution with {\sc
  emacss} led to a close match in final conditions (with $<\, 8\%$
difference in both mass and half-mass radius) losing slightly more
mass during its evolution, and reaching smaller final radii. For these
three clusters the final radii are smaller, because the cluster did
not expand as much as in the MOCCA simulation. For M4 the radius did
expand up to 2.87\,pc in the evolution with {\sc emacss}, but near the
end of the simulation the cluster already started to contract due to
stellar evaporation. For NGC 6397 we have almost an exact match in
final half-mass radius ($<\, 1\%$ difference), but the amount of
mass loss is less, leading to a difference of $\sim
21\,\%$ in linear mass with respect to the dedicated
study.\\ \\ For G1 the cluster modeled with {\sc
    emacss} more mass than the cluster modeled with a scaled $N$-body
  simulation ($\sim 6\, \%$ difference in mass), but it also expanded
  more ($\sim 25\, \%$ difference in radius). However, we note that in
  calculating the scaled-up version of the initial mass and half-mass
  radius of G1 found by \citet{2003ApJ...589L..25B}, we assumed that
  the same amount of mass loss and stellar evolution induced expansion
  occured as in the small scale model. This does not have to be the
  case. Therefore it could be that the scaled up initial condition
  from which we start evolving with EMACSS is different from the
  actual scaled up version of the initial condition of
  \citet{2003ApJ...589L..25B}, which would cause the differences in
  final mass and half-mass radius. For Pal 4 and Pal 14, we see that
on average the direct $N$-body modeling caused the modeled clusters to
lose slightly more mass than the clusters modeled with {\sc emacss}
leading to $< 15\, \%$ difference in mass for Pal 4 and $< 21\, \%$
difference in mass for Pal 14. For both Pal 14 and Pal 4, the {\sc
  emacss} clusters expanded a bit more ($< 4\, \%$
  and $< 2\, \%$ difference in radius, respectively).\\ \\ For
$\omega$ Cen the {\sc emacss} results compare less well to those
obtained from MOCCA. Their simulated cluster lost substantially more
mass (leading to a $\sim 59\,\%$ difference in
linear mass). We could not compare the final
half-mass radii, since \citet{2003MNRAS.339..486G} did not provide
this, as described in Section~\ref{par:InitialConditions}. It could be
that our assumption for calculating what the initial half-mass radius
for the best-fit of \citet{2003MNRAS.339..486G} (see
  Section~\ref{par:InitialConditions}) was incorrect and that their
initial half-mass radius was somewhat smaller or
larger. Starting at a different initial radius could
lead to a different amount of mass loss. However, we
have done several {\sc emacss} runs for $\omega$ Cen, starting at the
same initial mass, but with different initial half-mass radii in the
range 0.01 - 30\,pc and we saw that the amount of
mass loss changed only minimally. The
mass loss changed at most by a factor $\sim 1.08$
between two radii in this range. For initial half-mass radii $>$
20\,pc the amount of mass loss decreased. Only for
clusters initially more compact than 0.5\,pc will the amount
mass loss increase significantly,
but not as much as in the MOCCA simulation. It is also not likely that
$\omega$ Cen started out so compact, see
Section~\ref{subsec:Independent_2D_results}. There is also
poor agreement between {\sc emacss} and the direct
$N$-body results for the open cluster M67. Again, we see that the
cluster simulated with {\sc emacss} lost substantially less
mass. Furthermore, the {\sc emacss} cluster expanded by almost a
factor two during its evolution, whereas the final half-mass radius of
the cluster modeled with direct $N$-body integration is even slightly
smaller than the initial one. These differences require some
explanation.\\ \\ Besides starting from the same
  initial total mass and half-mass radius, we kept the conditions of
  our simulation as similar as possible to those of the dedicated
  studies. However, there are a number of
parameters that could not be taken equal between the
codes. One of these parameters is the initial number
of stars. Since {\sc emacss} is currently tested
against $N$-body simulations for only one value of the initial mean
mass ($\tilde{m} = 0.64$\,M$_{\odot}$) we always
used this value for all our simulations. This means that once we set
the initial total mass of the cluster, we immediately set the initial
total number of stars as well, which was different from the initial
number of stars in each of the dedicated models, see
  Table~\ref{table:results_2D}. This lead to different initial half-mass relaxation time scales, $t_{\rm rh}$,  with a larger $t_{\rm rh}$ for larger $N$. The initial half-mass relaxation time in the simulations of the dedicated studies for M67, $\omega$ Cen, Pal 4 and Pal 14 respectively was a factor 1.23, 1.47, 1.28 and 1.12 respectively smaller than the EMACSS simulations. One would thus intuitively expect the clusters with shorter initial half-mass relaxation times to dissolve quicker, and this would lead to a relatively quicker mass loss in the first phase of its life \citep{2010MNRAS.409..305L}. Furthermore, \citet{2003gmbp.book.....H} explain that in larger models, i.e. models with larger $N$, the escape rate per relaxation time is larger. This could explain why the dedicated studies on M67, $\omega$ Cen, Pal 4 and Pal 14 lost more mass, since they start with a larger number of stars initially. For the clusters M4, M22 and 47Tuc the evolution with EMACSS started out with larger number of stars, and thus larger initial half-mass relaxation times by a factor 1.09, 1.07 and 1.25 respectively. Here we thus see that the simulated clusters with EMACSS lost more mass, but not by much, see Figure~\ref{fig:direct_comparison}. For NGC 6397 and G1 the initial number of stars of the dedicated studies was not given.\\ \\
Also, the prescription of stellar evolution and the (mass limits of the) initial mass function used by
{\sc emacss} and by the dedicated studies and sometimes the Galactic
potential, setting the tidal field, were different. All this taken
together led only to moderate differences in the final cluster total
mass and half-mass radius for the clusters NGC 6397, M4, M22, Pal 4,
47 Tuc, Pal 14 and G1, but to more significant difference for $\omega$
Cen and M67. For M67, for example, it led to a larger
  initial Jacobi radius ($r_{\rm J}$ = 37.6\,pc) in EMACSS than the
  (similar) tidal radius ($r_{\rm t}$ = 31.8\, pc) in the N-body
  simulation.\footnote{Note again that $r_{\rm J} =
      r_{\rm t}$ for the type of potential we use here.} This, in
  part, explains the smaller amount of mass loss in the simulation
  with EMACSS and hence the overall larger amount of
  expansion. However, for $\omega$ Cen the
  difference in mass loss cannot be explained by
  this, since we had an initial Jacobi radius similar to the tidal
  radius in the MOCCA simulation ($r_{\rm J}$ = 89.9\,pc; $r_{\rm t}$
  = 90 pc).  \\ \\ Another difference is that all dedicated studies
(except for \citealt{2003MNRAS.339..486G}) include some direct
prescription for binaries, whereas the version of {\sc emacss} used
for this study does not. For each globular cluster in our sample, the
dedicated studies had only a small ($<\, 1\%$) or no primordial binary
fraction, so the effects that binaries have on the evolution of global
cluster parameters such as total mass and half-mass radius are
expected to be small here. Thus for validation purposes, binaries are
not expected to cause major differences in the results between {\sc
  emacss} and the dedicated studies for this sample of globular
clusters. For open cluster M67, however, a large initial binary
fraction of $50\%$ was assumed in \citet{2005MNRAS.363..293H} and they
showed that this fraction even increased throughout the 4 Gyr of
evolution, also due to the evaporation of single stars. So for M67 we
do expect binaries to play an important role in at least the cluster
dynamics and this could certainly result in a very different evolution
when compared to the evolution without binaries. This
therefore also contributes to the difference between
     {\sc emacss} and the direct $N$-body simulation of
     M67. Just like mass loss due to stellar
       evolution, hard binaries are expected to cause expansion of the
       half-mass radius \citep{2011MNRAS.410.2698G}, and so
     based on this type of binaries alone one would
     not expect the cluster evolved with {\sc emacss} to have expanded
     more. However, it is not obvious what to expect
       in terms of expansion of the half-mass radius for a steady
       $50\%$ fraction of primordial and dynamically formed, hard and
       soft binaries. An interesting side note is that when we
     continued the evolution for M67 for another
       $\sim 5\,$Gyr, it reaches a mass and
     half-mass radius within 4\% of those of
     \citet{2005MNRAS.363..293H}. This might suggest
       that the presence of binaries accelerates the evolution. To
       test this hypothesis, we are running an N-body simulation
       similar to the one conducted by \citet{2005MNRAS.363..293H},
       but without binaries, of which we will show the results in
       forthcoming work.\\ \\ The fact that {\sc emacss} is designed
     for evolving globular clusters instead of open clusters is not in
     itself a reason for the sparse agreement. The majority of open
     clusters reside in the disk and they are more likely to undergo
     encounters with giant molecular clouds, which have dramatic
     effects on the cluster evolution
     \citep{2006MNRAS.371..793G}. However, even though {\sc emacss}
     does not include these effects, neither did the simulations of
     \citet{2005MNRAS.363..293H}, so for the comparison between the
     two codes for the validation, this should not matter. Applying
     our method to an open cluster for actually constraining its
     initial conditions, this will become important and some
     prescription taking into account these disruptive events is
     required.

\subsection{Independent {\sc emacss-mcmc} runs}\label{subsec:Independent_2D_results}
\begin{figure*}
      \includegraphics[width=1.08\textwidth]{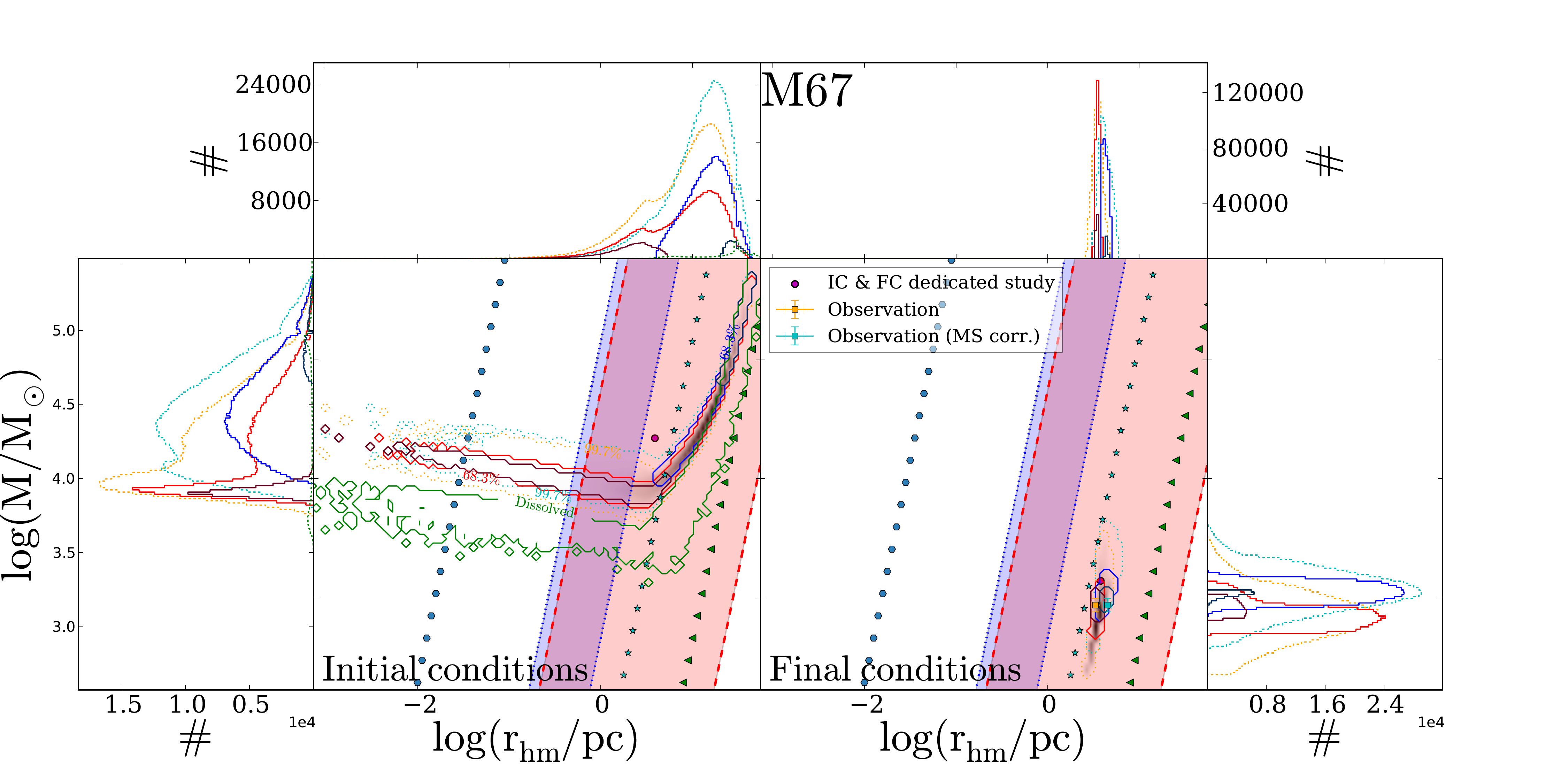}
      \centering
      \caption{The same as in Figure~\ref{fig:simple_results}, but for the cluster M67 and with more details: the contours and histograms of the $99.7\%$ confidence level (dotted), the $68.3\%$ (solid) confidence level and the $p > 0.9p_{\rm max}$ (solid) initial conditions for both the simulations without (yellow, red and brown, respectively) and with (cyan, blue and dark blue, respectively) a correction for mass
          segregation. The green contour and histograms show the
          clusters which dissolved before reaching the age of the
          cluster, $\tau_{\rm obs}$. The shaded regions, the lines depicted by green triangles, cyan stars and turquoise hexagons are as in Figure~\ref{fig:direct_comparison}.\label{fig:IC_and_FC1}}
\end{figure*}

\begin{figure*}
      \includegraphics[width=1.08\textwidth]{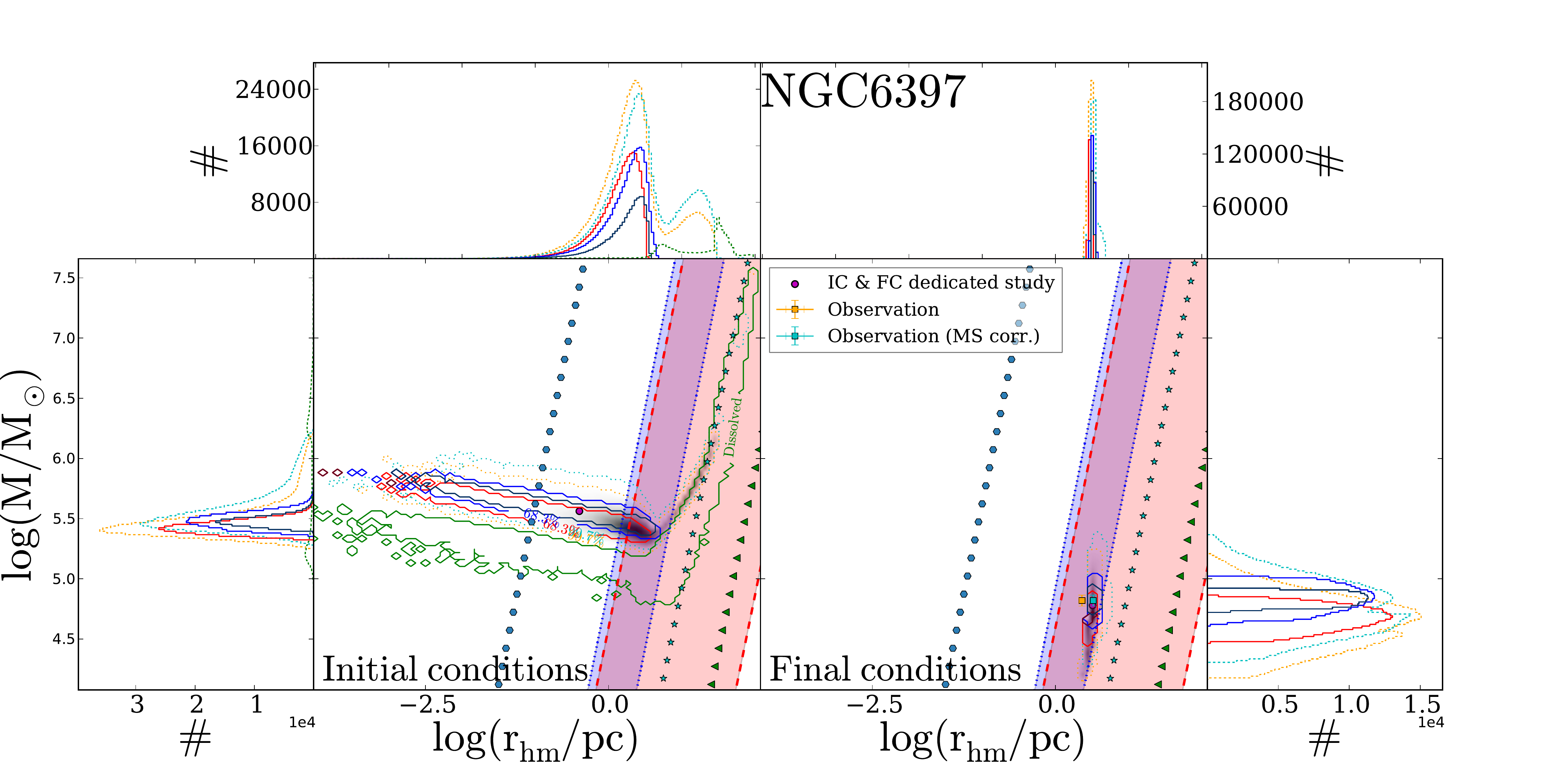}
      \centering
      \caption{The same as Figure~\ref{fig:IC_and_FC1}, but for the cluster NGC6397.\label{fig:IC_and_FC2}}
      \includegraphics[width=1.08\textwidth]{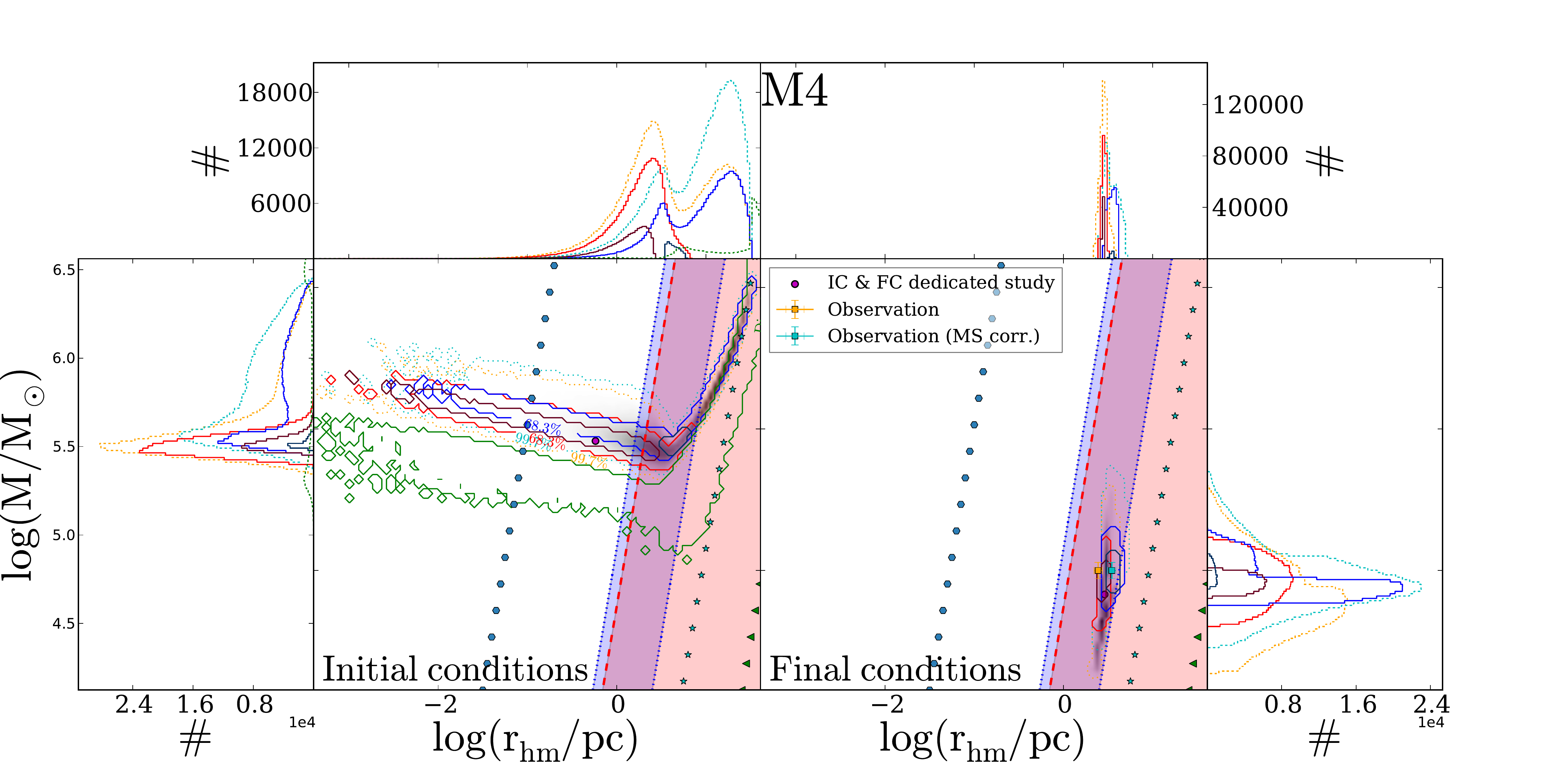}
      \centering
      \caption{The same as Figure~\ref{fig:IC_and_FC1}, but for the cluster M4. \label{fig:IC_and_FC3}}
\end{figure*}

\begin{figure*}
      \includegraphics[width=1.08\textwidth]{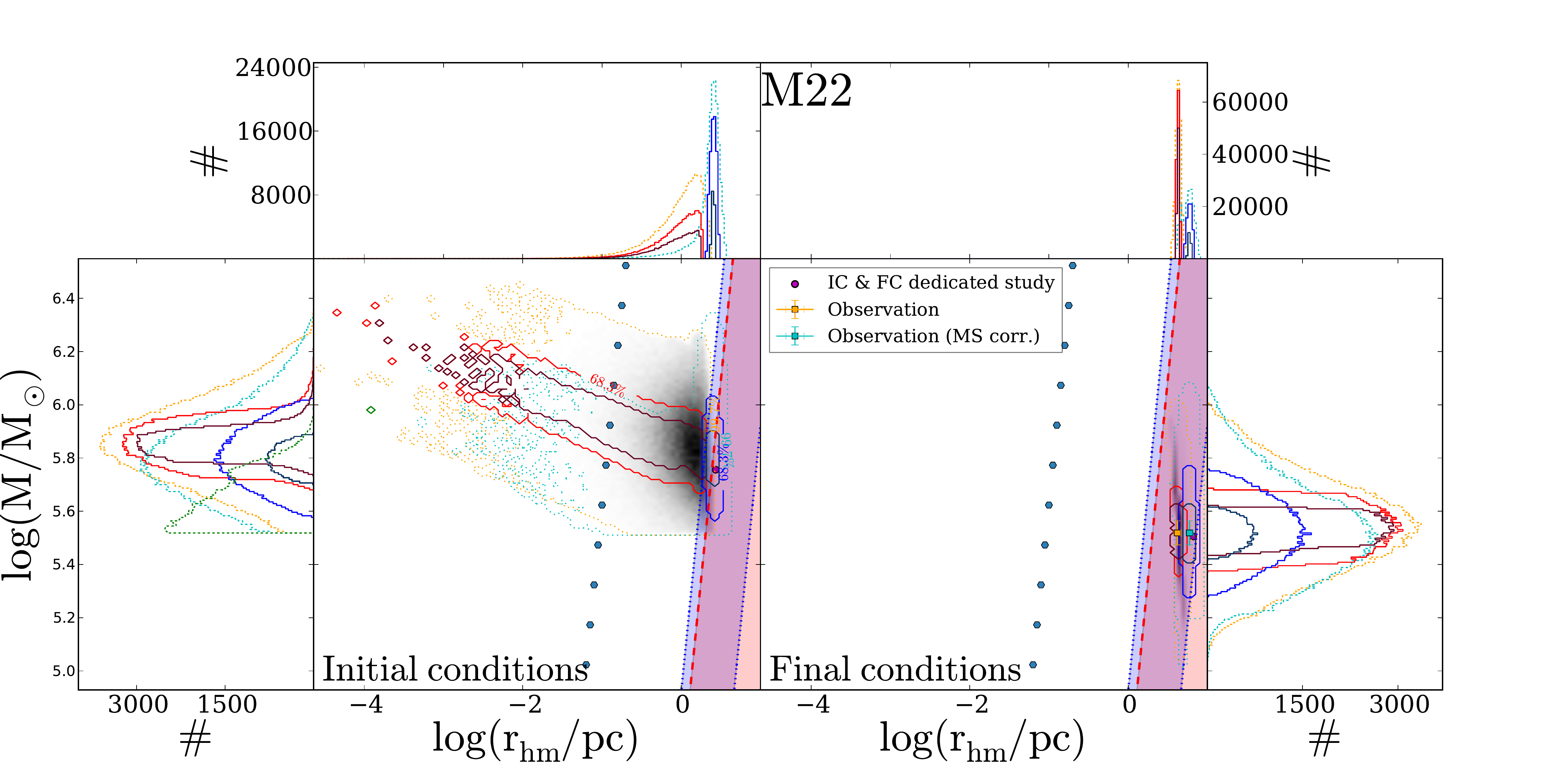}
      \centering
      \caption{The same as Figure~\ref{fig:IC_and_FC1}, but for the cluster M22.\label{fig:IC_and_FC4}}
      \includegraphics[width=1.08\textwidth]{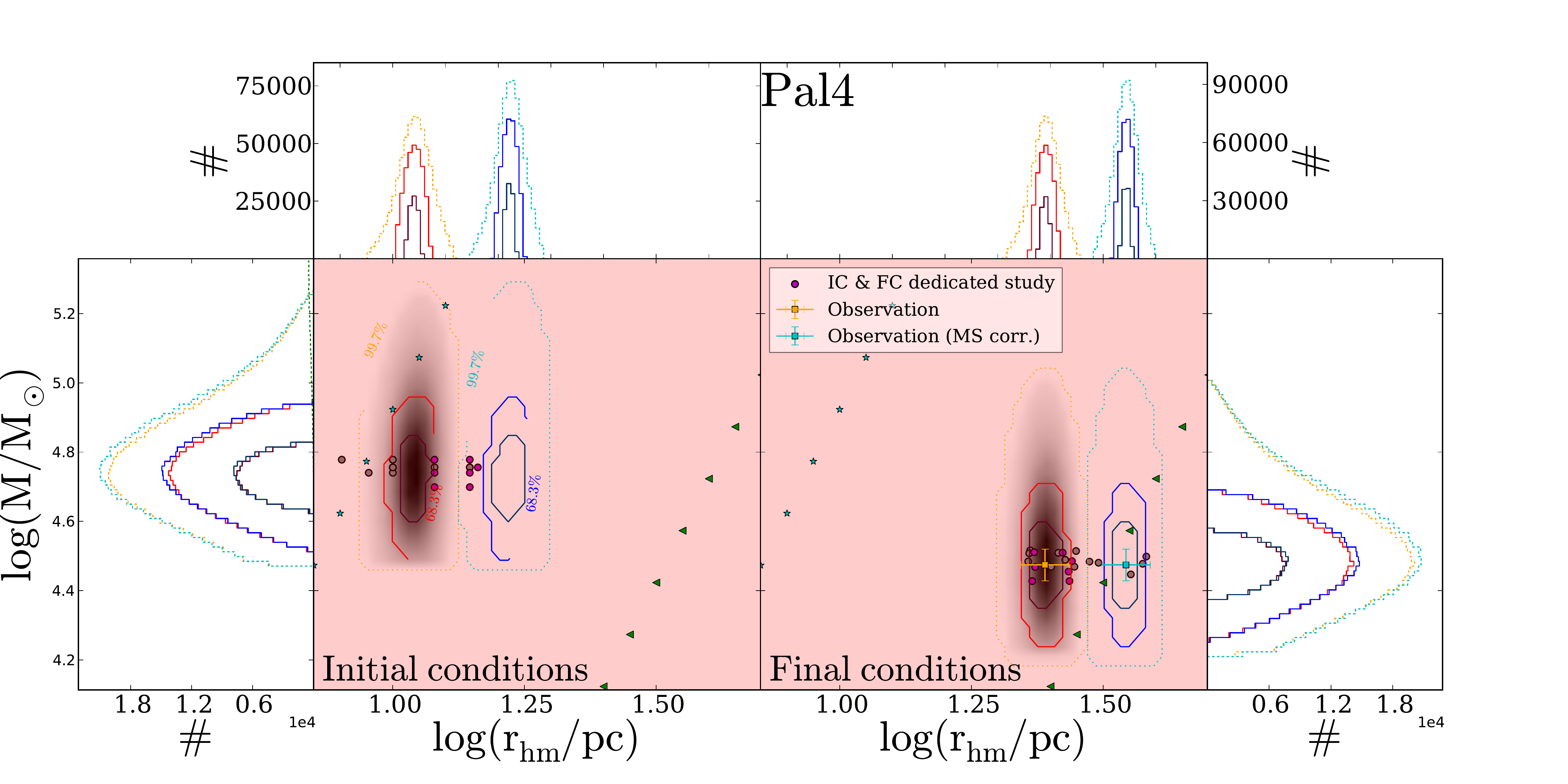}
      \centering
      \caption{The same as Figure~\ref{fig:IC_and_FC1}, but for the cluster Pal4.\label{fig:IC_and_FC5}}
\end{figure*}

\begin{figure*}
      \includegraphics[width=1.08\textwidth]{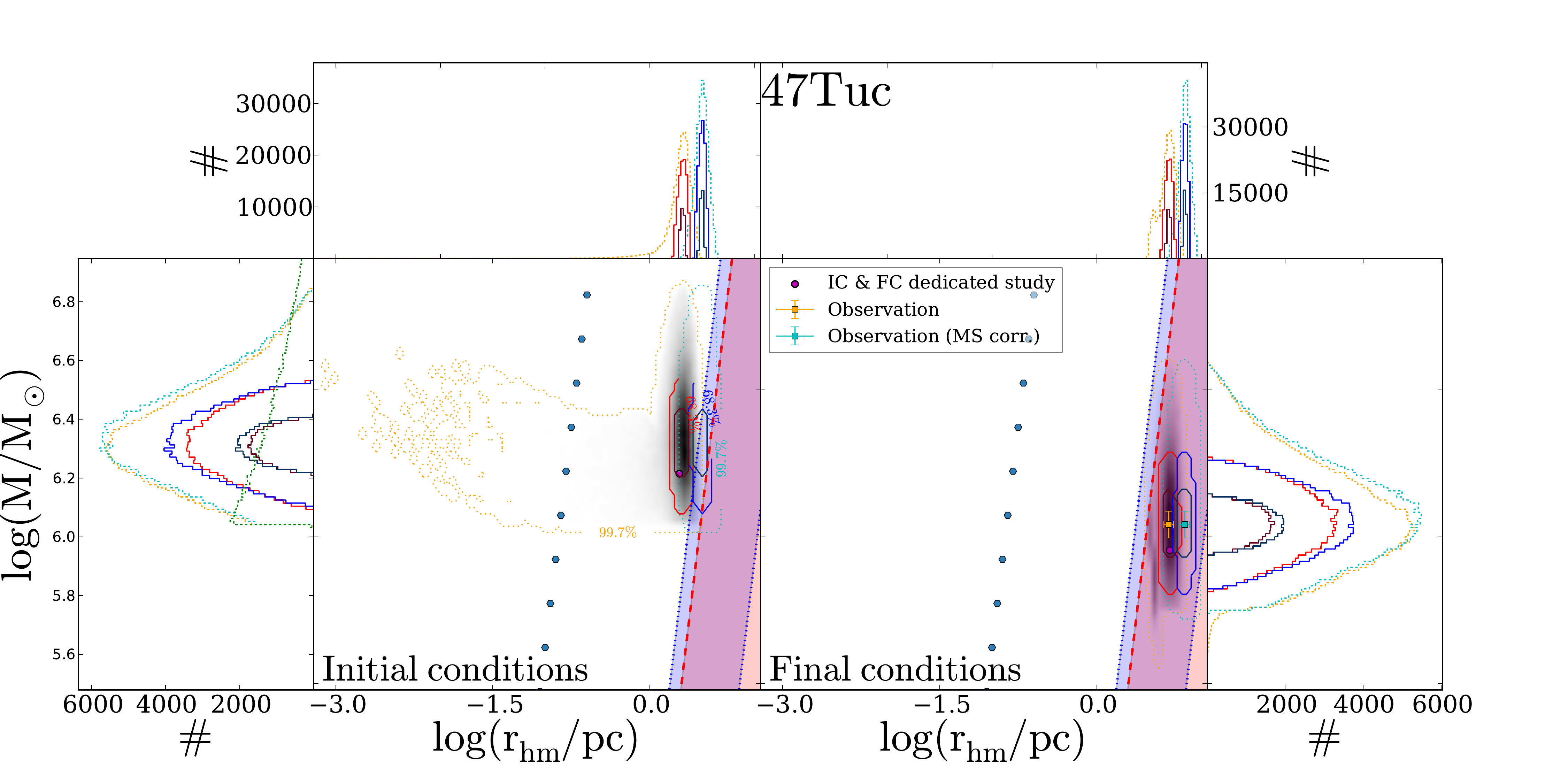}
      \centering
      \caption{The same as Figure~\ref{fig:IC_and_FC1}, but for the cluster 47Tuc.\label{fig:IC_and_FC6}}
      \includegraphics[width=1.08\textwidth]{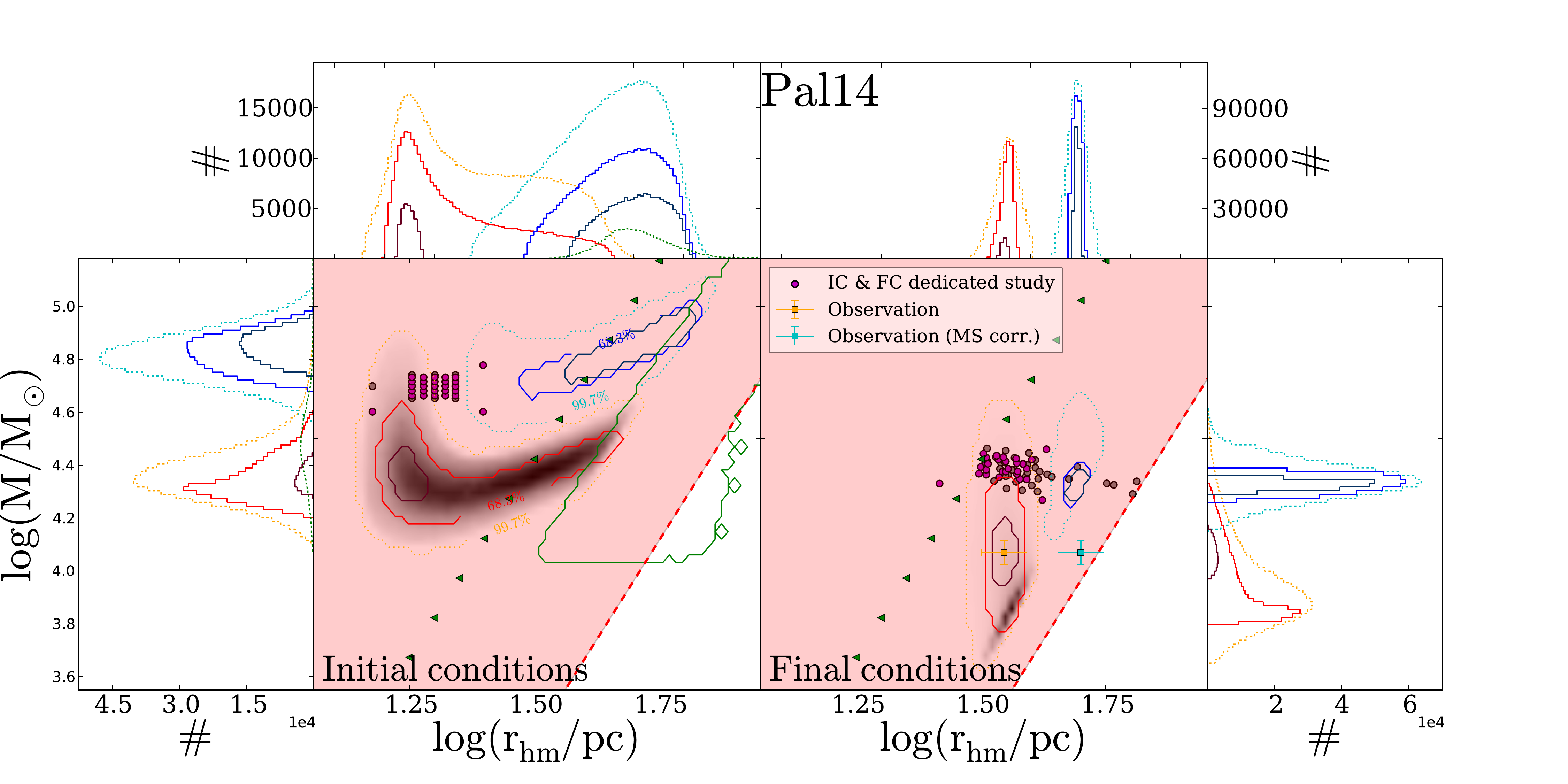}
      \centering
      \caption{The same as Figure~\ref{fig:IC_and_FC1}, but for the cluster Pal14.\label{fig:IC_and_FC7}}
\end{figure*}

\begin{figure*}
      \includegraphics[width=1.08\textwidth]{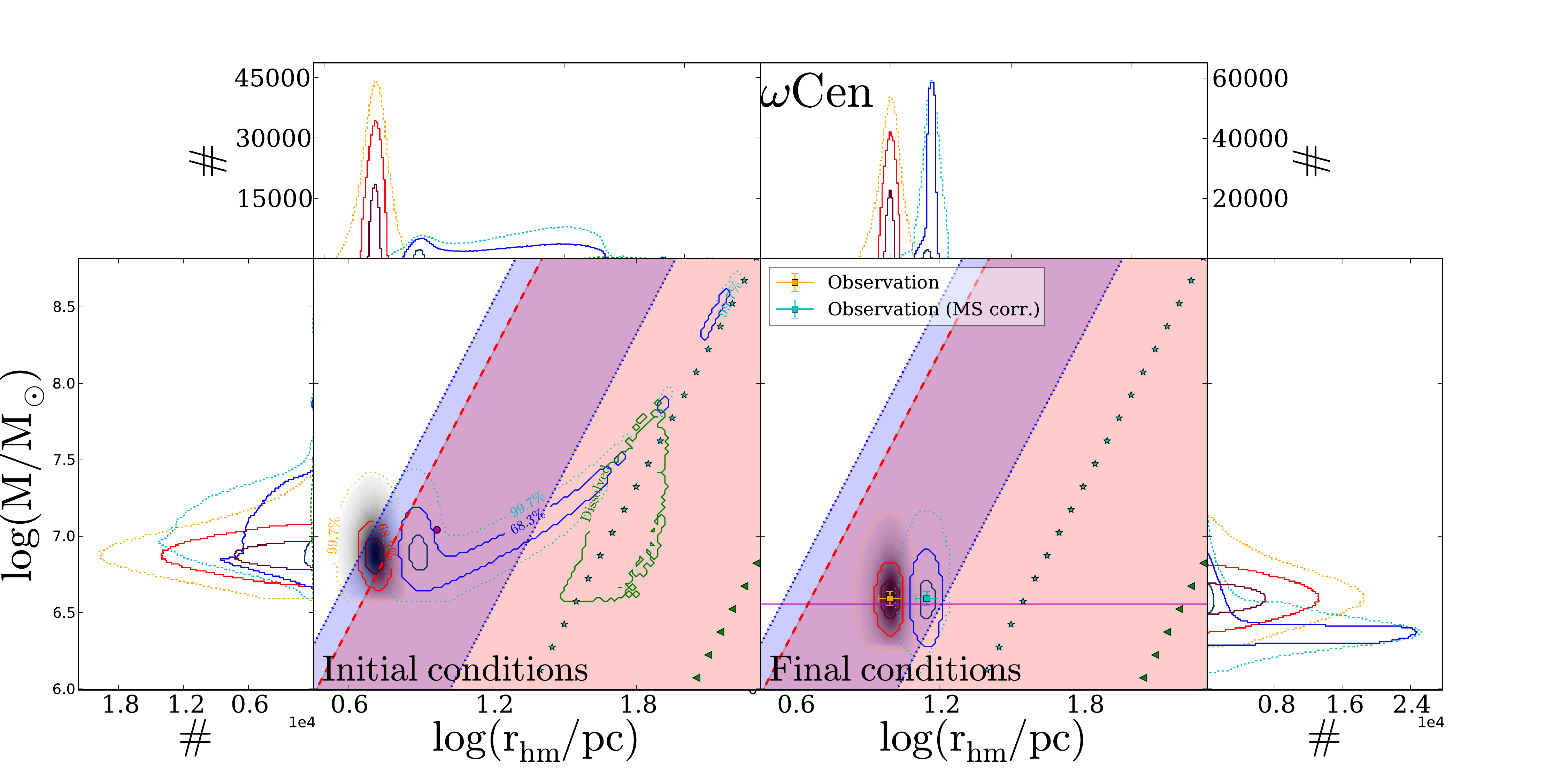}
      \centering
      \caption{The same as Figure~\ref{fig:IC_and_FC1}, but for the cluster $\omega$ Cen.\label{fig:IC_and_FC8}}
      \includegraphics[width=1.08\textwidth]{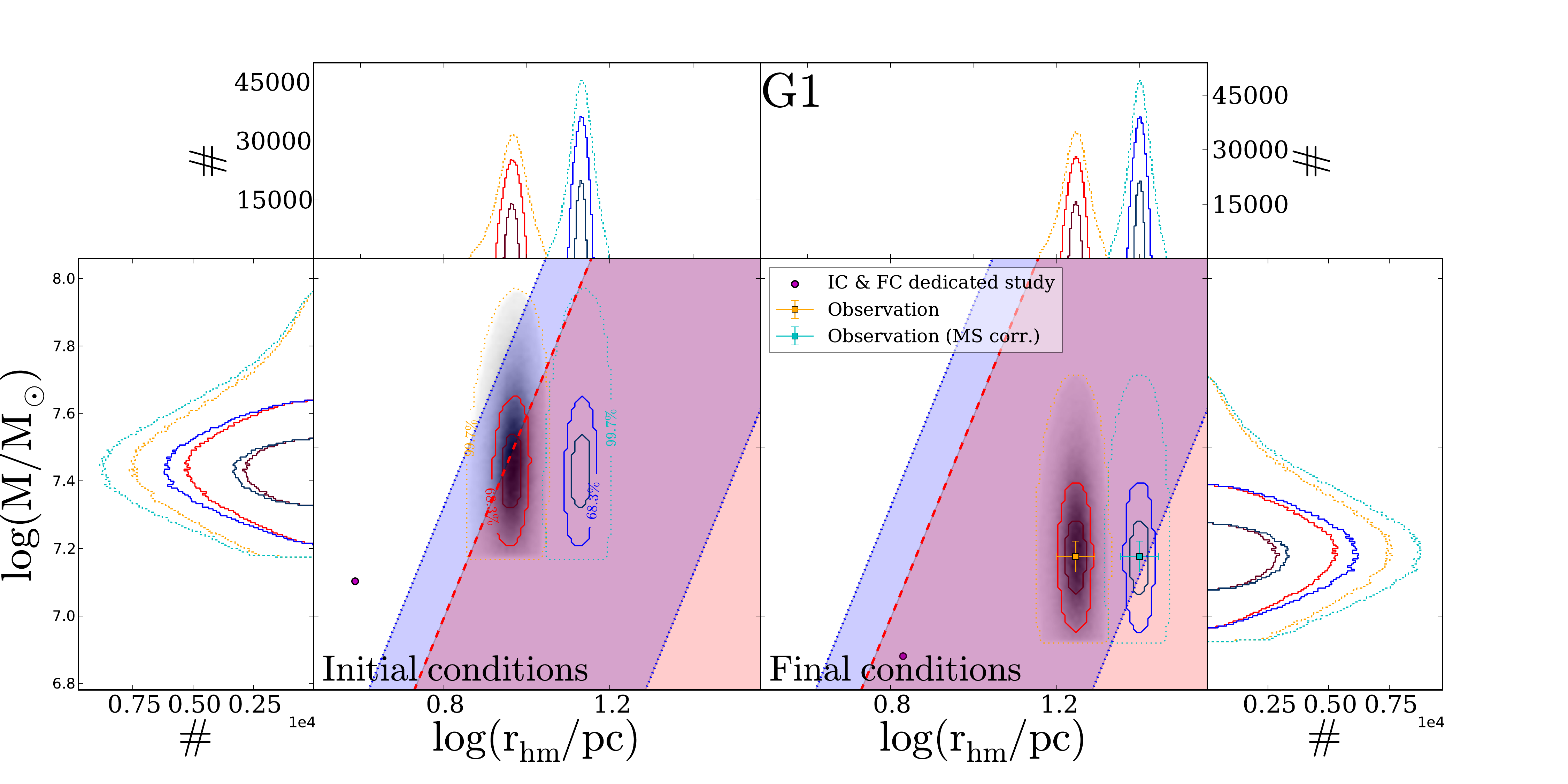}
      \centering
      \caption{The same as Figure~\ref{fig:IC_and_FC1}, but for the cluster G1.\label{fig:IC_and_FC9}}
\end{figure*}

The results of the independent {\sc emacss-mcmc} runs
  in 2D are shown in Figures~\ref{fig:IC_and_FC1} to
  \ref{fig:IC_and_FC9} and are summarized in
  Table~\ref{table:results_2D}, where we compare the observations and
  {\sc emacss}\rq{} best-fit results for the simulations where we did
  not correct the observations for mass segregation (see
  Section~\ref{subsec:Conversion}) and {\sc emacss}\rq{} best-fit
  results for the simulations where we did correct for mass
  segregation, in row one, four and five, respectively, per cluster.
  We see that for
 some clusters (M67, NGC 6397, M4, M22 and 47 Tuc) the initial
 conditions extend to small half-mass radii and high initial half-mass
 densities. These initial conditions correspond to an average distance
 between the cluster stars on the order of 10 AU, and thus these
 densities are too high to be physical. In
 Figures~\ref{fig:IC_and_FC1} to \ref{fig:IC_and_FC9} we show the
 range of observed half-mass densities for globular clusters,
 $\rho_{\rm hm} \sim 10^{-2-3.7}$ M$_{\odot}$pc$^{-3}$, calculated according to
\begin{eqnarray}
\rho_{\rm hm} = 3M/(8 \pi r_{\rm hm}^3), \label{eq:density}
\end{eqnarray}
 see also Figure 1 of \citet{2010ARA&A..48..431P}, using the
 \lq{}half-mass radii\rq{} of \citet{2010arXiv1012.3224H} corrected
 for mass segregation (see Section~\ref{subsec:Conversion}) and masses
 calculated by using the mass-to-light ratios, $M/L$, from
 \citet{2005ApJS..161..304M} - or a constant $M/L \sim 1.45$
 \citep{2000ApJ...539..618M} for clusters which do not have an
 observational value - and the absolute V-band magnitudes of
 \citet{2010arXiv1012.3224H} (red shaded region). The high-$\rho_{\rm
   hm}$ initial conditions mentioned above are significantly larger
 than the ones observed today and this is attributable to the small
 initial half-mass radius the MCMC sampled for these clusters. One
 could argue that the current half-mass densities need not be
 representative for the initial half-mass densities, and (globular)
 clusters may have had larger initial half-mass densities at younger
 ages. However, from Figures~\ref{fig:IC_and_FC1} to \ref{fig:IC_and_FC4} we can see that the range of present-day observed
 half-mass densities for clusters younger than 10 Myr is still a few orders of magnitudes less
 dense than the initial densities for M67, NGC 6397, M4 and M22 in our
 calculations. One could also argue that the precursors of the old
 (globular) clusters might have been very different from the young
 clusters today. High densities might even be essential to enable
 runaway mergers as a pathway to produce intermediate black holes
 (IMBHs) in globular clusters
 \citep{2008CQGra..25k4007M}. \citet{2008CQGra..25k4007M} gives the
 example that a mean density of $\sim 5 \cdot 10^7$
 M$_{\odot}$pc$^{-3}$ would be required for a cluster with a half-mass
 radius of 0.2 pc. Moreover, \citet{2014ApJ...794..147P} ran semi-analytical models for the formation of star clusters and show that the central cluster area can have stellar densities of $\sim 4 \cdot 10^5$ M$_{\odot}$pc$^{-3}$ at the moment of gas expulsion; see Figure 2 of \citet{2014ApJ...794..147P}. However, densities greater than $\sim 10^{10}$
 M$_{\odot}$pc$^{-3}$ seem too extreme. Moreover, it is important to
 keep in mind that the initial conditions we derive in this work are
 the conditions of a star cluster after residual gas expulsion and
 re-virialisation. Since all star clusters expand due to residual gas
 expulsion \citep{2007MNRAS.380.1589B}, the clusters are expected to
 be even denser, i.e. more massive and more compact, directly after
 cluster formation.\\ \\ The reason the MCMC code selected these
 initial conditions is that we have not build in a criterium for only
 selecting initial conditions below a maximum allowed initial
 half-mass density, simply because we do not know what this upper
 limit should be. We also considered it to be better not to limit the
 MCMC code, but to include possible density limits only in the
 analysis phase. Furthermore, for the four clusters (M67, NGC 6397, M4
 and M22) where probable initial conditions are found in the high
 half-mass density regions of parameter space, we see that the
 probable initial conditions are degenerate (see
   Section~\ref{subsubsec:Morphology}). From the brown contours in Figures~\ref{fig:IC_and_FC1} to \ref{fig:IC_and_FC4} we can conclude that lower density initial conditions are practically
 equally probable.\\ \\ The results of the 5D simulations are shown in
 Figure~\ref{fig:IC_and_FC_5D1} and in the figures in the
 Appendix~\ref{sec:Appendix}.  Table~\ref{table:results_5D}
 furthermore shows the characteristics of our best-fit model without a
 correction for mass segregation. From the brown contours in Figure~\ref{fig:IC_and_FC_5D1} we can once again see that the most probable
 $0.1\%$ of the initial conditions are found in the most sampled
 region, but that a probable initial condition does not have to have
 each parameter to originate from the most sampled region.


\begin{table*}
\centering
 \begin{tabular*}{1.02\textwidth}{ l l l l l l l l l }\hline
 Cluster 	& Source 					& N$_i$	& M$_i$ 	& r$_{hm,i}$	& N$_f$	& M$_f$ 	& r$_{hm,f}$ 		& CC?$^a$\\
 		& 	 					& 	& [10$^5$ M$_{\odot}$]& [pc]& 	& [10$^4$ M$_{\odot}$]& [pc] &\\ \hline \hline

 M67 		& Observation 				& - 		& - 	 	& - 		 	& -	 	&  0.140 	&  3.35  /  4.52       & - \\
   		& dedicated study		& 36.0\,k	&  0.187  	&  3.90 	 	& 3.52\,k	 	&  0.204  	&  3.80 	   & no \\ 
   		& EMACSS from IC of DS  	&  29.2\,k 	&  0.187  	&  3.90 	 	&  19.0 	&  0.820  	&  6.51 	   & no \\ 
   		& EMACSS best-fit	&  25.2\,k 	&  0.162  	&  0.006 	 	&  2.64 	&  0.141  	&  3.39   	    & yes \\
   		& EMACSS best-fit (MS corr.)	&  347\,k 	&  2.22  	&  43.0 	 	&  2.39 	&  0.149  	&  4.49 	    & no \\ \hline 
 NGC 6397 		& Observation 				& - 		& - 	 	& - 		 	& -	 	&  6.60 	&  2.31  /  3.29       & yes \\
   		& dedicated study		& -		&  3.65  	&  0.400 	 	& -	 	&  6.03  	&  3.22 	   & yes \\ 
   		& EMACSS from IC of DS  	&  570\,k 	&  3.65  	&  0.400 	 	&  143\,k 	&  7.28  	&  3.21 	   & yes \\ 
   		& EMACSS best-fit	&  1.21\,M 	&  7.77  	&  1.18 $\cdot 10^{-4}$ 	 	&  102\,k 	&  4.40  	&  2.89   	    & yes \\ 
   		& EMACSS best-fit (MS corr.)	&  435\,k 	&  2.79  	&  2.78 	 	&  138\,k 	&  6.60  	&  3.29 	    & yes \\ \hline 
 M4 		& Observation 				& - 		& - 	 	& - 		 	& -	 	&  6.30 	&  2.44  /  3.48       & no \\
   		& dedicated study		& 485\,k		&  3.40  	&  0.580 	 	& 86.0\,k 	&  4.61  	&  2.89 	   & yes \\ 
   		& EMACSS from IC of DS  	&  531\,k 	&  3.40  	&  0.580 	 	&  79.1\,k 	&  4.54  	&  2.69 	   & yes \\ 
   		& EMACSS best-fit 	&  1.20\,M 	&  7.66  	&  1.19 $\cdot 10^{-3}$	 	&  121\,k 	&  5.87  	&  2.78   	    & yes \\ 
   		& EMACSS best-fit (MS corr.)	&  495\,M 	&  3.17  	&  4.08 	 	&  125\,k 	&  6.28  	&  3.48 	    & no \\ \hline 
 M22 		& Observation 				& - 		& - 	 	& - 		 	& -	 	&  33.0 	&  4.17  /  5.95       & no \\
   		& dedicated study		& 832\,k		&  5.70  	&  2.72 	 	& -	 	&  32.0  	&  6.60 	   & no \\ 
   		& EMACSS from IC of DS  	&  891\,k 	&  5.70  	&  2.72 	 	&  811\,k 	&  29.9  	&  6.28 	   & no \\ 
   		& EMACSS best-fit 	&  1.17\,M 	&  7.50  	&  0.358 	 	&  849\,k 	&  33.1  	&  4.19   	    & yes \\ 
   		& EMACSS best-fit (MS corr.)	&  978\,k 	&  6.26  	&  2.53 	 	&  898\,k 	&  33.1  	&  5.95 	    & no \\ \hline 
 Pal 4 		& Observation 				& - 		& - 	 	& - 		 	& -	 	&  2.98 	&  24.5  /  35.0       & no \\
   		& dedicated study		& 100\,k		&  0.500-0.570  	&  12.0-14.5 	 	& -	 	&  2.68-3.24  	&  23.2-27.6 	   & no-no \\ 
   		& EMACSS from IC of DS  	&  78.1-87.3\,k 	&  0.500-0.570  	&  12.0-14.5 	 	&  76.5-87.2\,k 	&  2.78-3.18  	&  26.6-31.1 	   & no \\ 
   		& EMACSS best-fit	&  83.7\,k 	&  0.535  	&  10.9 	 	&  82.1\,k 	&  2.98  	&  24.5   	    & no \\ 
   		& EMACSS best-fit (MS corr.)	&  83.8\,k 	&  0.536  	&  16.6 	 	&  81.9\,k 	&  2.98  	&  35.0 	    & no \\ \hline 
 47 Tuc 		& Observation 				& - 		& - 	 	& - 		 	& -	 	&  110 	&  4.87  /  6.94       & no \\
   		& dedicated study		& 2.04\,M		&  16.4  	&  1.91 	 	& 1.87\,M	 	&  90.0  	&  4.96 	   & no \\ 
   		& EMACSS from IC of DS  	&  2.56\,M 	&  16.4  	&  1.91 	 	&  2.38\,M 	&  87.4  	&  4.57 	   & no \\ 
   		& EMACSS best-fit	&  3.19\,M 	&  20.4  	&  2.04 	 	&  3.01\,M 	&  110  	&  4.87   	    & no \\ 
   		& EMACSS best-fit (MS corr.)	&  3.22\,M 	&  20.6  	&  3.12 	 	&  3.00\,M 	&  110  	&  6.93 	    & no \\ \hline 
 Pal 14 		& Observation 				& - 		& - 	 	& - 		 	& -	 	&  1.1725 	&  35.2  /  50.2       & no \\
   		& dedicated study			& 70.0-100 k&  0.400-0.600 &  15.0-25.0 	 	&  -	 	&  1.86-2.89  	& 26.1-42.8 	   		& - \\ 
   		& EMACSS from IC of DS  	&  62.5-93.8\,k 	&  0.400-0.600  	&  15.0-25.0 	 	&  59.4-87.9\,k 	&  2.17-3.22  	&  31.5-47.9 	   & no - no \\ 
   		& EMACSS best-fit 	&  35.6\,k 	&  0.228  	&  17.6 	 	&  31.5\,k 	&  1.17  	&  35.2   	    & no \\ 
   		& EMACSS best-fit (MS corr.)	&  136\,k 	&  0.869  	&  64.1 	 	&  56.2\,k 	&  2.11  	&  49.3 	    & no \\ \hline 
 $\omega$ Cen 		& Observation 				& - 		& - 	 	& - 		 	& -	 	&  390 	&  9.89  /  14.1       & no \\
   		& dedicated study		& -		&  110  	&  9.33 	 	& -	 	&  360  	&  - 	   & no \\ 
   		& EMACSS from IC of DS  	&  17.2\,M 	&  110  	&  9.33 	 	&  15.7\,M 	&  570  	&  16.5 	   & no \\ 
   		& EMACSS best-fit 	&  11.7\,M 	&  74.7  	&  5.14 	 	&  10.6\,M 	&  390  	&  9.89   	    & no \\ 
   		& EMACSS best-fit (MS corr.)	&  11.9\,M 	&  76.2  	&  7.84 	 	&  10.7\,M 	&  390  	&  14.1 	    & no \\ \hline 
 G1 		& Observation 				& - 		& - 	 	& - 		 	& -	 	&  1.50 $\cdot 10^3$ 	&  17.6  /  25.1       & - \\
   		& dedicated study		& -		&  127  	&  3.86 	 	& -	 	&  760 	&  6.76 	   & no \\ 
   		& EMACSS from IC of DS  	&  19.8\,M 	&  127  	&  3.86 	 	&  19.8\,M 	&  713  	&  8.5 	   & no \\ 
   		& EMACSS best-fit 	&  41.7\,M 	&  267  	&  9.20 	 	&  41.7\,M 	&  1.50 $\cdot 10^3$  	&  17.6   	    & no \\ 
   		& EMACSS best-fit (MS corr.)	&  41.7\,M 	&  267  	&  13.5 	 	&  41.7\,M 	&  1.50 $\cdot 10^3$  	&  25.1 	    & no \\ \hline 
 \end{tabular*}
  \caption{Results for the validation in three
      significant figures. For each validation cluster mentioned in
      the first column seven parameters are compared between the
      observation in the first sub-row, the best-fit results of the
      dedicated study (DS, see Table~\ref{table:Validation_clusters}
      for the references) in the second sub-row, {\sc emacss}\rq{}
      result from the initial condition of the DS in the third
      sub-row, {\sc emacss-mcmc}\rq{} best-fit results
      without and with a correction for mass
        segregation in sub-rows four and five respectively (see
      Section~\ref{subsec:Conversion}). The parameters are the initial
      number of stars $N_i$ in column three, the initial mass M$_i$
      (in 10$^5$\,M$_{\odot}$) in column four, the initial half-mass
      radius $r_{\rm hm,i}$ (in\,pc) in column five, the final number
      of stars $N_f$ in column six, the final (current) mass M$_f$ (in
      10$^4$\,M$_{\odot}$) in column seven, the final (current)
      half-mass radius $r_{\rm hm,f}$ (in\,pc) in column eight and the
      question whether mass segregation has occurred in column
      nine. The observed value for the current half-mass
      radius in column eight shows two values: the first one
      without a correction for mass segregation and
      the second one with a correction for mass
        segregation, see Section~\ref{subsec:Conversion} and
      Table~\ref{table:Validation_clusters2}.\\ $^a$ The results
      whether a cluster had undergone core-collapse according to
      observations are taken from \citet{1995AJ....109..218T}, except
      for G1, where we adopt what \citet{2003ApJ...589L..25B} argued
      about the core-collapse state of these two clusters.\\ $^b$ The
      authors studying this cluster with their dedicated study find
      multiple initial conditions; we only list their first model
      here.\label{table:results_2D}}
 \end{table*}

\begin{figure*}
      \centerline{\includegraphics[width=215mm]{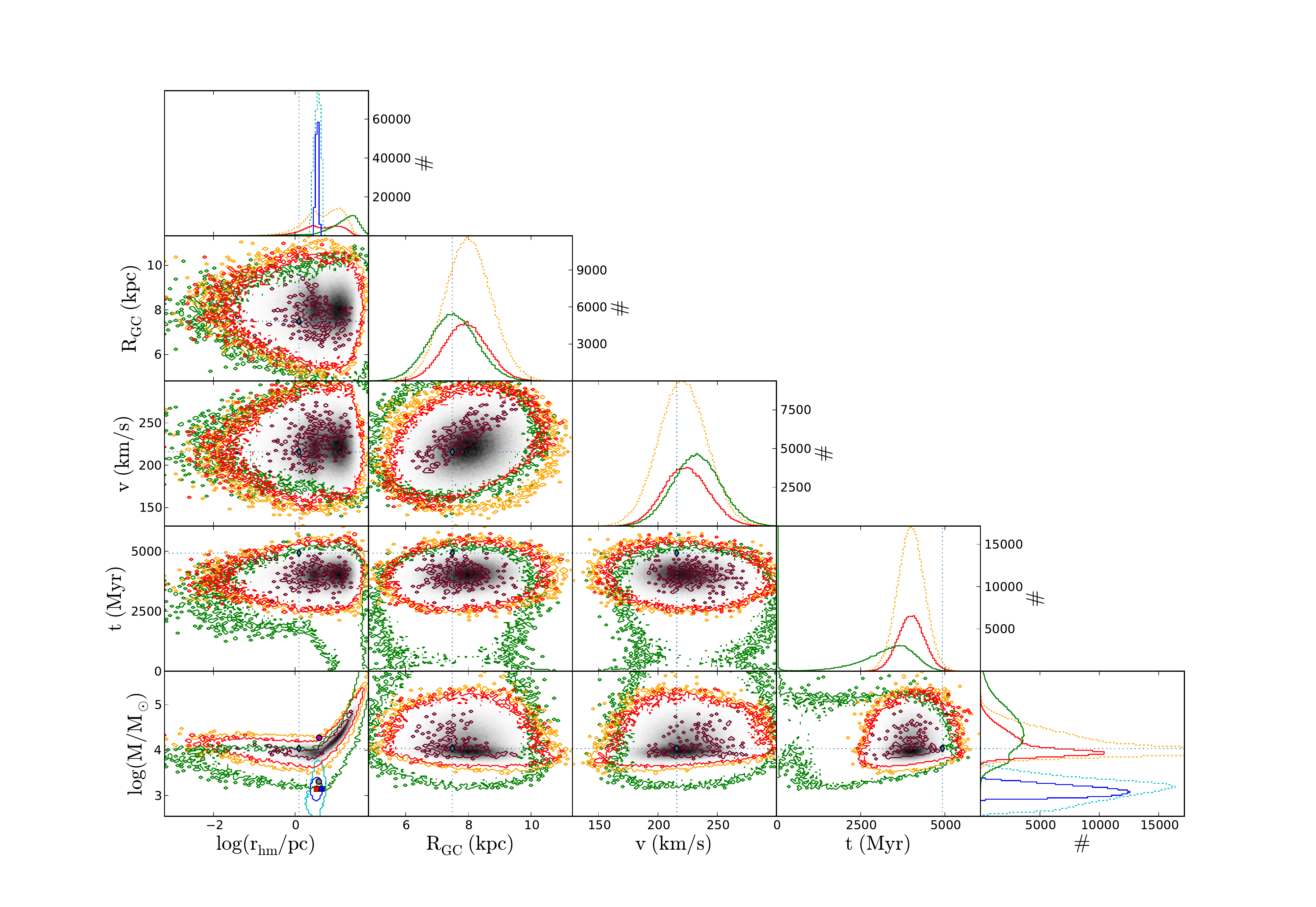}}
      \centering
     \caption{Initial (and final) condition
         distributions for cluster M67 from the 5D MCMC simulations
         with fitting parameters mass and half-mass radius, and
         nuisance parameters Galactocentric radius, orbital velocity
         and age, each with $10\%$ errors for the simulations which
         fit the observables without a correction for mass
         segregation. Numbered from top to bottom, from left to right,
         in panels two, four, five, seven, eight, nine, eleven,
         twelve, thirteen and fourteen show two-dimensional histograms
         in a black-white density plot of two of the five parameters
         initial mass, initial half-mass radius, initial
         Galactocentric radius, initial orbital velocity and (final)
         age against one another: the darker the area, the more
         initial conditions in this area were sampled. Over-plotted on
         these two-dimensional histograms are the outer contours of
         the $99.7\%$ (yellow) and $68.3\%$ (red) confidence contours
         for the simulation fitted to observations without a
         correction for mass segregation. The cyan and blue contour in
         panel eleven show the $99.7\%$ and $68.3\%$ confidence
         contours of the final conditions in total mass vs half-mass
         radius, respectively. The brown contours show the most probable region with $p > 0.999p_{\rm max}$. The green contours show the clusters
         which dissolve before reaching the age of the cluster,
         $\tau_{\rm obs}$. The red square with error bars show the
         observation of the cluster's current mass and half-mass
         radius when no correction for mass segregation is made, and
         the blue square with error bars shows the observable with a
         correction. The pink filled circle denotes the initial
         condition used by the dedicated study, which evolves to the
         final condition shown by the grey filled circle in the right
         panel. The corresponding projected histograms are shown in
         panels one, three, six, ten and
         fifteen. \label{fig:IC_and_FC_5D1}}
\end{figure*}

\begin{table*}
\centering
 \begin{tabular*}{0.73\textwidth}{ l l l l l l l  l l l l }\hline
 Cluster      & N$_i$	& M$_i$                & r$_{hm,i}$ & R$_{\rm RG}$ & $v$  & N$_f$   & M$_f$                & r$_{hm,f}$ & $t$\\
              &       	& [10$^5$ M$_{\odot}$] & [pc]       & kpc          & km/s &         & [10$^4$ M$_{\odot}$] & [pc]       & Gyr\\ \hline \hline
 M67          & 17.0\,k  & 0.108                & 1.24       & 7.49         & 216  & 2.18\,k & 0.139k               & 3.35		& 4.93\\ 
 NGC 6397     & 579\,k	& 3.70                 & 1.52       & 1.51         & 265  & 99.5\,k & 6.37                 & 2.37		& 9.89\\ 
 M4           & 820\,k 	& 5.25                 & 0.179      & 1.45         & 244  & 98.3\,k & 6.29                 & 2.45		& 11.5\\ 
 M22          & 988\,k	& 6.32                 & 1.65	       & 4.59         & 197  & 516\,k  & 33.0                 & 4.17		& 10.8\\
 Pal 4        & 83.4\,k	& 0.534                & 11.1       & 96.6         & 219  & 46.6\,k & 2.98                 & 24.5		& 10.6\\ 
 47 Tuc       & 3.21\,M 	& 20.5                 & 1.97       & 3.57         & 219  & 1.72\,M & 110                  & 4.87		& 13.8\\ 
 Pal 14       & 34.8\,k	& 0.222                & 16.8	       & 88.3         & 236  & 18.3\,k & 1.17                 & 35.2		& 12.1\\ 
 $\omega$ Cen & 11.7\,M	& 74.9                 & 5.12       & 1.23         & 218  & 6.09\,M & 390                  & 8.89		& 12.4\\ 
 G1           & 41.2\,M	& 264                  & 9.38       & 42.8         & 234  & 23.4\,M & 1501                 & 17.6		& 10.3\\ \hline
 \end{tabular*}
  \caption{Best-fit parameters for the 5D simulations fitting the
    observations without a correction for mass
      segregation. For each validation cluster mentioned in the first
    column nine parameters of the best-fit model given. The parameters
    are the initial number of stars $N_i$ in column two, the initial
    mass M$_i$ (in 10$^5$\,M$_{\odot}$) in column three, the initial
    half-mass radius $r_{\rm hm,i}$ (in\,pc) in column four, the
    Galactocentric radius $R_{\rm GC}$ (in kpc) in column five, the
    orbital velocity $v$ (in km/s) in column six, the final number of
    stars $N_f$ in column seven, the final mass M$_f$ (in
    10$^4$\,M$_{\odot}$) in column eight, the final half-mass radius
    $r_{\rm hm,f}$ (in\,pc) in column nine and the age $t$ of the
    cluster (in Gyr) in column ten.\label{table:results_5D}}
 \end{table*}

\subsubsection{Correcting observations for mass segregation}\label{subsubsec:Correcting_MS}
The dedicated studies to which we compare our results did not correct
their observations for mass segregation. This is because it is
difficult to determine the amount of mass segregation that the cluster
experienced and hence the ratio $r_{\rm hm}/r_{\rm phl}$
at $t = \tau_{\rm obs}$, i.e. the factor $c_{\rm
  MS}$ in Section~\ref{subsec:Conversion}. However, many globular
clusters show signs of mass segregation. Even the outer halo cluster
Pal 14 has recently been shown to be (primordially) mass segregated
\citep{2014arXiv1407.2937F}. For this reason we
investigated what the influence of observing a mass
segregated cluster, but not correcting the half-mass radius for it,
could have on the determined initial conditions of the cluster. We did
this for a fixed ratio $r_{\rm hm}/r_{\rm phl}$, see
Section~\ref{subsec:Conversion}, but the outcome will illustrate the
importance of correcting for mass segregation on the interpretation of
the evolution of the observed cluster.\\ \\ The
first thing we can see from
Figures~\ref{fig:IC_and_FC1} - \ref{fig:IC_and_FC9}
  is that correcting the observations for mass segregation or not,
  affects the shape and location of the corresponding $99.7\%$
confidence region in the initial conditions. In other words,
it changes which initial conditions are considered
probable. In some cases these changes are mild, such that the $99.7\%$
confidence regions still show some overlap, e.g. for M67, NGC 6397,
M4, M22; in Section~\ref{subsubsec:Morphology} we
  will see that this has to do with the degeneracy due to
  core-collapse. But in the other cases the shapes or location of the
$99.7\%$ confidence regions are different. And in
  almost all cases, the contours of highest probability are completely
  separated. It even turns out that the factor
$c_{\rm MS} = 1.9$ we used for the globular clusters
\citep{2007MNRAS.379...93H} is typical for clusters of low density,
whereas this factor seems to increase for more dense clusters (Mirek
Giersz, private communication). For denser clusters we thus expect the
differences in the initial conditions for the corrected and
non-corrected simulations to be even more distinct. However, many
studies do not aim to (just) fit structural parameters such as
half-mass radius and total cluster mass, but aim to (also) have a good
fit in the surface brightness profile, velocity dispersion profile and
luminosity profile at multiple radii (\citealt{2003MNRAS.339..486G,
  2003ApJ...589L..25B, 2005MNRAS.363..293H, 2008MNRAS.389.1858H,
  2009MNRAS.395.1173G, 2011MNRAS.410.2698G, 2014arXiv1401.3657H}); see
Section~\ref{subsec:Validation}. It will therefore be interesting to
investigate whether a cluster that will be modeled with $N$-body or
Monte Carlo simulations starting from one of the best-fit initial
conditions with a correction for mass segregation, can produce the
three above-mentioned profiles that match the observed ones equally
well as a cluster starting at the best-fit initial conditions without
this correction.\\ \\ For further validation purposes, we focus on the
comparison between our simulations without a
  correction for mass segregation to the results of the dedicated
studies.

\subsubsection{Dissolving clusters}\label{subsubsec:Dissolving_clusters}
From the Figures~\ref{fig:IC_and_FC1} to
  \ref{fig:IC_and_FC9}, we see that for each of the nine validation
clusters, the modeled clusters which do not survive up to the cluster
age are in most cases relatively larger (and
  sparser) and/or less massive than the clusters which do
survive. Sometimes these clusters dissolved too
  sparse initial conditions were chosen by the MCMC code, which led to
  immediate (within a few Myr) dissolution. E.g. this is the case for
  all dissolving clusters for Pal 14, which have $\rho_{\rm hm} > 0.1$
  M$_{\odot}$pc$^{-3}$. In other cases these clusters underwent
  relatively quick mass loss, dissolving before reaching the observed
  cluster's age. The surviving and dissolving initial conditions are
well separated, except at the borders of the two distributions, which
is to be expected. From
Figure~\ref{fig:IC_and_FC_5D1} we see that
for the 5D runs the dissolution regions projected
onto a two-dimensional plane are not well separated from the surviving
(high posterior probability) regions anymore. See e.g. panel 11 of
Figure~\ref{fig:IC_and_FC_5D1}, where the $99.7\%$
  confidence contour largely overlaps with the dissociation region
contour. The reason for this overlap is simply because that figure is
a projection of a five-dimensional space of initial conditions onto a
two-dimensional space. In other words, a suitable initial condition
($r_{\rm hm,i}, M_{\rm i}$) for one combination of the nuisance
parameters ($R_{\rm GC}, v_{\rm}$, $\tau$), could be
dissolving for another combination of these nuisance parameters. If we
were to take 2D projections of this five-dimensional space at distinct
combinations of ($R_{\rm GC}, v_{\rm}$, $\tau$), the
dissociation regions should again be well separated from the surviving
regions, except at borders where we expect to have an overlap again.

\subsubsection{Degeneracies: dynamical age and morphology}\label{subsubsec:Morphology}
\begin{figure}
 \includegraphics[width=0.5\textwidth]{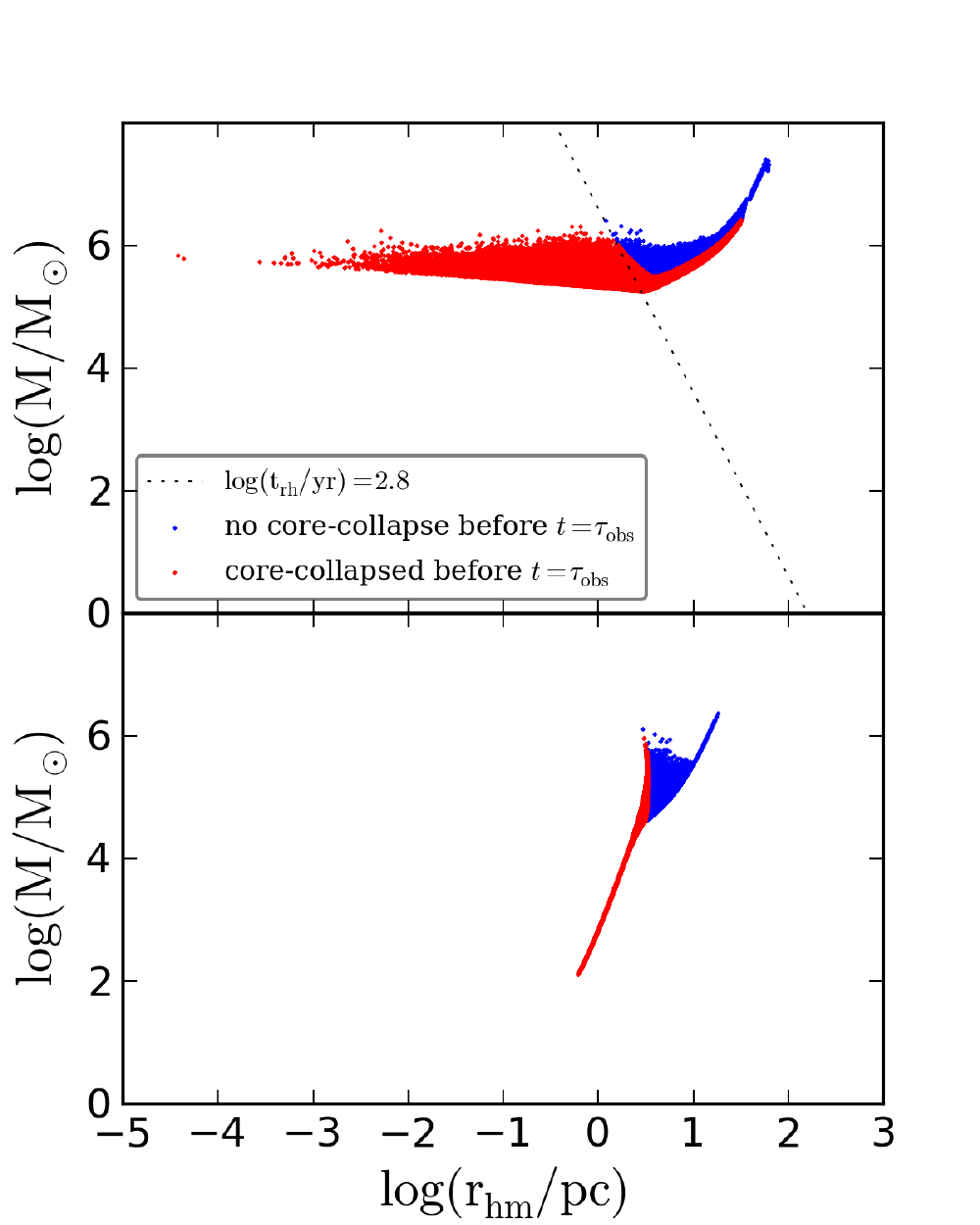}
 \centering
 \caption{The distinction between the initial
   conditions (top panel) and final conditions
   (bottom panel) of the model clusters which
   undergo a core-collapse during their evolution
   (red) and those which do not (blue) for the 2D run for the cluster
   M4 without a correction for mass
     segregation). \label{fig:ccvsnocc}}
\end{figure}

We can divide the surviving clusters in two types of
  clusters: 1) the clusters which underwent a core-collapse and are
  now in the third and \lq{}balanced\rq{} phase of their evolution and
  2) the clusters which did not undergo a core-collapse (yet) and are
  still in the second and \lq{}unbalanced\rq{} phase of their
  evolution (see Section~\ref{subsec:EMACSS} for a definition of these
  evolutionary phases). See e.g. Figure~\ref{fig:ccvsnocc} where we
  make a distinction between the initial and final conditions of the
  model clusters which undergo a core-collapse during their evolution
  (red) and those which do not (blue) for the 2D run for the cluster
  M4 without a correction for mass segregation. We can see that the
  clusters which undergo a core-collapse in our simulations generally
  start out with smaller half-mass relaxation times, $log(t_{\rm rh})
  \la 2.8$ in this example for M4; these clusters also have relatively
  small radii. For those clusters with $log(t_{\rm rh}) \ga 2.8$ we
  see that for a fixed radius the clusters that did not yet undergo a
  core-collapse are more massive. This is what one would expect from
  the fact that the core-collapse time, $t_{\rm cc}$, scales with the
  half-mass relaxation time, $t_{\rm rh}$, which increases for large
  $r_{\rm hm}$ and large $N$. The lower mass clusters with $log(t_{\rm
    rh}) \ga 2.8$ also form a degeneracy towards larger radii and
  masses. With EMACSS we keep track of the evolution of the derivative
  of the half-mass radius, $dr_{\rm hm}/dt$. It turns out that the
  degeneracy towards large radii is produced by the fact that those
  clusters experienced a phase of contraction. This is consistent with
  the fact that once clusters have reached a certain relatively large
  radius (due to expansion), the escape of stars over the Jacobi
  radius becomes dominant \citep{2008MNRAS.389L..28G}. The large
  amount of mass loss of these clusters corresponds to the decrease of
  their half-mass relaxation time, such that they will still undergo a
  core-collapse. This degeneracy is limited on the low density side by
  the dissolution region. \\ \\
\begin{figure}
 \includegraphics[width=0.5\textwidth]{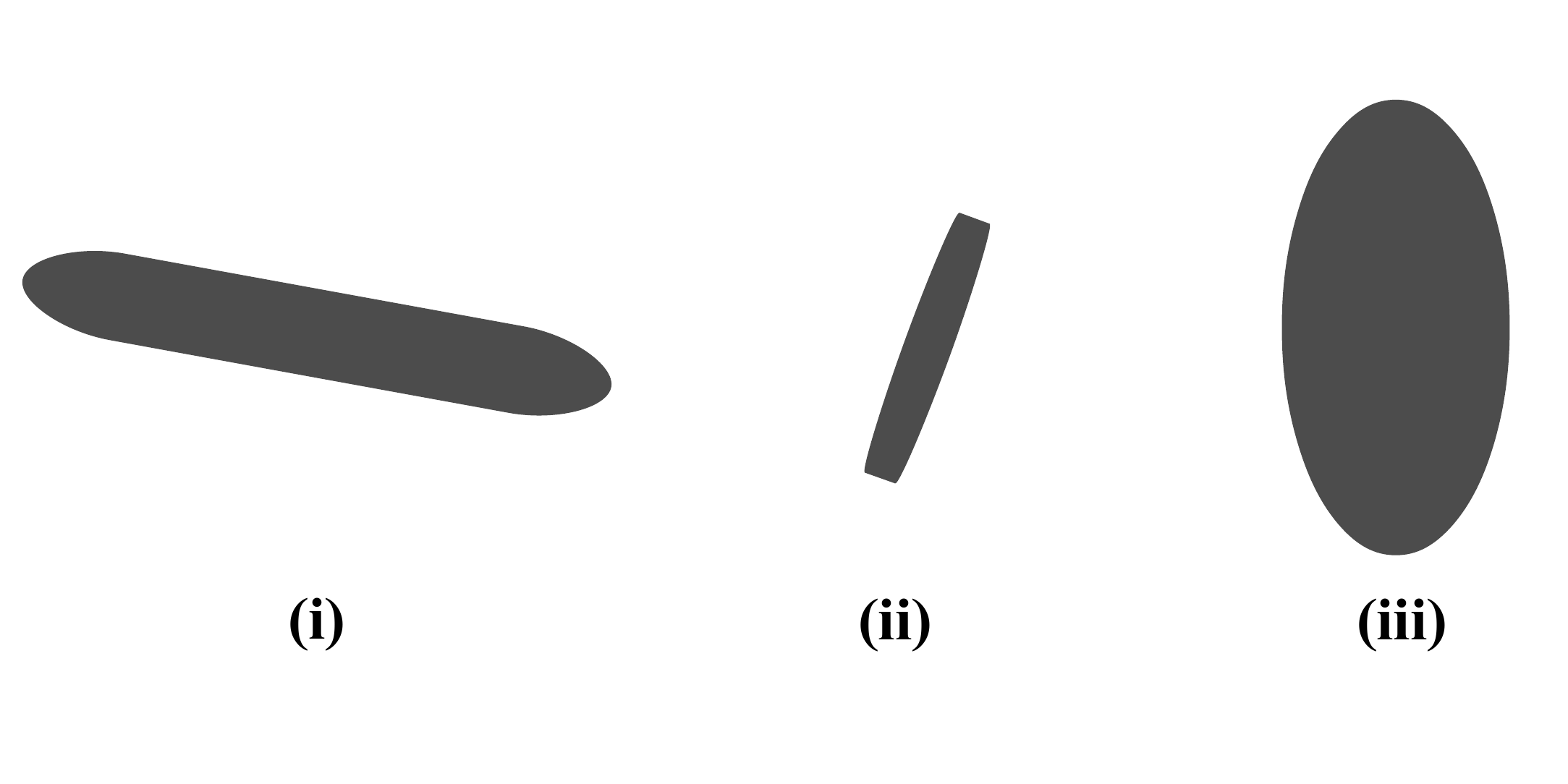}
 \centering
 \caption{A schematic overview of the three morphologies: (i) a roughly horizontal distribution with a
   negative slope towards small half-mass radii; (ii) a distribution with a positive slope reaching to large
   radii; (iii) a two-dimensional gaussian-shaped distribution at relatively
   larger initial half-mass radii than the previous
   case. \label{fig:morphology}}
\end{figure}

When we examine Figures~\ref{fig:IC_and_FC1} to \ref{fig:IC_and_FC9},
we notice that the $99.7\%$ confidence regions in the initial
conditions are built-up out of three characteristic shapes (see also Figure~\ref{fig:morphology}):
\begin{enumerate}
 \item a roughly horizontal distribution with a
   negative slope towards small half-mass radii. This distribution has
   high posterior probabilities for a large spread in half-mass radii
   and a moderate spread in mass. The $99.7\%$ confidence regions of
   the clusters M67, NGC 6397, M4 and M22 have this
     morphology for both our models with and without a correction for
     mass segregation (MS). This distribution
     corresponds to the clusters that undergo a core-collapse within
     their evolution time, $\tau_{\rm obs}$.
 \item a distribution with a positive slope reaching to large
   radii. This characteristic shape is seen in the initial conditions
   for the clusters M67, NGC 6397, M4 and Pal 14
   for the models with and without a correction for
     MS and for $\omega$ Cen for
     the models with a correction for MS. This
     distribution corresponds to the clusters that experienced a phase
     of contraction.
 \item a two-dimensional gaussian-shaped distribution at relatively
   larger initial half-mass radii than the previous
   case. The $99.7\%$ confidence regions of the
     clusters Pal 4 and G1 have purely this morphology for the models
     with and without a correction for MS. This shape is also seen in
     the $99.7\%$ confidence regions of the clusters M22, 47 Tuc, Pal
     14 and $\omega$ Cen for the models with and without a correction
     for MS. The initial conditions from this distribution
   correspond to clusters that do not undergo a
   core-collapse within the evolution time. In this
   case the initial conditions are better constrained.
\end{enumerate}

We see the morphology of $99.7\%$ confidence regions are built-up out
of one or more of the three shapes above. The
  $99.7\%$ confidence regions of the clusters M67, NGC 6397 and M4 are
  built-up out of morphology (i) and (ii) for models with and wqithout
  a correction for MS. The $99.7\%$ confidence regions of the
clusters M22 for both MS-corrected and not
  MS-corrected models and 47 Tuc for the not
  MS-corrected models have a morphology consisting of shapes (i) and
(iii), which shows that both core-collapsed and not
core-collapsed clusters could fit the observed mass
and half-mass radius. The $99.7\%$ confidence region of the clusters
Pal 14 for both MS-corrected and not MS-corrected
  models and $\omega$ Cen for models with a
  correction for MS consist of morphological shapes (ii) and
(iii).\\ \\ From the morphologies
(i) and (iii) in
  Figures~\ref{fig:IC_and_FC1} to \ref{fig:IC_and_FC9}, we see that
the morphology roughly changes along the sequence (i) $\rightarrow$
(i) $\&$ (iii) $\rightarrow$
(iii). Since the clusters are mentioned in the order
of increasing current half-mass relaxation time, which is a proxy for
the dynamical age of a star cluster, we thus see a connection between
the morphology of probable initial conditions and the dynamical age of
a cluster. The connection is quite intuitive. Firstly, the morphology
is connected to the core-collapse state of the
cluster ((i) - core-collapsed;
(iii) - not core-collapsed). When
a cluster undergoes core-collapse, the half-mass
radius increases, because energy is injected into the halo
\citep{1980MNRAS.191..483L,2002MNRAS.336.1069B}. The amount by which
the half-mass radius increases depends on many different cluster
characteristics, such as the binary fraction, as well as on the
interactions taking place. Two clusters with the same initial
conditions, but with a different initial distribution of positions and
velocities of the stars, will have different pathways to
core-collapse and the halt thereof, also known as
statistical fluctuations \citep{2009MNRAS.395.1173G}. It is therefore
practically impossible to sharply constrain the initial half-mass
radius of the cluster, because we do not know the amount by which it
increased during core-collapse. So there are many
different possible initial half-mass radii and hence we see a
degeneracy in the initial distribution of half-mass radii. When on the
other hand the cluster has not yet undergone core collapse, there is
only a narrow range of possible initial half-mass radii and thus the
initial half-mass radii are better constrained.\\ \\ Secondly, when
the half-mass relaxation time is small, so is the
core-collapse time, since $t_{\rm cc} \propto t_{\rm
  hm}$. So an observed cluster with low $t_{\rm hm}$ is more likely to
have undergone a core-collapse and hence its initial
condition distribution will have morphology (i). If an observed
cluster has a higher $t_{\rm hm}$, there is a probability that the
cluster comes from an initial condition which will undergo
core-collapse during its evolution up to the
observed clusters age, but also probability that the cluster comes
from an initial condition which will not. This cluster will thus have
an initial condition distribution made up of morphology
(i) and (iii). The clusters with the largest
half-mass relaxation times will not have initial
conditions distributions with morphology (i), but
  rather (iii).\\ \\ Morphology (ii) is associated
  with the contraction of the cluster, or, in other words, a decrease
  of the $r_{\rm hm}$. And just like with the degeneracy associated
  with core-collapse, it is practically impossible to constrain the
  half-mass radius sharply, because we do not know by which amount the
  cluster contracted. So many different evolutionary paths, each with
  different amounts of contraction, could have produced the current
  observables and therefore we have this degeneracy. To test our
hypotheses for the connection between morphology of the initial
condition distribution and the dynamical age of the
  cluster, we need to apply our method on a larger sample of
clusters, which we are doing in forthcoming work. If the connection
between the morphology and the dynamical age is indeed robust,
then the initial conditions provided by our method
  gives an independent hint onto the core-collapse state of a
  cluster.\\ \\ We will now compare our findings to
  the observations. For M67 we find that our best-fit
model without a correction for MS underwent a
core-collapse. To the best of our knowledge M67 is
not classified as a post-core-collapse cluster and
neither have we come across any other claims of open clusters going
through core-collapse. However,
\citet{2005MNRAS.363..293H} do mention that for binary-rich clusters
there is no clear core-collapse, at least not in the
way that is witnessed in simulations without primordial binaries,
where high core densities need to be reached in order to form binaries
(see e.g. \citealt{2004MNRAS.355.1207H}). The fact that we find a
post-core-collapse best-fit cluster for M67, is most
likely a direct result of not directly taking into account binaries
within {\sc emacss}, because it is the high binary fraction which
causes open clusters to not undergo a core-collapse
in a way that is found for globulars
\citep{2005MNRAS.363..293H}.\\ \\ \citet{1995AJ....109..218T}
classified $\omega$ Cen, M4, 47 Tuc and M22 as King profile clusters:
these clusters were interpreted as clusters which have not undergone
core-collapse yet. NCG 6397 was classified as a
post-core-collapse
cluster. Defining the term \textit{dynamically old
    (young)} for clusters with a large (small) age over half-mass
  relaxation time ratio, $\tau_{\rm obs}/t_{\rm rh}$, the cluster NGC
  6397 is dynamically old given its current half-mass relaxation time
  of 0.4 Gyr \citep{2010arXiv1012.3224H}, see also
  Table~\ref{table:Validation_clusters2}. Dynamically old clusters are
  more likely to have already ondergone core-collapse, which is
  consistent with the classification. 47 Tuc and $\omega$ Cen are
  dynamically young given their current half-mass relaxation times of
  3.55 Gyr and 12.3 Gyr respectively \citep{2010arXiv1012.3224H}, and
  are thus both far from core-collapse \citep{1995AJ....109..218T}.
However, for M22 and M4 \citet{2010arXiv1012.3224H}
  lists relatively short half-mass relaxation times, 1.7 Gyr and 0.85
Gyr respectively, such that these clusters have undergone a number of
relaxation times. Moreover, \citet{2008MNRAS.389.1858H} for the first
time claimed that M4 is a post-core-collapse
cluster, based on the observed behavior of the core radius, $r_c$, of
their model of M4 (see their Figure 5). Therefore, it is interesting
to see that our best-fit models without a correction
  for mass segregation for both M4 and M22 have also undergone a
core-collapse, and, as expected, the best-fit model
for $\omega$ Cen did not. 47 Tuc is an interesting case, because we
find that there are initial conditions of model
  clusters within the 97\% confidence region which underwent a
core-collapse, and initial
  conditions of model clusters which did not, see
Section~\ref{subsubsec:Morphology}. However, our models show a favor
for the not core-collapsed solutions, consistent
with the observations \citep{1995AJ....109..218T}.\\ \\ Both Palomar 4
and 14 are classified as King profile clusters, so they are
interpreted as clusters which have not yet reached the
post-core-collapse phase
\citep{1995AJ....109..218T}. This is to be expected
  for dynamically young clusters with their current half-mass
relaxation times, $t_{\rm rh}$, of 2.63 Gyr and 10.47 Gyr respectively
\citep{2010arXiv1012.3224H}. Our simulations also show that our
best-fit models had not undergone core-collapse
yet. And also for G1 we find that the best-fit conditions do not
undergo core-collapse within its evolution up to its
observed age. As argued in Section~\ref{sec:Validation}, given
G1\rq{}s half-mass relaxation time, it is reasonable to assume that G1
has not undergone core-collapse yet
\citep{2003ApJ...589L..25B}.

\subsubsection{Accuracies on observables}\label{subsubsec:Accuracies_of_observables}

\begin{figure*}
  \includegraphics[width=1.0\textwidth]{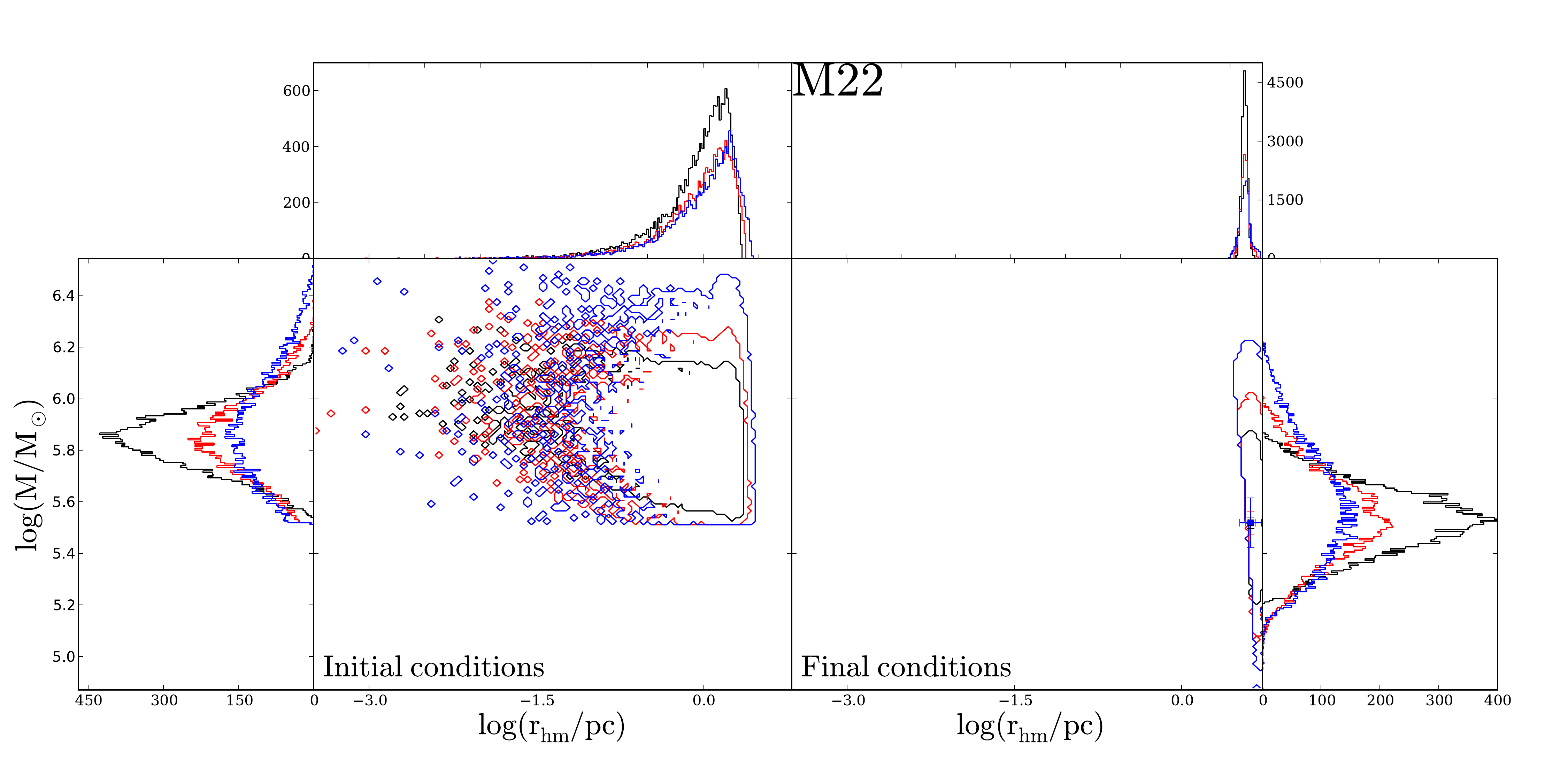} 
 \centering
  \caption{This figure is similar to the lower panels of
    Figure~\ref{fig:IC_and_FC4}, but this time we show the $99.7\%$
    confidence contours for the simulation fitted to observations
    without a correction for mass segregation, using a $5\%$ (black),
    $10\%$ (red) and $20\%$ (blue) error on the
    observables.\label{fig:IC_and_FC_error}}
\end{figure*}

\begin{figure*}
      \includegraphics[width=\textwidth]{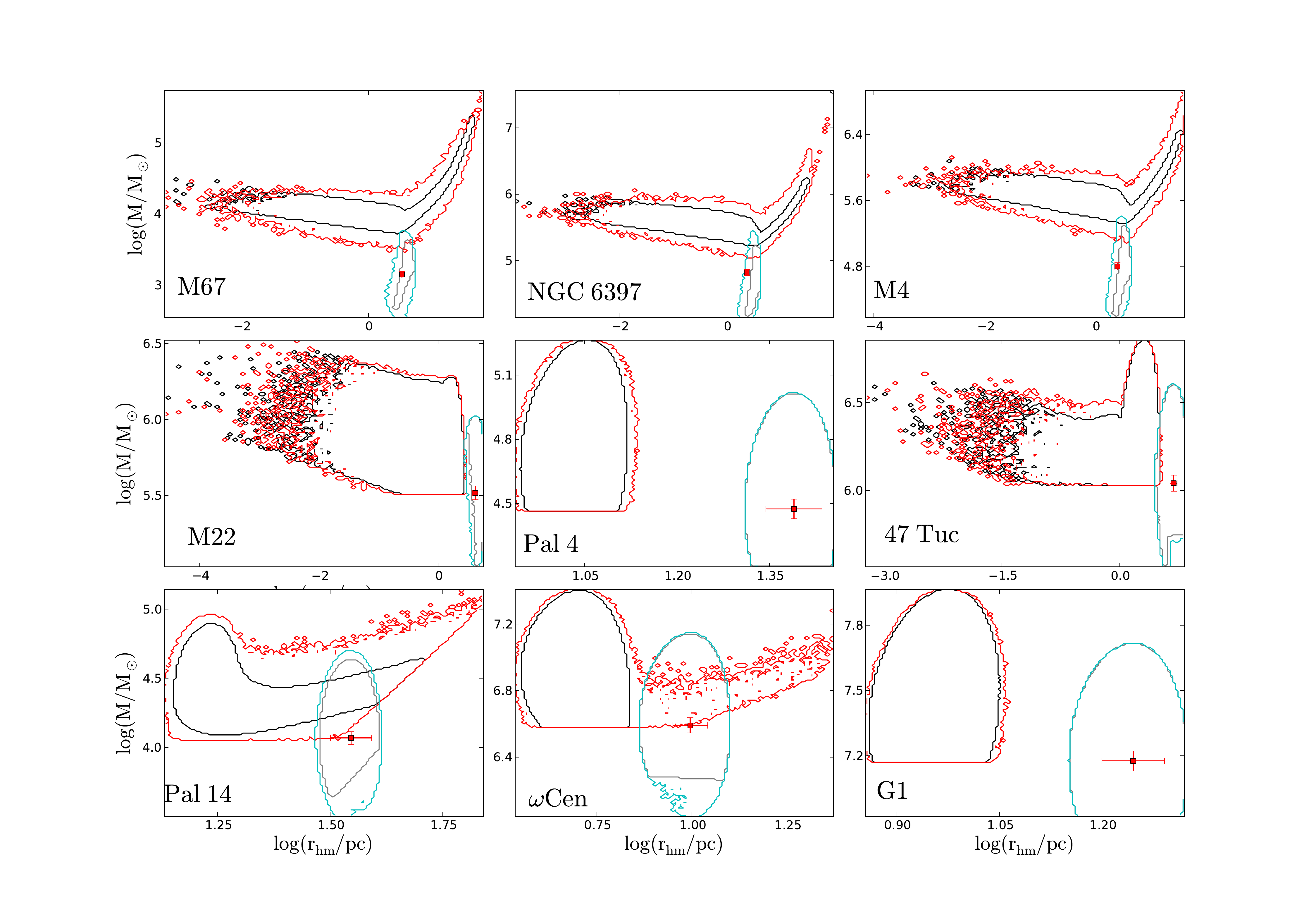}
      \centering
      \caption{A comparison of the 2D and the 5D results to infer the stability of the initial conditions against observational errors in the Galactocentric radius, orbital velocity and age. The black (grey) lines show
          the $99.7\%$ confidence contours of the initial (final)
          conditions of the 2D results. The red (yellow) lines show the $99.7\%$
          confidence contours of the initial (final) conditions when including the Galactocentric radius, orbital velocity and age as nuisance parameters, i.e. the 5D results.\label{fig:max_diff}}
\end{figure*}

In Figure~\ref{fig:IC_and_FC_error} we study how
including smaller and/or larger errors on the observed mass and
half-mass radius will effect the determined initial
  conditions. The figure shows the initial and final condition
distributions in half-mass radius versus total mass for the cluster
M22 for three different error percentages. We chose
M22 for its characteristic initial condition morphology consisting of
shapes (i) and (iii) and we wanted to see if this
  morphology is conserved if the observables had larger or smaller
  errors. Decreasing/increasing
the errors also decreases/increases the size of the $99.7\%$
confidence region, but the degeneracy in the initial
  half-mass radius and the overall morphology
  remain the same. This makes it more likely that the result that M22
could both be a core-collapsed cluster and not
core-collapsed cluster is robust and that it is not
a consequence of not being able to distinguish between the two due to
observational errors. We furthermore see that the lower the
observational errors in the parameters mass and half-mass radius, the
better one can especially constrain the initial mass
of that cluster, but also slighty the half-mass
  radius.\\ \\
In Figure~\ref{fig:max_diff} we compare our 2D simulations with our 5D simulations to infer the stability of the probable initial conditions against observational errors in the galactocentric radius, orbital velocity and age - in essence combining the results from figures like Figures~\ref{fig:IC_and_FC1} and \ref{fig:IC_and_FC_5D1} per cluster. We can see that including distance, age and orbital velocity as nuisance parameters broadens the $99.7\%$ confidence regions for each of the clusters, and that this broadening is most visible for the clusters with a degeneracy (morphologies (i) or (ii)) in the high probability part of their initial conditions. This is because these clusters undergo either
  expansion or contraction and especially these processes are
  sensitive to the parameters setting the tidal field ($R_{\rm GC}$
  and $v$) and the time the time that it could have
  expanded/contracted ($\tau$). The morphology is conserved for all but one cluster; in these cases the $99.7\%$ regions are just broadened upward and downward. Upward if the variation in one of the
  nuisance parameters increases the strength of the tidal field (smaller
  $R_{\rm GC}$ and larger $v$); in these cases the clusters must have
  started out out more massive to withstand the larger amount of mass
  loss. An upward shift can also be caused if the sampled age is larger than $\tau_{\rm obs}$. This
  is because the cluster spends a longer time losing mass and thus
  must have started out with a higher mass to reproduce the same
  observed mass. In the opposite cases (decreasing the tidal field strength or smaller ages) we see a downward shift. See especially the clusters M67, NGC 6397, M4, Pal 14 and 47 Tuc in Figure~\ref{fig:max_diff}. M22 also has a degeneracy towards small initial half-mass
  radii, but the effects mentioned above are much weaker here. This
  could be due to the lower sampling of initial radii $\log(r_{\rm
    hm}/pc) < 0.0$.\\ \\ 
  We also observe a less prominent, but still
  visible shift in the initial conditions of the degeneracy-free
  clusters Pal 4 and G1 in Figure~\ref{fig:max_diff}. There
  we see that a stronger tidal field, respectively a shorter evolution
  time, causes the initial condtions distribution to shift to slightly
  larger half-mass radii, indicating that slightly larger clusters are
  more probable to reproduce the observables in these cases. For the cluster $\omega$ Cen this shift due to the increase of tidal field strength even goes that far, that the morphology changes: (iii) $\rightarrow$ (ii) $\&$ (iii). This indicates that initially larger (and more massive) clusters which contract during their lifetime can also reproduce the observables for this cluster.
All taken together this shows that for some clusters accurate observations
  (error bars $< 10\%$ of the mean of the observable) are required for
  determining the initial conditions, whereas for the other clusters
  smaller accuracies will be sufficient.

\subsubsection{Star cluster evolutionary tracks}\label{subsubsec:Tracks}
\begin{figure*}
 \centering
  \centerline{\includegraphics[width=225mm]{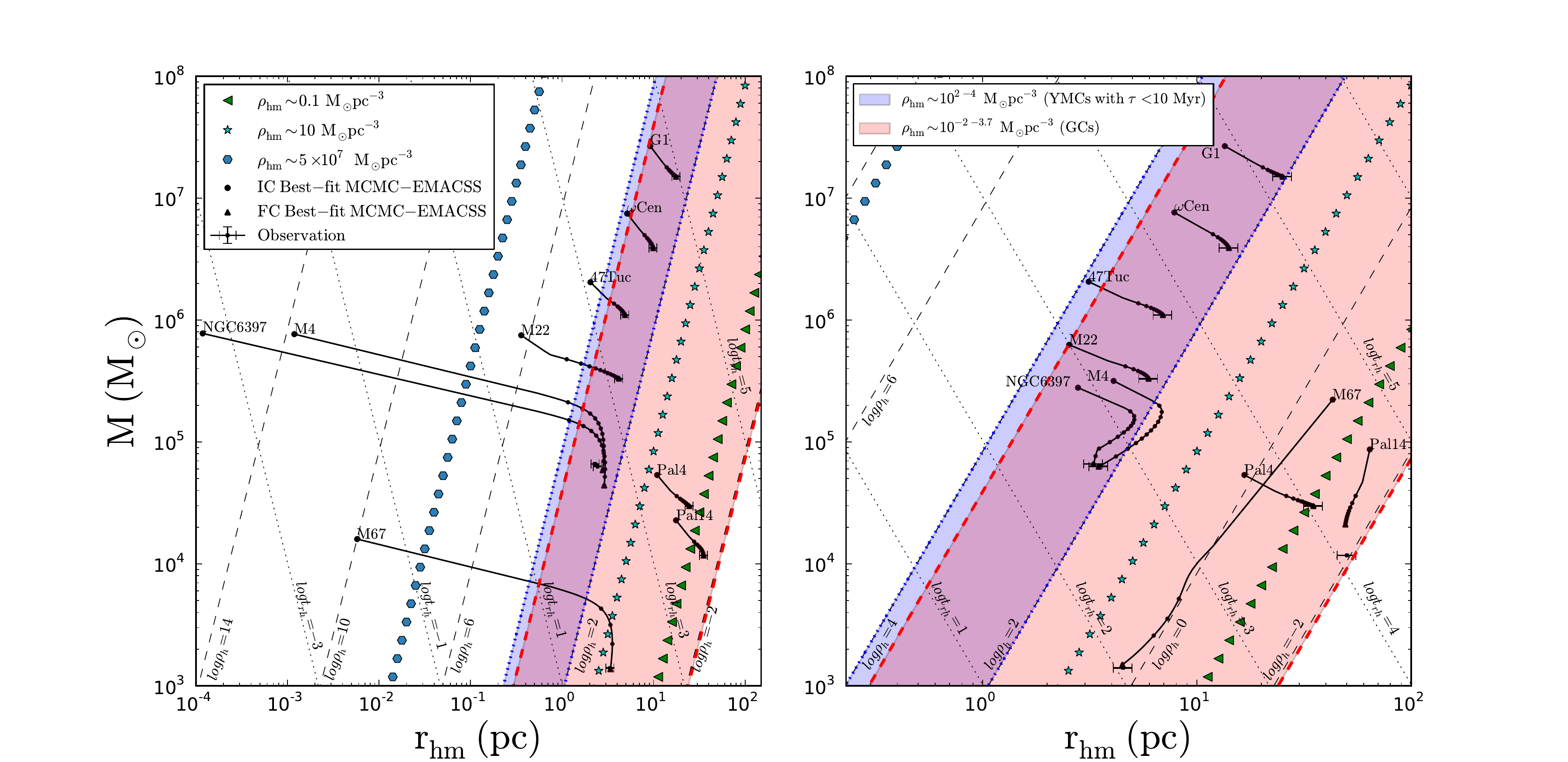}}
 \caption{The star cluster evolution tracks in a total mass vs
   half-mass radius diagram for our best-fit model for each of the
   nine validation clusters in the case that we did not correct the
   observations for mass segregation (left panel)
   and the case where we did right panel. For each
   cluster the square with error bars shows the observed mass and
   half-mass radius with $10\%$ errors in both parameters, the large
   solid circle shows the initial condition for our best-fit model and
   the solid triangle shows the final condition. The solid line shows
   the evolution of the best-fit model in mass and half-mass radius
   and the small solid circles over-plotted on this line mark a passed
   Gyr of evolution, such that M4 has twelve small over-plotted solid
   circles and M67 has four. The dotted lines show lines of constant
   half-mass relaxation time and the dashed lines show lines of
   constant half-mass density, corresponding to the values indicated. The shaded regions, the lines depicted by green triangles, cyan stars and turquoise hexagons are as in Figure~\ref{fig:direct_comparison}.\label{fig:summary}}
\end{figure*}

For our best-fit models with or without a correction
  for mass segregation (see Section~\ref{subsec:Conversion}) we make
cluster evolution tracks by plotting the mass and half-mass radius at
different equally spaced time intervals in a mass versus half-mass
radius diagram, see Figure~\ref{fig:summary}. Just before submitting
we noticed the recently submitted paper by
\citet{2014ApJ...794..147P} in which the authors also
  make evolutionary tracks for young massive clusters covering the
  first 10 Myr. These authors simulate the formation and subsequent
  expansion due to redisual gas expulsion. The evolution of a cluster
  in the mass-radius diagram during the first 20 Myr is shown in their
  Figure 4. Considering the fact that we model star cluster evolution
  after all redisual gas has been removed and the cluster is back in
  virial equilibrium, our evolutionary tracks would, in principle,
  start some time after their tracks end, assuming it would still take
  some time for the cluster to re-virialise. Another paper where the
  evolution of star clusters is studied by means of evolutionary
  tracks is \citet{2008MNRAS.389..889K}, where the authors construct a
  dynamical temperature-luminosity diagrams and show that most of
  their investigated cluster families share a common sequence in this
  diagram.\\ \\ We additionally plot the lines of constant half-mass
relaxation time and constant half-mass density. We
  calculate the half-mass relaxation time according to
Eq. (17) of \citet{2010ARA&A..48..431P}, using
$r_{\rm v} = r_{\rm hm}$ for virial radius, $r_{\rm v}$, and
ln($\Lambda$) = 10 for Coulomb logaritm, ln($\Lambda$) = ln($\gamma
N)$. This is a reasonable assumption for all of the clusters using
$\gamma = 0.02$ \citep{1996MNRAS.279.1037G} and having $N$ vary in the
range $10^3 - 10^7$.\\ \\
From Figure~\ref{fig:summary} we see that most best-fit cluster models
start their evolution with a rapid mass loss and
expansion phase, due to stellar evolution. After that, most of the
clusters continue to expand their half-mass radius and lose mass, but
at a much slower rate. As a logical consequence the
  half-mass density of these clusters decreases, whereas their
  half-mass relaxation time increases. Some clusters continue at this
pace for the rest of their evolution (G1, $\omega$ Cen, 47 Tuc, M22
and Pal 4 in both cases, and Pal14 when we did not
  correct for mass segregation).\\ \\ When we did
  correct for mass segregation, the best-fit models for M67 and Pal
14 are solely contracting during their life time. The interesting
feature here is that both clusters seem to contract approximately
along a line of constant log($\rho_{\rm hm}$). The clusters exhibit
mass loss due to stellar evaporation over the tidal
radius leading to the decrease of the half-mass radius
\citep{1961AnAp...24..369H,2011MNRAS.413.2509G} and, as these authors
mention, this contraction happens at (roughly) constant half-mass
density. The best-fit models with a correction for
  mass segregation for NGC 6397 and M4 start out more than three
orders of magnitude larger, but at lower mass, compared to
the models without a correction for mass
  segregation. Their evolution also starts out with a phase of rapid
mass loss due to stellar evolution and the associated expansion of the
half-mass radius. This is followed by a tidally-limited contraction
phase, with the decrease of the half-mass radius while $log(\rho_{\rm
  hm})$ remains close to 2.\\ \\ We see that the
  best-fit models without a correction for mass segregation for M67,
NGC 6397 and M4 start out being very compact (see the
  beginning of Section~\ref{subsec:Independent_2D_results} for a
  discussion on their high initial densities). This is followed by a
relatively slower expansion of the half-mass radius and a slower
mass loss phase. After about 1 Gyr, for M67, and 3
Gyr, for NGC 6397 and M4, these clusters enter a phase where the
amount of expansion decreases and comes to a halt. From this point on
the track bends towards the contraction of the half-mass radius. It
seems that the models of $\omega$ Cen, 47 Tuc and
Pal 14 without a correction for mass segregation are
  just about to enter that phase and the models for
  $\omega$ Cen, 47 Tuc and M22 with a correction for mass
  segregation.

\subsubsection{Determining initial conditions for observed star cluster}\label{subsubsec:Observables}
With our {\sc emacss-mcmc} method we are able to reproduce the cluster
observables well, i.e. reaching a maximum posterior probability
$p>0.99$, for all of the 5D runs and most of the 2D runs. However, the
2D runs for NGC 6397 and M4 without a correction for
  mass segregation and for Pal 14 with a correction
  for mass segregation reach a maximum posterior probability of
  $\sim0.34$, $\sim0.78$ and $\sim0.46$, respectively; see also
Figures~\ref{fig:IC_and_FC2}, \ref{fig:IC_and_FC3}
  and \ref{fig:IC_and_FC7} where we notice that the observation data
points are not included in the $99.7\%$ a confidence regions in the
the mentioned cases. Here we see that for M67 the best-fit cluster
model evolves to a slightly smaller cluster than observed, and for NGC
6397 and M4 to slightly larger clusters, but with similar masses as
the three observed clusters. For Pal 14 our best-fit
  cluster without a correction for mass segregation evolves to a
final radius in the observed range, but significantly more massive
than the observations. But, we see similar offsets in final mass
and/or final half-mass radius for the dedicated study for these
clusters, see Table~\ref{table:results_2D}.\\ \\ The reason for not
being able to reproduce the cluster parameters can be indicative of
two things: some of the modeling assumptions are incorrect or the
observables are poorly constrained. The first point is usually what is
going on for modeling star clusters, since the model at hand always
uses a number of assumptions. The second point is also interesting,
nevertheless. For example, if the assumptions made are proven to be
reasonable based on the observations, e.g. that the cluster\rq{}s
orbit is observed to be very close to circular, and
we know that one of the observables, e.g. the orbital velocity, $v$,
is determined with a large error bar. If we then run
the method and are not able to reproduce the cluster observables, one
can run the method with $v$ as a nuisance parameter, probing different
values of the orbital velocities within the error bars. When the
second run \textit{is} able to reproduce the cluster observables, this
shows that the orbital velocities for which this better match to the
observations is found, are more suitable values for the observed
orbital velocity within the observed errors.\\ \\ We see something
similar happening for the above mentioned clusters, where we are not
able to reproduce the cluster observables in the two-dimensional run,
but are able to do so in the five-dimensional
runs. However, we can not exclude the model
assumptions to be responsible to the mismatch in the 2D
runs. Moreover, we also see that this mismatch only occurs for one of
the cases only, e.g. M4\rq{}s observables are not reproduced without a
correction for mass segregation, but are if we do include the
correction.

\subsubsection{Comparison to dedicated studies}\label{subsubsec:Comparison}
We now determine how probable the best-fit initial conditions of the
dedicated studies are according to our independent {\sc emacss-mcmc}
results, by checking whether these initial conditions are included in
our $99.7\%$ and/or $68.3\%$ confidence regions for the simulations
fitted to observations without a correction for mass
  segregation, see Section~\ref{subsec:Conversion}).\\ \\ For M67 the
best-fit initial condition of \citet{2005MNRAS.363..293H} is clearly
outside our confidence regions, but the differences and similarities
of the models are not univocal. Our modeling shows a degeneracy, more
significant in the initial half-mass radius, but also in the initial
total mass. Therefore, when we consider our models with the same
initial half-mass radius as the best-fit N-body model, our models are
significantly less massive. When we consider the similar initial total
mass, our models are either more compact or more extended. However,
this wide range of initial conditions allowed by our modeling all
evolve to a small region in half-mass radius, but with a spread in
final mass. The final condition of \citet{2005MNRAS.363..293H} is
found in this confidence region as well. Compared to our least massive
models, this again shows the more prominent mass
  loss in the $N$-body simulation compared to the evolution with {\sc
  emacss}. But compared to our more extended initial clusters, it
shows a larger amount of expansion, since it started out with a
smaller initial half-mass radius, but end up with similar final
half-mass radii.\\ \\ For NGC 6397, M4 and 47 Tuc the best-fit initial
and final conditions of GH03-14 are included in our $99.7\%$ and
$68.3\%$ confidence regions and thus agree well with the dedicated
studies. For M4 and NGC 6397 our method shows, though, that there is a
wide range of possible initial radii and some spread in initial mass,
which are also probable. For 47 Tuc our modeling shows that the
most probable initial conditions well constrained in
a two-dimensional gaussian-shaped region, consistent with the best-fit
model of \citet{2011MNRAS.410.2698G}. For M22 the best-fit initial
condition of \citet{2014arXiv1401.3657H} is on the high initial
half-mass radius border of our $99.7\%$ confidence region, but in the
same mass range. Our methods thus favors smaller initial clusters for
M22.  For $\omega$ Cen the best-fit initial condition of
\citet{2003MNRAS.339..486G} is not included in our $99.7\%$ confidence
regions. Even though the mass range is comparable, our method shows a
favor for smaller initial half-mass radii. The final mass of
\citet{2003MNRAS.339..486G} and our final mass are in the same
range.\\ \\ For Pal 4 three out of seven initial conditions of the
model we compare to of \citet{2014MNRAS.440.3172Z} are included in our
$99.7\%$ confidence region. This shows that both models agree in the
initial mass of the cluster, but our modeling favors smaller initial
half-mass radii. However, the final conditions of all seven initial
conditions are within the $99.7\%$ confidence region of the final
conditions. As seen in Section~\ref{subsec:Direct_comparison_results},
this seems to indicate that the N-body modeling of Pal 4 led to less
expansion than the EMACSS modeling.\\ \\ For Pal 14 a fraction of the
initial conditions of \citet{2011MNRAS.411.1989Z} are included in the
high mass tail of our $99.7\%$ confidence regions. In this case both
models agree on the initial half-mass radius, which shows to be
degenerate in our modeling, but our method favors a lower initial mass
and allows even more extended initial conditions. Since most of the
final conditions of \citet{2011MNRAS.411.1989Z} agree quite well with
our final conditions, albeit on the high-mass end, this again shows
what seems to be a general trend: the clusters modeled by direct
N-body integration lose more mass and expand less.\\ \\ For G1 the
best-fit initial conditions of \citet{2003ApJ...589L..25B} are clearly
outside our $99.7\%$ confidence regions; both their initial mass and
half-mass radius is smaller than those in of our confidence regions
with or without mass segregation. This is consistent
with the results from
Section~\ref{subsec:Direct_comparison_results}. Our model favors more
massive and more extended initial conditions than the best-fit
(scaled) initial condition of \citet{2003ApJ...589L..25B}.

\subsection{Validation}\label{subsec:Validation}

One could wonder how useful it is to constrain the initial conditions
based on only two parameters \citep{2008MNRAS.389.1858H}.  The reason
is that constraining the cluster initial conditions based on just mass
and half-mass radius is in principle not enough to truly pin-point
\textit{the} initial conditions of an observed cluster, which we also
show with the degenerate shapes (i)a and (i)b in
Section~\ref{subsubsec:Morphology}. Additionally good fits to the
surface brightness, the velocity dispersion and the luminosity
profiles (at a few different radii) are required
\citep{2008MNRAS.389.1858H}. But if one wishes to obtain a reasonable
set of probable initial conditions that, given the
assumptions, contain \textit{the} initial condition of a particular
observed cluster with $99.7\%$ confidence, to have a good starting
point for follow-up modeling and if one wants to get a first
understanding this cluster must have evolved and how that depends on
several input parameters, our method provides a
decent way to do this.\\ \\ We see that the {\sc
  emacss-mcmc} method does a remarkably satisfying job in finding
initial conditions for observed star clusters. In the direct
comparison we found good agreement for most of the clusters. For the
two clusters $\omega$ Cen and M67 we found poor respectively no
agreement, but we argue that this comes from the differences in the
underlying physics of {\sc emacss} and the codes used in these
dedicated studies. In the independent {\sc emacss-mcmc} runs we were
able to evaluate whether the best-fit initial conditions found by the
dedicated studies were also good according to our method or whether
they could be improved. And that is the main strength of our method:
being able to evolve a distribution of initial conditions, study
degeneracies and get a good grasp on which set of initial conditions
are appropriate for a given observed cluster, such that these can be
followed up with more detailed modeling, such as Monte Carlo or
$N$-body simulations.\\ \\ Moreover, we have shown with our method
that we are able to add more dimensions to the initial conditions
space while preserving the performance and speed of the simulation. We
have shown this for a five-dimensional initial condition space to
sample from, see Section~\ref{subsec:Independent_2D_results}, but one
could obviously add more parameters to the initial condition\rq{}s
space to fit for, but one could also add nuisance parameters over
which one marginalizes; this something we want to explore more in
future work.

\section{SUMMARY}\label{sec:Summary}

In this paper we presented our {\sc emacss-mcmc} method with which we
are able to derive a distribution of initial conditions that, after
evolution up to the cluster's current age, evolves to the currently
observed conditions. We validate our method by applying it to a set of
star clusters which have been studied in detail numerically with
$N$-body simulations and Monte Carlo methods
(hereafter: the dedicated studies): the Galactic
globular clusters M4, 47 Tuc, NGC 6397, M22, $\omega$ Centauri,
Palomar 14 and Palomar 4, the Galactic open cluster M67, and the M31
globular cluster G1. As in the dedicated studies, our
  derived initial conditions are the conditions of a star cluster
  after residual gas expulsion and re-virialisation.  We conclude
that the results of our method are in agreement with the dedicated
studies for the majority of clusters. For the two clusters $\omega$
Cen and M67 we find little agreement, but we argue that this due to
the differences in the underlying physics of {\sc emacss} and the
codes used in these dedicated studies. For example not having a direct
treatment for binaries in the parametrized code, which is especially
important for the evolution of an open cluster. \\ \\ We have
furthermore discussed the following points:
\begin{itemize}
\item We have shown the importance of correcting the observations for
  mass segregation. If one does not correct the observed radius for
  mass segregation and derives initial conditions that, after
  evolution up to the clusters age, match the uncorrected, and thus
  smaller, half-mass radius, this modeled cluster does not necessarily
  resemble the actual cluster of interest.
\item We have shown that the distribution of initial
  conditions can contain two types of degeneracies: a) a degeneracy
  towards smaller half-mass radii associated with core-collapse, and
  b) a degeneracy towards larger half-mass radii, associated with
  contraction.
\item We made star cluster evolutionary tracks for our best-fit models
  and discussed how the different phases of the cluster evolution are
  distinguishable in these tracks.
\item We find that there is a connection between the morphology of
  99.7$\%$ confidence region of initial conditions and the dynamical
  age of a cluster and that a degeneracy in the initial half-mass
  radius towards small radii is present for clusters which have
  undergone a core-collapse during their evolution
  time.
\end{itemize}
We conclude that our {\sc emacss-mcmc} method does a satisfying job in
finding initial conditions for observed star clusters which can be
used in follow up modeling. In forthcoming work are applying our
method to two groups of star clusters in order to provide initial
conditions which can be followed-up with more accurate methods:
Galactic and extragalactic globular clusters.

\section{ACKNOWLEDGEMENTS}
We would like to thank the anonymous referee for
  carefully reading our manuscript and providing useful suggestions
  for improvement. We would also like to thank
Douglas Heggie and Mirek Giersz for valuable
discussions, and Holger Baumgardt and Akram Hasani
  Zonoozi for kindly providing (information about) their simulation
  data for G1 and Pal 14 $\&$ Pal 4, respectively. JTP would like to
thank Carmen Martinez Barbosa for valuable help in
getting started with MCMC and Vincent Henault Brunet
  for useful discussion on MCMC.  This work was supported by the
Netherlands Research School for Astronomy (NOVA) and by the
Netherlands Research Council NWO (grants \#643.200.503, \#639.073.803
and \#614.061.608). MG acknowledges support from the ERC
(ERC-StG-335936, CLUSTERS) and Royal Society in the form of a
University Research Fellowship (URF).

\appendix

\section{Additional figures}\label{sec:Appendix}
\begin{figure*}
      \centerline{\includegraphics[width=215mm]{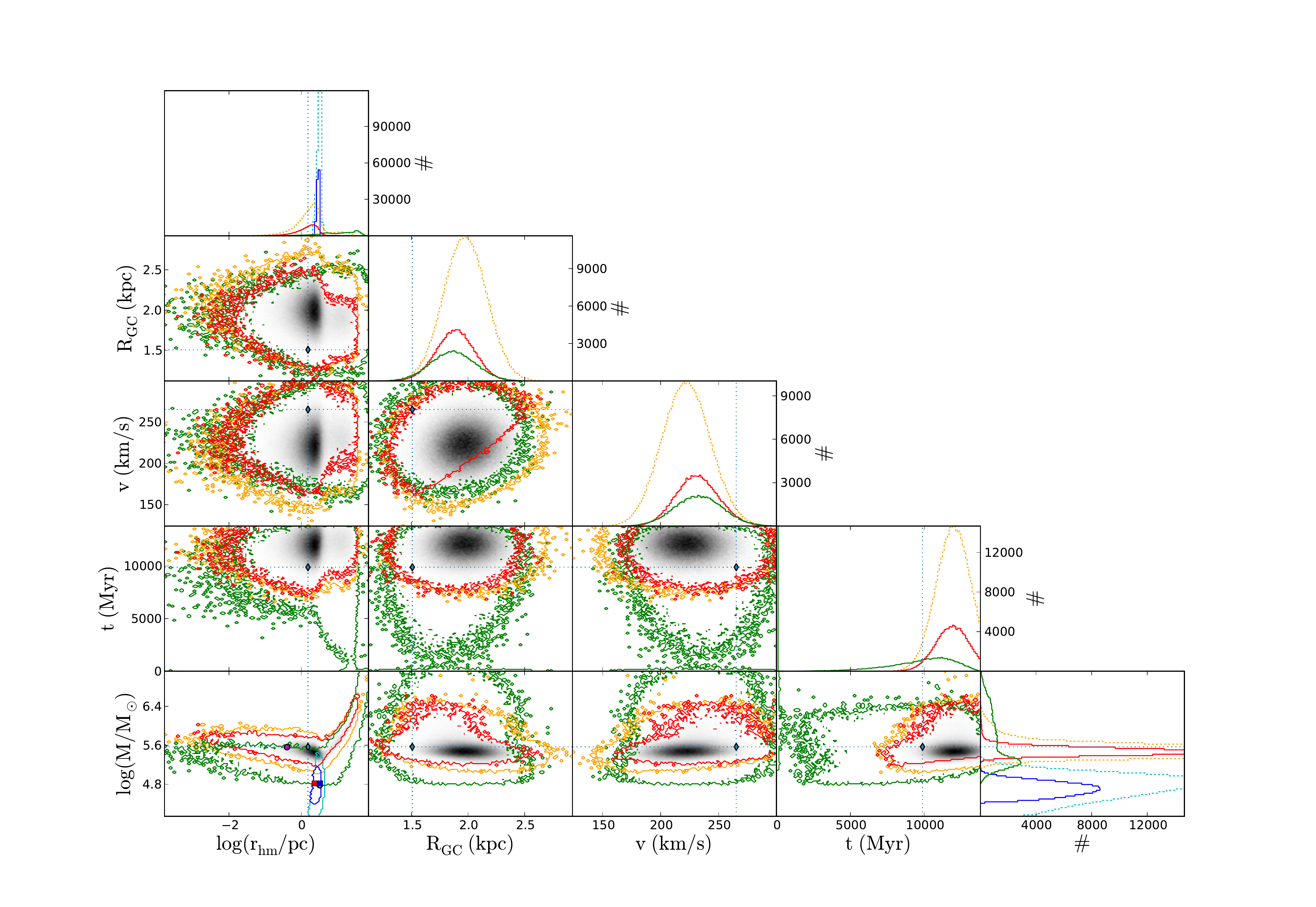}}
      \centering
      \caption{The same as Figure~\ref{fig:IC_and_FC_5D1}, but for the cluster NGC6397.\label{fig:IC_and_FC_5D2}}
\end{figure*}
\begin{figure*}
      \centerline{\includegraphics[width=215mm]{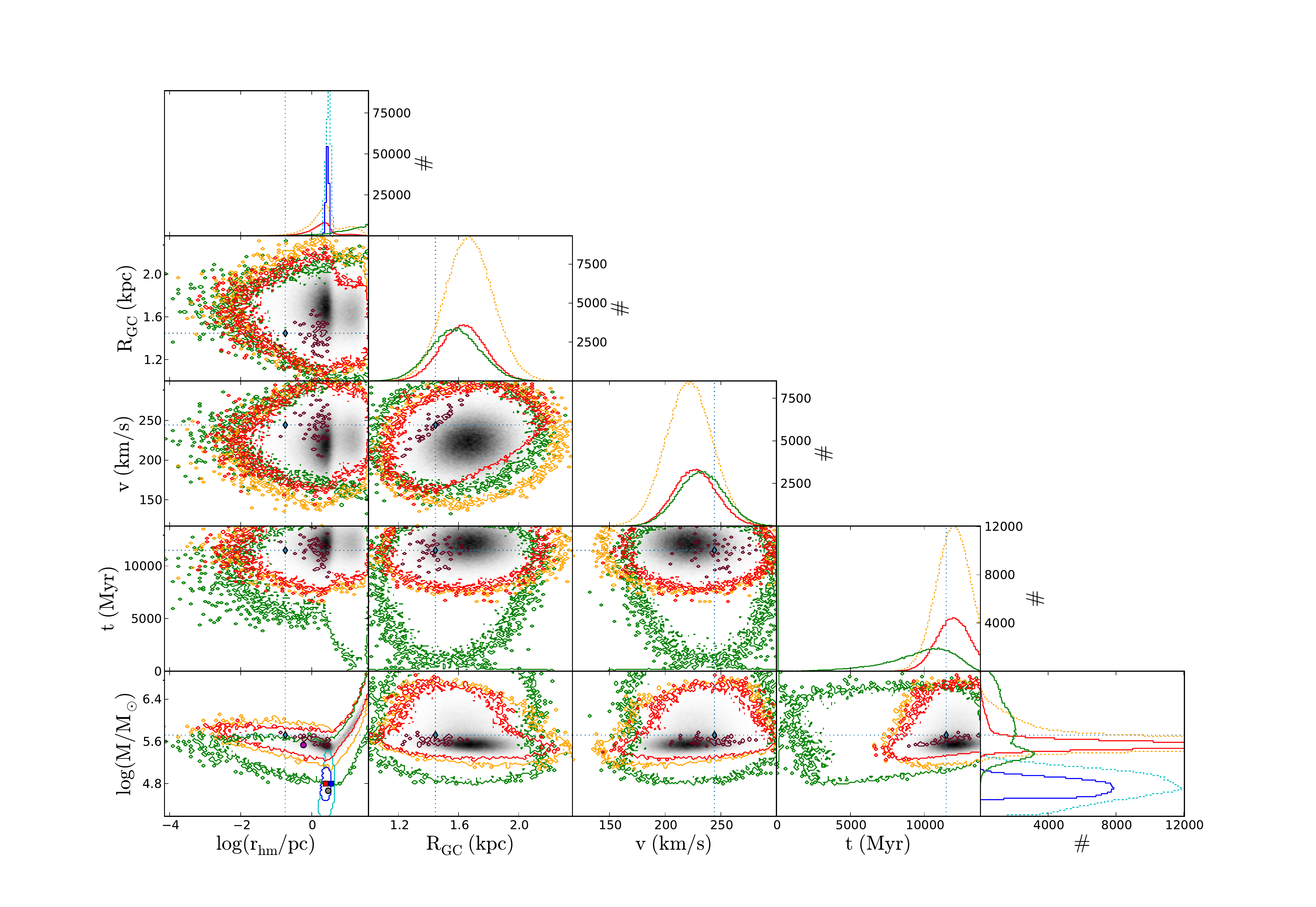}}
      \centering
      \caption{The same as Figure~\ref{fig:IC_and_FC_5D1}, but for the cluster M4.\label{fig:IC_and_FC_5D3}}
\end{figure*}
\begin{figure*}
      \centerline{\includegraphics[width=215mm]{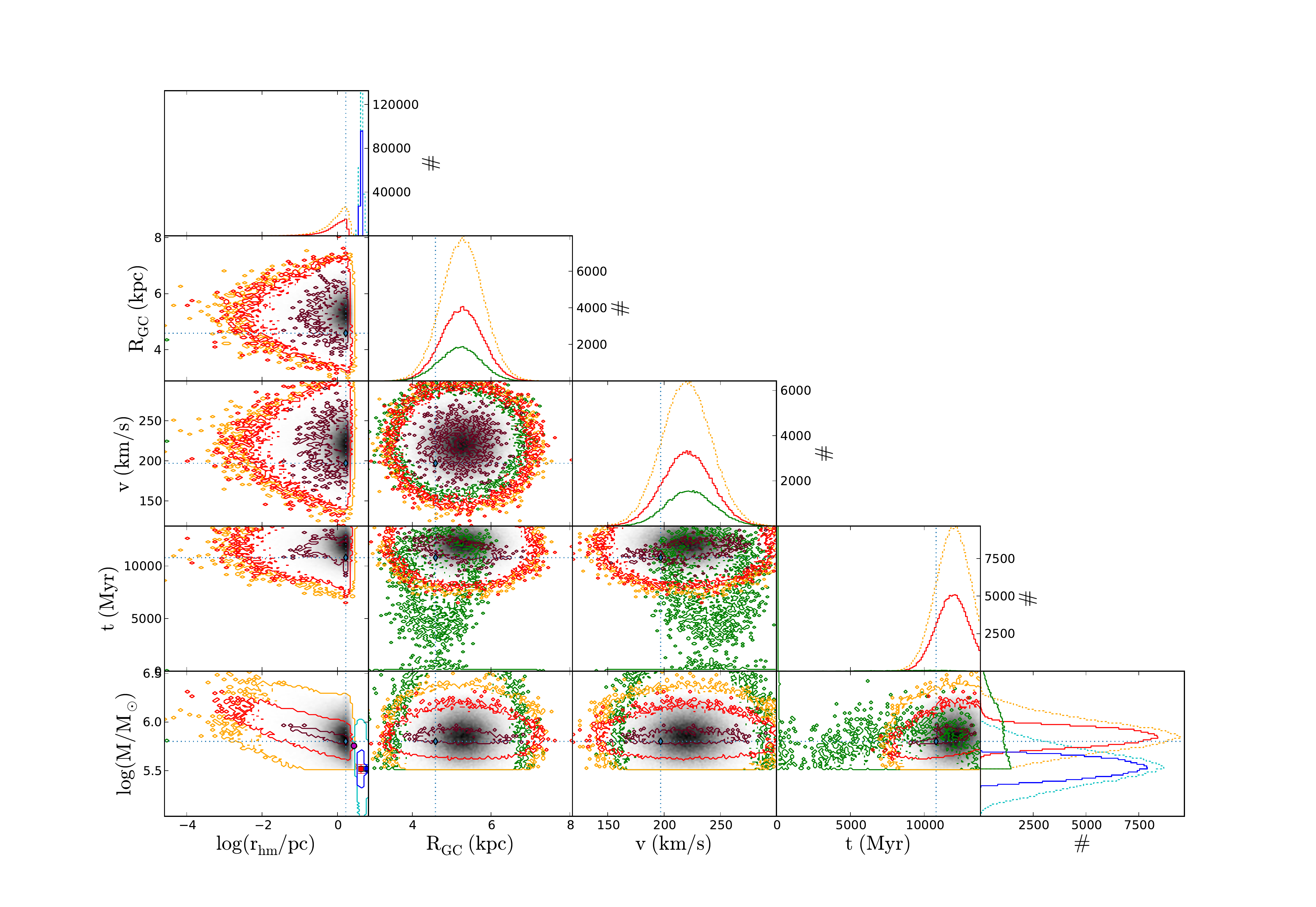}}
      \centering
      \caption{The same as Figure~\ref{fig:IC_and_FC1}, but for the cluster M22.\label{fig:IC_and_FC_5D4}}
\end{figure*}
\begin{figure*}
      \centerline{\includegraphics[width=215mm]{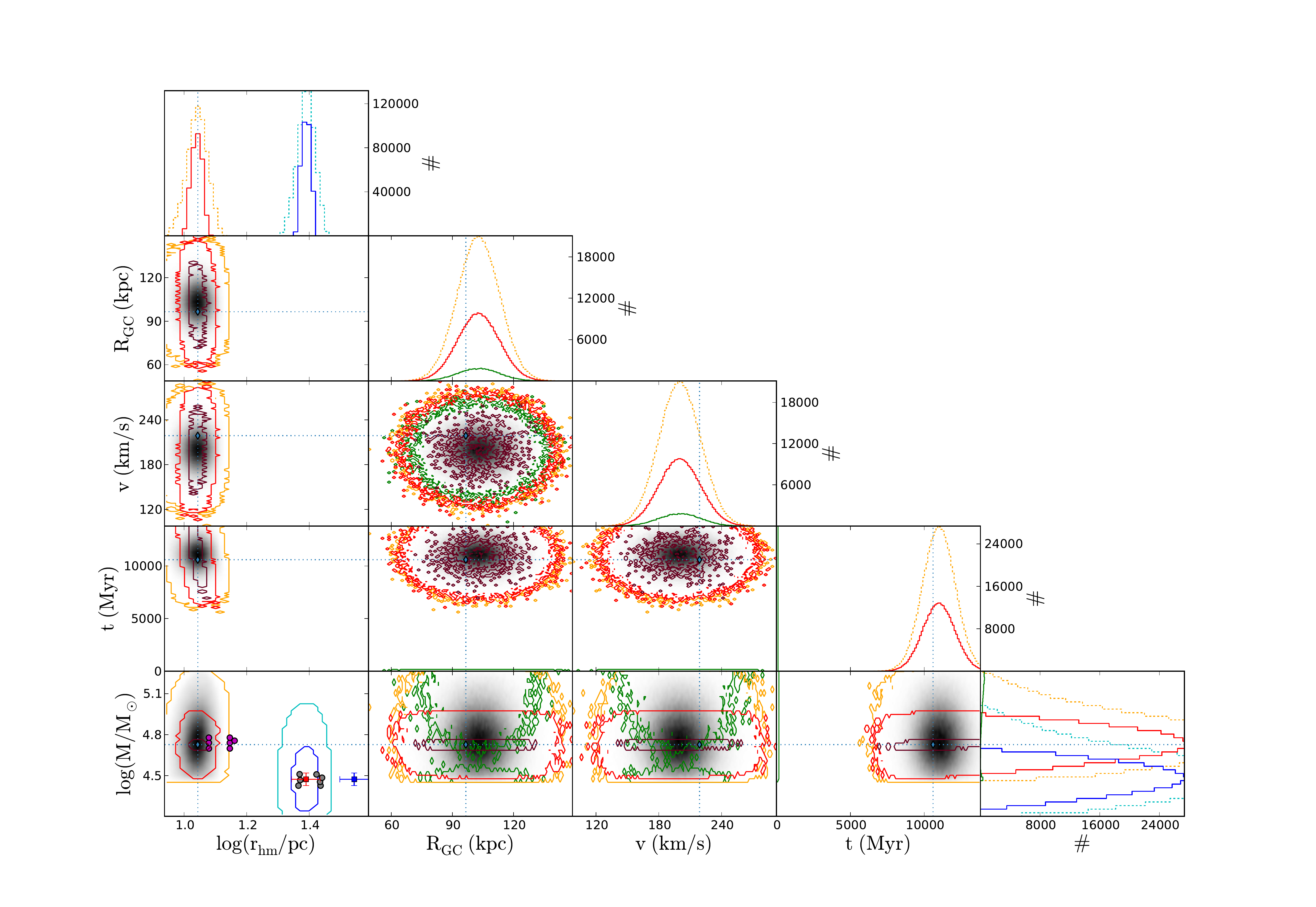}}
      \centering
      \caption{The same as Figure~\ref{fig:IC_and_FC_5D1}, but for the cluster Pal 4.\label{fig:IC_and_FC_5D5}}
\end{figure*}
\begin{figure*}
      \centerline{\includegraphics[width=215mm]{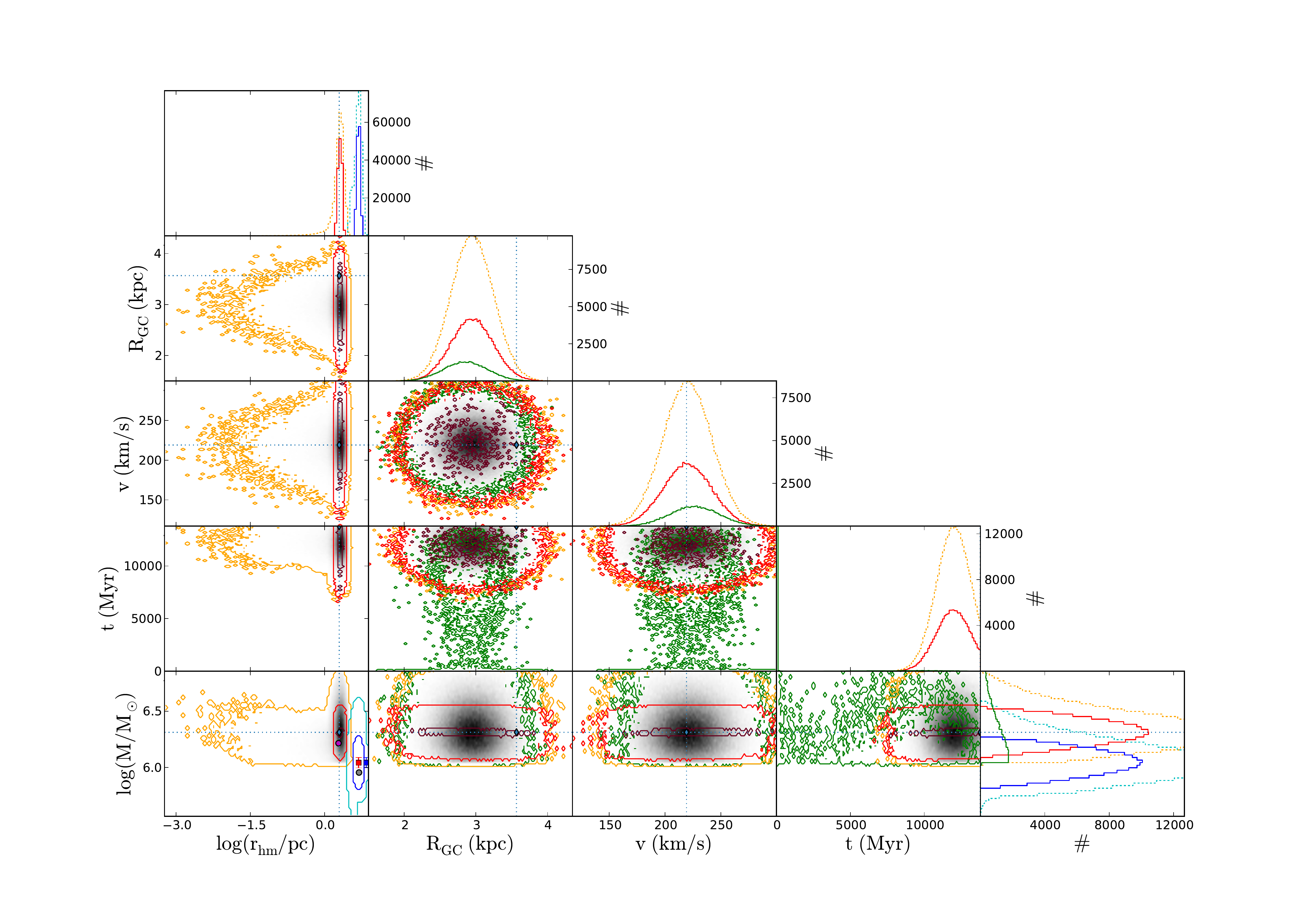}}
      \centering
      \caption{The same as Figure~\ref{fig:IC_and_FC_5D1}, but for the cluster 47 Tuc.\label{fig:IC_and_FC_5D6}}
\end{figure*}
\begin{figure*}
      \centerline{\includegraphics[width=215mm]{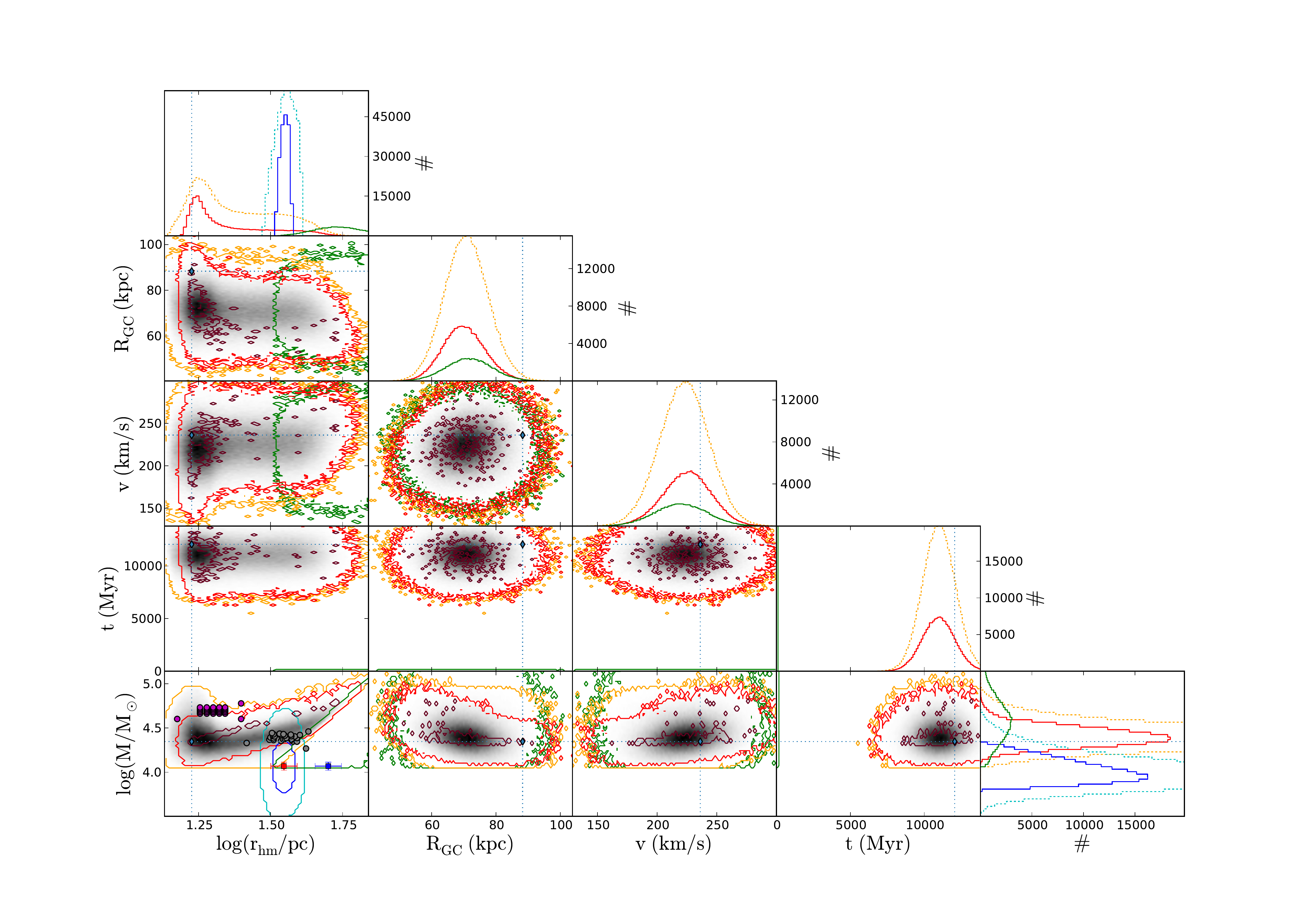}}
      \centering
      \caption{The same as Figure~\ref{fig:IC_and_FC_5D1}, but for the cluster Pal 14.\label{fig:IC_and_FC_5D7}}
\end{figure*}
\begin{figure*}
      \centerline{\includegraphics[width=215mm]{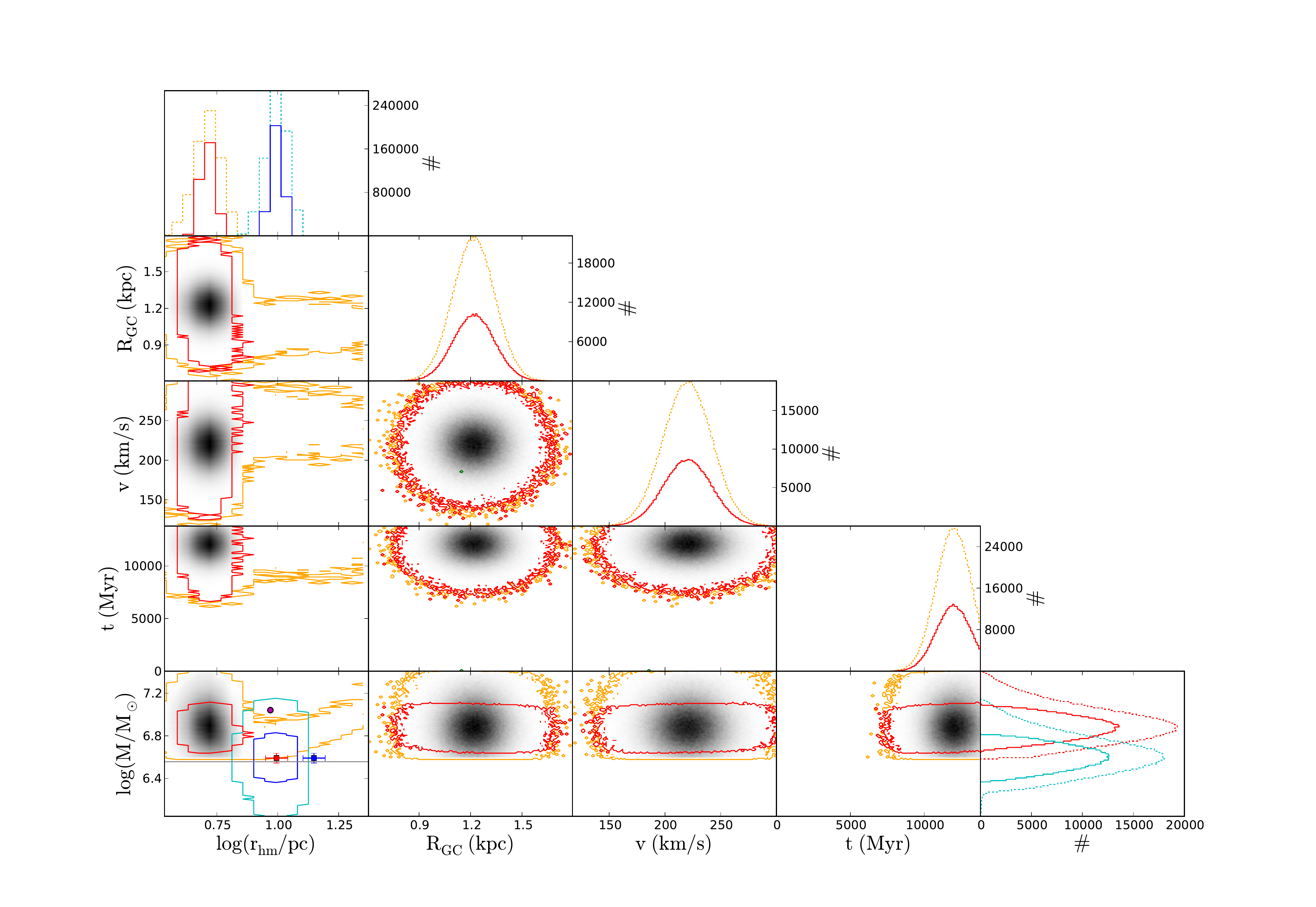}}
      \centering
      \caption{The same as Figure~\ref{fig:IC_and_FC_5D1}, but for the cluster $\omega$ Cen.\label{fig:IC_and_FC_5D8}}
\end{figure*}
\begin{figure*}
      \centerline{\includegraphics[width=215mm]{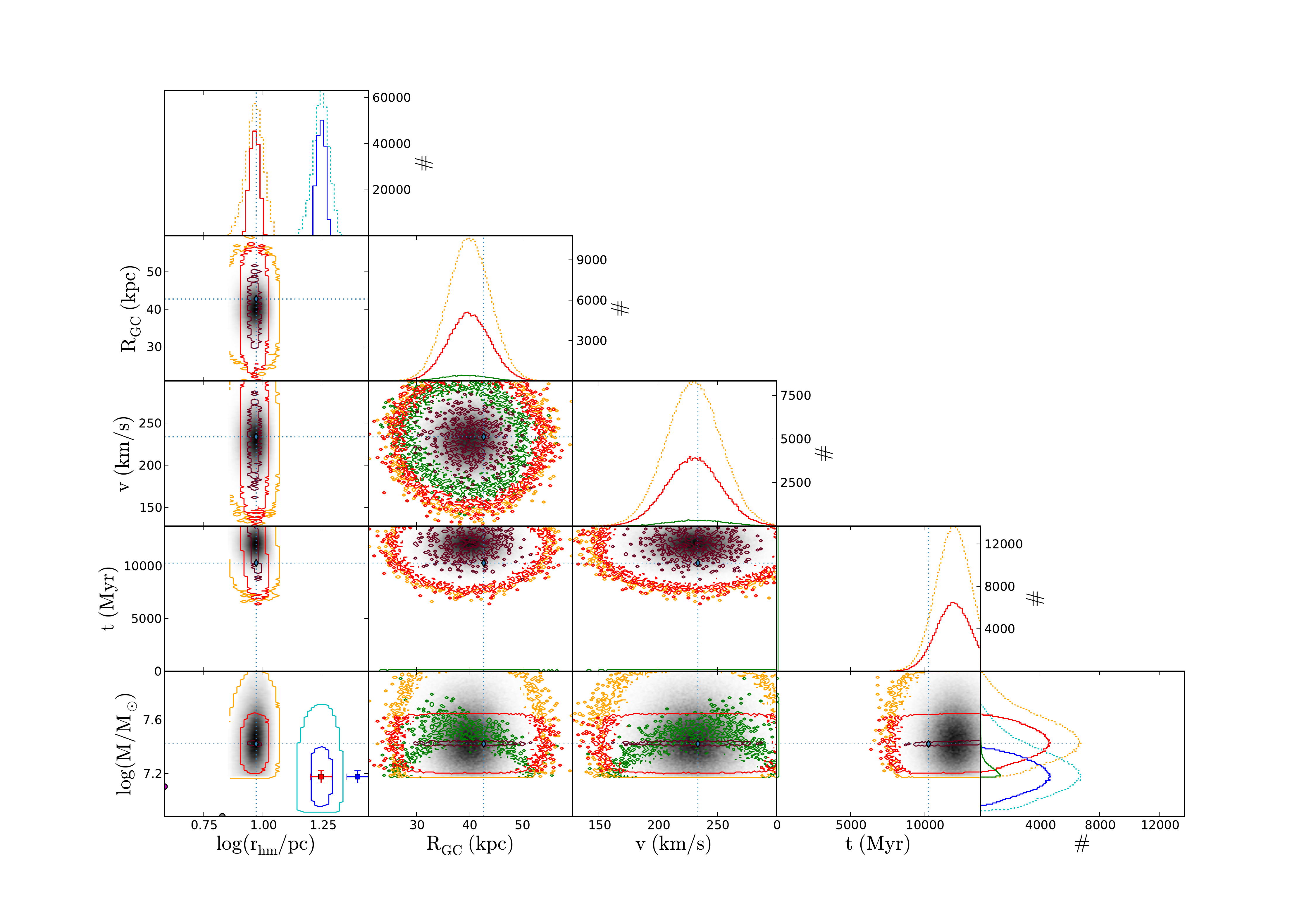}}
      \centering
      \caption{The same as Figure~\ref{fig:IC_and_FC_5D1}, but for the cluster G1.\label{fig:IC_and_FC_5D9}}
\end{figure*}

\bibliographystyle{mn2e}
\bibliography{emacss}

\label{lastpage}
\end{document}